%% file: paper.tex
\documentclass[12pt,letterpaper]{article}  
\pdfoutput=1 
\usepackage{jheppub}       

\usepackage{amsthm}
\usepackage{amsmath,soul}        
\usepackage{enumerate}  
\usepackage{graphicx}       
\usepackage{color}
\usepackage{float}
\usepackage{amsfonts}
\usepackage{caption}
\usepackage[dvipsnames]{xcolor} 
\usepackage{subcaption} 

\newcommand{\eg}{{\it e.g.,}\ } 
\newcommand{\ie}{{\it i.e.,}\ }
\newcommand{\reef}[1]{(\ref{#1})}
\newcommand{\beq}{\begin{equation}}
\newcommand{\eeq}{\end{equation}}
\newcommand{\beqa}{\begin{eqnarray}}
\newcommand{\eeqa}{\end{eqnarray}} 
\newcommand{\beqar}{\begin{eqnarray*}}
\newcommand{\eeqar}{\end{eqnarray*}}
\newcommand{\ssc}{\scriptscriptstyle}
\newcommand{\bbb}{\phi_{0}}
\newcommand{\aaa}{\phi_{1}}

\newcommand{\hyper}{{}_2\text{F}_1}  
\newcommand{\hyperf}[4]{{}_2\text{F}_1\!\left( #1 , #2 \,\text{;}\, #3 \,\text{;}\, #4 \right)}
\newcommand{\bspliteq}{\begin{equation}\begin{split}}
\newcommand{\espliteq}{\end{split}\end{equation}}
\newcommand{\comment}[1]{\textcolor{red}{\bf [[[#1]]]}}
\newcommand{\blue}[1]{\textcolor{blue}{\{\{#1\}\}}}  
  
\newcommand{\todd}[1]{\textcolor{BrickRed}{[[#1]]}}
\newcommand{\pl}{\ell_p}

\newcommand{\bvect}[1]{\boldsymbol #1}
\newcommand{\freq}{\mathfrak w}
\newcommand{\labell}[1]{\label{#1}}  

\newcommand{\bs}{\boldsymbol}
\newcommand{\mcO}{\mathcal{O}}
\newcommand{\cO}{\mathcal{O}}

\newcommand{\tij}{I_{\mu\nu}}
\newcommand{\bthesis}{\iffalse}

\newcommand{\alphamax}{\alpha_2}
\newcommand{\alphasource}{\alpha_1}
\newcommand{\up}[1]{^{\scriptscriptstyle(#1)}}

\newcommand{\W}{\Omega_n}
\newcommand{\cjjo}{C_{JJO}}
\newcommand{\join}[1]{:\! #1\! :}
\newcommand{\sing}{{\big|_{\text{sing.}}}}

\newif\ifthesis
\thesisfalse
\newif\ifblue
\bluetrue
\title{A Holographic Model For Quantum Critical Responses}{\tiny }

\author[a]{Robert C. Myers,}
\author[a,b]{Todd Sierens}
\author[c]{and William Witczak-Krempa}

\affiliation[a]{Perimeter Institute for Theoretical Physics,\\
Waterloo, Ontario N2L 2Y5, Canada}
\affiliation[b]{Department of Physics \& Astronomy and
Guelph-Waterloo Physics Institute,\\
University of Waterloo, Waterloo, Ontario N2L 3G1, Canada}
\affiliation[c]{Department of Physics, Harvard University,\\ Cambridge, MA 02138, USA}

\emailAdd{rmyers@perimeterinstitute.ca}
\emailAdd{tsierens@perimeterinstitute.ca}
\emailAdd{wkrempa@physics.harvard.edu} 

\abstract{
We analyze the dynamical response functions of strongly interacting quantum critical states described by conformal field theories (CFTs). We construct a self-consistent holographic model that
incorporates the relevant scalar operator driving the quantum critical phase 
transition. Focusing on the finite temperature dynamical conductivity $\sigma(\omega,T)$, we study its 
dependence on our model parameters, notably the scaling dimension of the relevant operator.
It is found that the conductivity is well-approximated by a simple ansatz proposed in \cite{katz} 
for a wide range of parameters. We further dissect the conductivity at  
large frequencies $\omega\gg T$ 
using the operator product expansion, and show how it reveals the spectrum of our model CFT.
Our results provide a physically-constrained framework to study the analytic continuation of 
quantum Monte Carlo data, as we illustrate using the O(2) Wilson-Fisher CFT. Finally,
we comment on the variation of the conductivity as we tune away from the quantum critical point,
setting the stage for a comprehensive analysis of the phase diagram near the transition.
}

\begin{document}

\maketitle
%\newpage

\section{Introduction}
A quantum critical (QC) system can be broadly defined as a quantum many-body system with a gapless energy spectrum,
and generically taken to be interacting. Some of the best understood instances are described 
by conformal field theories (CFTs).  
A canonical example of a CFT is the QC phase transition at zero temperature in the quantum Ising model
in 1+1 or 2+1 spacetime dimensions \cite{book}, 
which results from tuning the transverse magnetic field across a critical value.    
QC systems essentially come in two flavors: QC phase transitions or QC phases.
The former fundamentally necessitate tuning, such as the QC point in the quantum Ising model 
which results from tuning the transverse magnetic field across a critical value. In contrast,
a QC phase exists without fine-tuning. A simple example is a two-component Dirac fermion in 2+1 dimensions. 
A mass term breaks time-reversal symmetry and is thus forbidden if we demand that the symmetry be preserved.
(One could turn on a chemical potential to obtain a metal but this is not the type of tuning we are referring to,
as we shall see). In contrast, the mass term $\varphi^2$ of the scalar $\varphi^4$-theory, describing the QC Ising
transition, is invariant under all the symmetries of the theory and thus needs to be fine-tuned
to reach the quantum phase transition point. 
\begin{figure}   
\centering
\includegraphics[scale=.44]{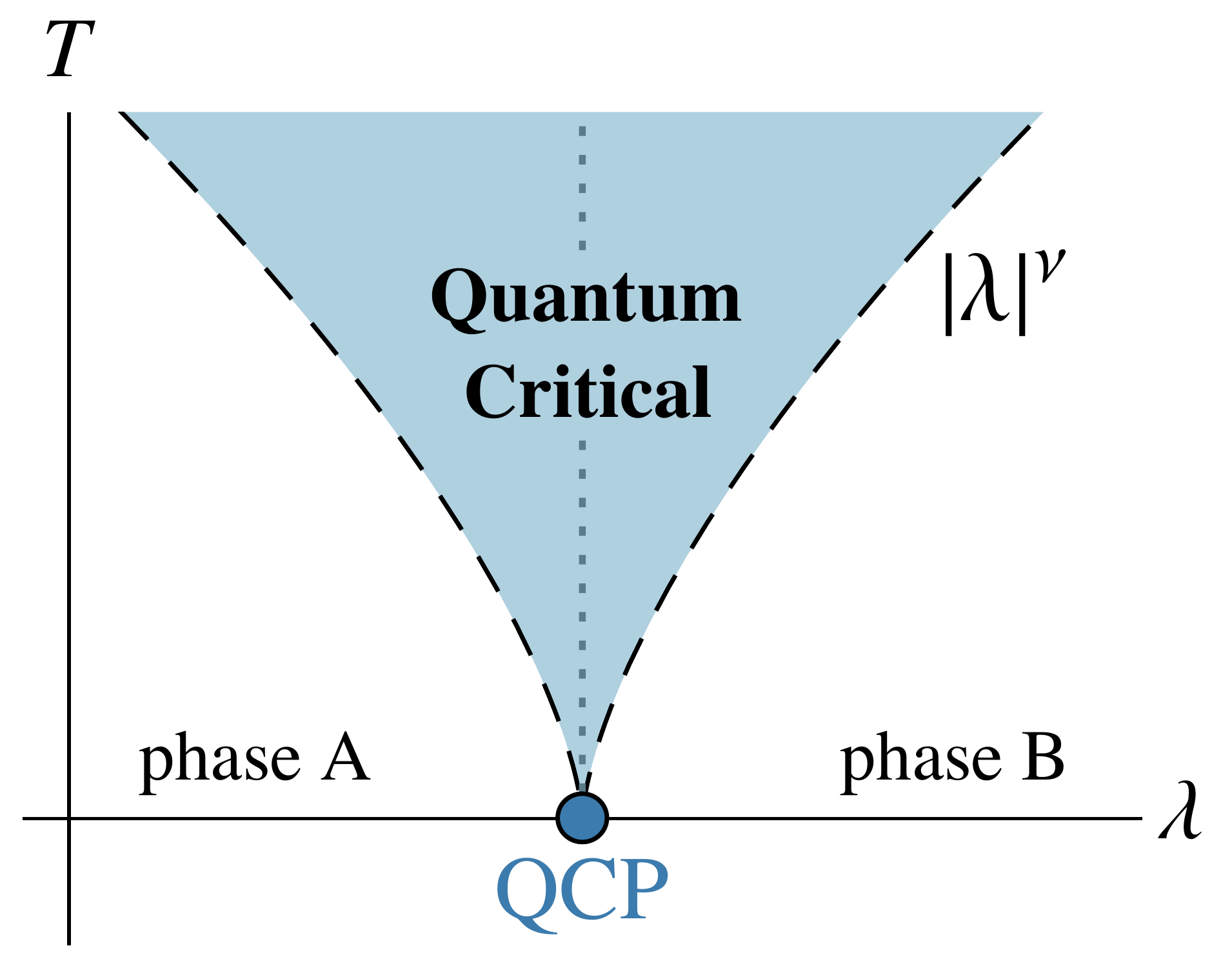}     
\caption{\label{fig:1} Phase diagram near a quantum critical point (QCP).
The physics in the shaded region (``fan'') is dominated by the thermally excited theory of the QCP.
The transition is driven by a relevant operator with coupling $\lambda$ and scaling dimension $\Delta=d-1/\nu$, where $d$ is the spacetime dimension, and $\nu$ the ``correlation length'' critical exponent. This paper mainly focuses on the $\lambda\!=\!0$ 
line (dotted); for detuning effects see section \ref{non critical} and fig.~\ref{critical cond}. 
}   
\end{figure}    
  
An important challenge in the study of CFTs/QC systems is to understand their real-time
dynamics \cite{fisher1990}, especially at finite temperature \cite{damle}. In the linear response regime, important examples   
are the frequency-dependent conductivity $\sigma(\omega)$ and dynamical shear viscosity $\eta(\omega)$.
Because the corresponding theories are strongly interacting, perturbative QFT methods are of limited use 
in analyzing the dynamics. At the same time, nonperturbative
quantum Monte Carlo simulations suffer from the perennial problem of analytically continuing
Euclidean data to real time. In contrast, holography yields real-time results for
strongly interacting systems lacking quasiparticles. However, in the context
where the duality is best understood, these CFTs correspond to large-$N$ gauge theories \cite{Maldacena}.
It is thus important to identify which of their dynamical properties are generic,
and which are special to the holographic regime.   

Progress in applying holography and general non-perturbative CFT methods to these questions was recently made 
in \cite{selfdual,Myers:2010pk,sum-rules,ws,ws2,natphys,katz,chen,will-hd}. For instance, new sum rules for the dynamical conductivity  
of (conformal) QC systems were first discovered using holography \cite{sum-rules,ws,ws2} 
(see also \cite{justin2} in the context of doped holographic SCFTs), 
and subsequently proved for a large class of CFTs \cite{katz}, including the Wilson-Fisher CFTs.    
Further, references \cite{natphys,katz} constructed holographic models which allowed comparison with 
quantum Monte Carlo (QMC) results for the dynamical conductivity in the O(2) Wilson-Fisher fixed-point theory. In particular, ref.~\cite{katz} recognized that the relevant scalar operator that needs         
to be tuned to reach the QC phase transition plays an important role in determining the dynamics. 
Hence the holographic studies in \cite{katz} incorporated this operator in an essential way. However, a shortcoming of their construction was that the dual of the relevant operator in the boundary theory was not  incorporated as a dynamical field in the bulk gravity theory. Our primary goal in this paper then is to construct a new holographic model where the relevant boundary operator is incorporated in a self-consistent way. The key feature, which distinguishes our holographic model from previous models, is that it incorporates a natural bulk interaction which ensures that the relevant operator acquires a thermal expectation value. Further, as shown in figure \ref{variousX}, it allows us to easily study the dynamical conductivity $\sigma(\omega)$ for a wide range of conformal dimensions $\Delta$ and of the two holographic parameters, $\alphasource$ and $\alphamax$ (which are proportional to the OPE coefficients, $C_{TTO}$ and $C_{JJO}$, respectively -- see further explanation in section \ref{setup}).    
Our model also provides a holographic framework where we can examine the response functions as we tune away from the quantum critical point. 
Although we focus on the dynamical conductivity in 2+1 dimensions, our analysis can be extended to
treat other response functions, such as the shear viscosity $\eta(\omega)$, in arbitrary dimensions. 
\begin{figure}[!htb]
\centering
\begin{subfigure}{0.45\textwidth}{\centering
\includegraphics[width=\textwidth]{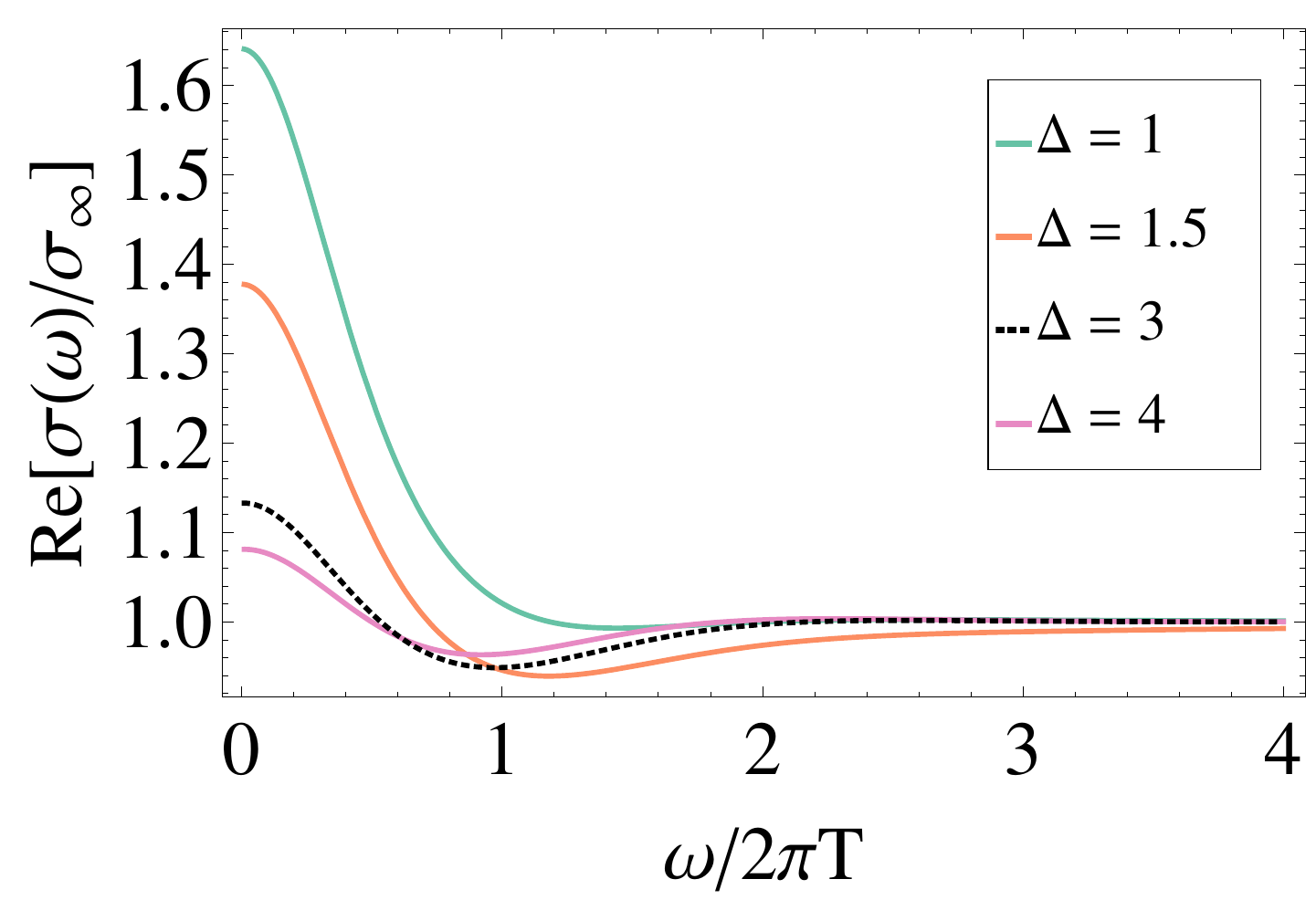}} 
\end{subfigure}\hspace{0.4cm} 
\begin{subfigure}{0.45\textwidth}{\centering
\includegraphics[width=\textwidth]{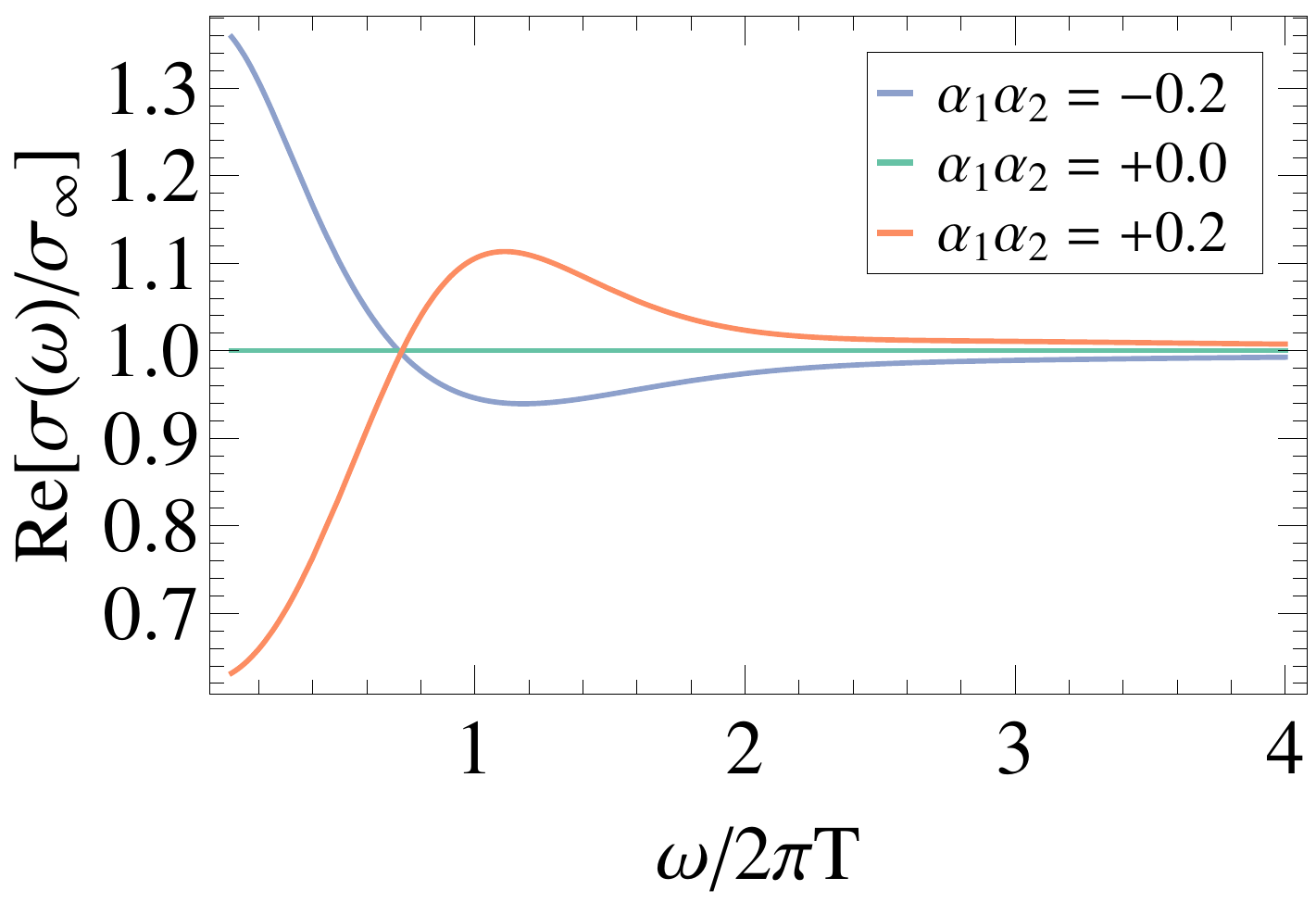}}
\end{subfigure} 
\caption{{\bf A demonstration of the holographic model}: real part of the conductivity as a function of frequency for various values 
of the scaling dimension of the scalar operator $\Delta$ with $\alphasource \alphamax=0.1$ (left), and for various choices of $\alphasource \alphamax$ with $\Delta=1.5$ (right). Note that $\alphasource,\,\alphamax$ are proportional to the OPE coefficients
$C_{TTO},\,C_{JJO}$, respectively, of the boundary CFT (see Table~\ref{table}). \labell{variousX}}   
\end{figure}   

The paper is organized as follows: in section \ref{sec:ingredients}, based on general CFT considerations, we present the key ingredients  
that a holographic model will need to describe QC response functions. In section \ref{setup}, we present our holographic
model and focus on evaluating the dynamical conductivity. We then analyze in detail the large-frequency asymptotics of the conductivity 
in section \ref{main}, and compare the results with those predicted by the operator
product expansion (OPE). We conclude in section \ref{discuss} with a brief discussion of our results and we also make some preliminary comments on the behaviour of the boundary theory when we detuned away from the QCP. This paves the way for the holographic study of observables in the entire phase diagram surrounding a QCP. We have four appendices to discuss certain technical details: Appendix \ref{oo sec} provides the details of calculating various vacuum correlators in the boundary CFT, which are used in section \ref{main}.  Appendix \ref{holosig}
describes some of the details for the calculation of the dynamical conductivity $\sigma(\omega)$ made in section \ref{hsigma}. In appendix \ref{special}, we consider the bulk scalar profile and conductivity for special cases of the conformal dimension of the relevant operator. Appendix \ref{lock} extends the high frequency expansion of the conductivity in section \ref{freqy} to second order in the $\alphamax$ expansion.

\section{Required ingredients: CFT analysis} \labell{sec:ingredients}

In our holographic study, we will be mainly concerned with canonical QC phase transitions
described by CFTs. These are realized by tuning a (single) coupling $\lambda$ 
to a specific value, which will be zero here:  
\begin{align}
 S= S_{\rm CFT} + \lambda\int\! d^dx\ \mathcal O(x) \,,  \labell{punt}
\end{align}
where the local scalar operator $\mathcal O$ is relevant, \ie 
its scaling dimension satisfies $\Delta<d$. Unitarity also requires that $\Delta\ge(d-2)/2$.  
At this point, it may be useful to recall the action of the $\varphi^4$ QFT in $d=2+1$:
  \begin{align} \label{WF}
  S = \int\! d^3x \, \Big[ \partial_\mu\varphi\cdot\partial^\mu\varphi   + u\, (\varphi\!\cdot\!\varphi)^2\Big]
+ \lambda \int\! d^3 x\,\, \varphi\!\cdot\!\varphi  \,, 
\end{align}
where $\varphi_a(x)$ is a real $N_s$-component vector. For all $N_s$, the RG fixed point at finite interaction $u$
corresponds to a non-trivial CFT, often called the O($N_s$) Wilson-Fisher (quantum critical) fixed point.
For the case of a single real scalar, $N_s=1$, this critical point corresponds to the Ising CFT. 
The relevant scalar $\mathcal O\sim \varphi\!\cdot\!\varphi$ here is the mass operator, and $\lambda$ the corresponding 
coupling that needs to be tuned to zero to reach the QCP.
In general, $\mathcal O$ is an important operator in the  
spectrum, and it is not surprising that it plays a key role in determining the quantum dynamics of 
various observables. In our holographic model, we must include a scalar field $\phi$ in the bulk gravity theory to be dual to $\mathcal O$ in the boundary theory.

At finite temperature, $\mathcal O$ typically acquires an expectation value:  
\begin{align} 
  \langle \mathcal O\rangle_T = B\, T^\Delta\,,
\labell{OT}
\end{align} 
where $B$ is a pure number determined by CFT data (scaling dimensions and OPE coefficients). 
Of course, the expectation value (\ref{OT}) vanishes at zero temperature since, by definition,  $\mathcal O$ is
not sourced at the QCP. That is, at $\lambda=0$, the vacuum of the corresponding CFT contains no scales and so the expectation value of all operators must vanish. The Wilson-Fisher CFT described above provides a simple example with this behaviour,
with the mass operator $\mathcal O\sim \varphi\!\cdot\!\varphi$ acquiring an expectation value as 
shown in eq.~\reef{OT} at finite $T$ \cite{katz}. However, not all CFTs describe QCPs (by the present definition), since in some cases there is no 
relevant scalar that is invariant under the full symmetry group of the CFT.
An elementary example is the free Dirac fermion CFT, where the mass operator breaks time-reversal symmetry.
As a consequence, it does not acquire a thermal expectation value, \ie $\langle\bar\psi \psi\rangle_T=0$.
Symmetry requirements alone are sufficient to set the mass to zero, so that the Dirac CFT 
does not need to be fine-tuned, unlike (\ref{WF}), and the theory describes a quantum critical \emph{phase}
not a point. Typical holographic theories that have been studied up to this point do not exhibit the behaviour shown in eq.~\reef{OT}. Rather, at finite temperature, only the stress tensor acquires a nonvanishing expectation value  in these models. Hence, a key ingredient of our holographic model will be a natural mechanism which ensures that eq.~\reef{OT} holds.

Finally, the large-frequency/momentum structure of two-point correlation functions
is determined by the OPE of the corresponding operators \cite{willprl}. For example, the conductivity is determined by the current-current correlator and hence the large-frequency structure is given by the $JJ$ OPE.
In this context, the first non-trivial operator in the
$JJ$ OPE is the relevant scalar $\mathcal O$ \cite{katz}.  
Hence to study the conductivity, we first need to introduce a bulk gauge field in our holographic model to match the current in the boundary theory. Further, we will need include appropriate bulk interactions to realize the property that the OPE coefficient corresponding to the fusion $JJ\rightarrow \mathcal O$ is non-zero in the boundary theory. Alternatively, the vacuum three-point function $\langle J J \cO\rangle$ must be non-zero, as will be illustrated in section \ref{main}. 

\section{Holographic model}  \labell{setup} 

Here we describe an explicit holographic model with all of the ingredients described in the previous section. We will be focusing our attention on three-dimensional CFTs and so in the bulk, we begin with four-dimensional Einstein gravity coupled to a negative cosmological constant, 
\begin{equation}
S_\text 0 = \frac{1}{2 \pl^2}\int d^4 x \sqrt {-g}  \left( R + \frac{6}{L^2}\right)\,.
\labell{gamma0}
\end{equation}
Here, $\pl$ is the Planck length, which is related to Newton's gravitational constant by $\pl^2 = 8 \pi G$. The vacuum solution is then simply the anti-de Sitter (AdS) geometry with the curvature scale $L$.  The ratio of these two scales determines the central charge of the boundary CFT, \eg see \cite{airport}: $C_T = \frac{24}{\pi^2}\,\frac{L^2}{\pl^2}$. Another useful solution, which will set the background geometry for our calculations, is the planar black hole:
\begin{equation} \label{metric1}
ds^2 = \frac{r^2}{L^2}\left(-f(r)dt^2 + dx^2 + dy^2 \right) + \frac{L^2 dr^2}{r^2 f(r)}\,,
\end{equation} 
with $f(r) = 1 - \frac{r_0^3}{r^3}$. The position of the event horizon is $r = r_0$ and taking $r_0\rightarrow 0$ yields the familiar Poincar\'e patch of AdS space. According to the usual AdS/CFT correspondence, this solution (\ref{metric1}) is dual to the CFT at finite temperature (and zero chemical potential), where the temperature is given by 
\begin{equation}
T = \frac{3 r_0}{4 \pi L^2}\,.\labell{temp}
\end{equation}
It will simplify our calculations to change to a dimensionless radial coordinate $u = r_0/r$, with which the metric becomes
\begin{equation}
ds^2 = \frac{r_0^2}{L^2 u^2} \left( -f(u) dt^2 + dx^2 + dy^2\right) + \frac{L^2 du^2} {u^2 f(u)}\,,\labell{metric}
\end{equation}
where $f(u) = 1- u^3$. In these coordinates, $u \rightarrow 0$ corresponds to the asymptotic AdS boundary and $u=1$ is the black hole horizon.

\begin{table}
  \begin{center}  
    \begin{tabular}{c|c|c|c}
      Bulk coupling & Bulk operator  & CFT correlator ($T\!=\!0$) & Observable \\
      \hline
      $L^2/\pl^2$ & {$R$}  & $\langle T_{\mu\nu}\, T_{\rho\delta} \rangle$ & $C_T$ \\ %,\ $\langle\cO\cO\rangle$\\
      \hline
      $1/g_4^2$ & $F_{ab}F^{ab}$  & $\langle J_\mu\, J_\nu \rangle$ & $\sigma_\infty$  \\
      \hline
    {$m^2L^2$} & $\phi^2$ & $\langle \cO\,\cO \rangle$ & $\Delta$ \\
      \hline\hline
      $\alphasource$ & $\phi\, C_{abcd}C^{abcd}$  & $\langle  T_{\mu\nu} T_{\rho\delta}\, \cO \rangle$ & $C_{TTO}$ \\
      \hline
      $\alphamax$ & $\phi\, F_{ab}F^{ab}$  & $\langle J_\mu J_\nu \,\cO \rangle$ & $C_{JJO}$
    \end{tabular}
  \end{center}
  \caption{The five dimensionless parameters which characterize the bulk gravity theory and the dual correlators in the boundary CFT which they control --- see appendix \ref{oo sec}.$^1$ 
\labell{table}}
\end{table}
\setcounter{footnote}{1}
\footnotetext{Note that the normalization of two-point function $\langle \mcO \mcO\rangle$ is also fixed by $C_T\propto L^2/\pl^2$.}
To ensure that the boundary CFT also contains a (conserved) current $J_\mu$ and a scalar operator $\cO$ with conformal dimension $\Delta$, we introduce the following bulk actions for a (massless) gauge field $A_a$ and a scalar field $\phi$ with mass $m^2 L^2 = \Delta (\Delta-3)$:\footnote{Latin (Greek) indices are used to indicate Lorentz vector or tensor quantities in the bulk (boundary).}
\begin{align}
  S_\phi &= -\frac{1}{2\pl^2} \int d^4x \sqrt{-g}\, \Big[ (\nabla_a\phi)^2 +m^2 \phi^2 -2\,\alpha_1\, L^2\phi\, C_{abcd}C^{abcd} \Big]\,, \labell{scalarX}\\ 
  S_A &= -\frac{1}{4g_4^2}\int d^4x \sqrt{-g}\, \Big(1+\alpha_2\phi \Big)\,F_{ab}F^{ab}\,, \labell{gaugeX}
\end{align}
where $F_{ab}$ is the field strength of $A_a$, and $C_{abcd}$ is the Weyl curvature tensor.
The scalar action \reef{scalarX} is normalized with a factor of $1/\pl^2$ to ensure that the scalar field $\phi$ is dimensionless, which will be convenient in the following calculations. The gauge field $A_a$ has the usual dimension of inverse length and so the  Maxwell coupling $g_4$ is dimensionless.
The scaling dimension $\Delta$ is taken to be above the unitary bound for $2+1$ dimensional CFTs, 
$\Delta_{\rm min}=1/2$. We further note that in the range $1/2\leq \Delta < 3/2$, the theory
will contain at least one other relevant scalar, which can be thought of as $\cO^2$. In this regime, 
the CFT dual thus describes a \emph{multicritical} point instead of a simple critical point; we refer
the reader to section \ref{finger} for further details.   
%rcm comment for point 2
Further, although our motivation in the previous section considered relevant operators with $\Delta<3$, the  following holographic analysis easily extends to irrelevant operators with $\Delta>3$ as well. However, certain technical issues arise for  $\Delta\ge 6$ --- see further comments in footnotes \ref{footy7} and appendix \ref{special}.

If we supplement eq.~\reef{gamma0} with the free actions in eqs.~\reef{scalarX} and \reef{gaugeX}, \ie with $\alpha_1=0=\alpha_2$, a thermal state (with vanishing chemical potential) in the boundary CFT is still described by the above black hole solution \reef{metric}. In particular, $\phi$ and $A_a$ would both vanish in the bulk
%rcm no hair footnote for point 1
solution.\footnote{The gauge field vanishes because we have assumed that the black hole is not charged, \ie the chemical potential vanishes in the boundary theory. If bulk scalar has a positive mass-squared, \ie $\Delta>3$, there are no hair theorems which ensure that $\phi$ vanishes, \eg \cite{Torii:2001pg}. However, with a negative mass-squared, \ie $\Delta<3$, stable black hole solutions can be found with nontrivial scalar hair, \eg \cite{Torii:2001pg,Winstanley:2002jt,Buchel:2007vy,Buchel:2013lla}. However, from a holographic perspective, the latter solutions involve turning on the (dimensionful) coupling constant for the corresponding operator in the boundary theory \eg \cite{Buchel:2007vy,Buchel:2013lla}. However, as explained below, we wish to focus on the critical theory in which this coupling vanishes and so we impose boundary conditions where the only black hole solutions have vanishing $\phi$ for
the free theory.} However, a key ingredient, which we wanted to include in our holographic model, is that the scalar operator should acquire a nonvanishing thermal expectation value. Therefore the dual scalar $\phi$ must be sourced to have a nontrivial profile in the black hole background. The latter is engineered by adding the new interaction in eq.~\reef{scalarX} which couples the scalar field to the Weyl curvature. The Weyl curvature vanishes in the vacuum AdS geometry since the latter is conformally flat and hence the vacuum of the boundary CFT remains stable. However, $C_{abcd}C^{abcd}$ provides a nontrivial source for the scalar in the black hole background \reef{metric} and as desired then,  $\langle \mathcal O \rangle_T\ne 0$ in the CFT. We show in appendix \ref{oo sec} that the (dimensionless) coupling $\alphasource$ is related to the CFT parameter controlling the vacuum three-point function $\langle TT \cO\rangle$. %$\langle T_{\mu\nu}T_{\kappa\sigma}\,\mathcal O\rangle$. 

Lastly, as described above, the three-point function $\langle JJ \cO\rangle$ %$\langle J_\mu J_\nu\, \mathcal O\rangle$ 
must be nonvanishing in the vacuum of the boundary theory. The simplest way to accomplish the latter is to add the $\phi\, F^2$ interaction in eq.~\reef{gaugeX}.  
The (dimensionless) coupling $\alphamax$ is then dual to the CFT parameter which controls the desired three-point function. The four dimensionless couplings which characterize the bulk gravitational theory and their role in the dual boundary CFT are summarized in table \ref{table}.

Now in principle, one would want to solve the full nonlinear equations of the total action to solve for a new black hole solution in which the scalar field has a nontrivial profile. However, in the present paper, we only approach this problem to leading order in a perturbative approach. In particular, we will construct the background perturbatively in the amplitude of the scalar field and in fact, we only perform the present calculations to leading order in this expansion. Alternatively, since the bulk scalar is sourced by the interaction in eq.~\reef{scalarX}, one can think that we are working to leading order in a small $\alphasource$ expansion.

Hence to leading order, the background geometry is given by eq.~\reef{metric}. Then from eq.~\reef{scalarX}, the scalar field equation becomes
\begin{equation}
\left(\nabla^2 - m^2\right)\phi + \alphasource\, L^2\, C_{abcd}C^{abcd}=0\,.\labell{scalar equation}
\end{equation}
Because the black hole background \reef{metric} is translation invariant in the boundary directions, $C_{abcd}C^{abcd}$ only depends on $u$. Hence we can solve eq.~\reef{scalar equation} with a simple ansatz $\phi=\phi(u)$, in which case the above equation reduces to 
\begin{equation}
u^4\,\partial_u\!\left(\frac{(1-u^3)}{u^2}\,\partial_u\phi(u)\right) + \Delta(3-\Delta)\,\phi(u) + 12 \alphasource u^6=0\,.\labell{scalar ode}
\end{equation}
This equation has an exact solution:\footnote{This representation of the solution is only valid for $\Delta<6$. In particular, the integral defining $g_\Delta(u)$ in eq.~\reef{solute} diverges for $\Delta\ge 6$ --- see further comments in appendix \ref{special}. Further, the two independent solutions presented in eq.~\reef{solution} are actually identical for  $\Delta=3/2$. Of course, the coefficients of $g_\Delta(u)$ and $h_\Delta(u)$ also diverge for this particular value of $\Delta$. The correct solution for $\Delta=3/2$ is presented in appendix \ref{special}. However, we note that the conductivity is still a smooth function of $\Delta$ at this special value and so where results are presented for $\Delta=3/2$ in the following, we have actually evaluated our expressions with a nearby value of the conformal dimension, \ie $\Delta=1.50001$. \labell{footy7}}
%todd fixed signs
\begin{equation}
\begin{split}
\phi(u) =& \hyperf{\frac\Delta 3}{\frac\Delta 3}{\frac{2\Delta} 3}{u^3}\ \left(\aaa - \frac{12\alphasource}{2\Delta-3} g_\Delta(u)\right)\,u^\Delta\\
&\quad+ \hyperf{1-\frac\Delta 3}{1-\frac\Delta 3}{2-\frac{2\Delta} 3}{u^3} \ \left(\bbb + \frac{12\alphasource}{2\Delta-3} h_\Delta(u)\right)\,u^{3-\Delta}\,, \labell{solution}
\end{split}
\end{equation}
where $\bbb$ and $\aaa$ are integration constants and $\hyper(z_1,z_2;z_3;z_4)$ denotes the standard hypergeometric function. Further,  $g_\Delta(u)$ and $h_\Delta(u)$ are given by 
\begin{equation}
\begin{split}
g_\Delta(u) = & \int_0^{u} dy\,y^{5-\Delta}\  \hyperf{1-\frac\Delta 3}{1-\frac\Delta 3}{2-\frac{2\Delta} 3}{y^3} \,, \\
h_\Delta(u) = & \int_0^{u} dy\,y^{2+\Delta}\ \hyperf{\frac\Delta 3}{\frac\Delta 3}{\frac{2\Delta} 3}{y^3} \,.
\labell{solute}
\end{split}
\end{equation}
%todd solution simplified (& footnote tweaked)
Given the definitions in eq.~(\ref{solute}), we have $g_\Delta(0) = h_\Delta(0) = 0$ at the AdS boundary. 

The above solution has the expected asymptotic behaviour for $u\to0$ with
\begin{equation}
\phi(u) = \bbb\, u^{3-\Delta}\, \Big(1+O(u^3)\Big)+\aaa\, u^\Delta\, \Big(1+O(u^3)\Big) \,. \labell{boots}
\end{equation}
Note that since we are using the dimensionless radial coordinate $u$ here, both of the coefficients, $\bbb$ and $\aaa$, are also dimensionless.
Recall the first term is the non-normalizable mode, and the coefficient $\bbb$ corresponds to the coupling $\lambda$ which deforms the boundary theory as in eq.~\reef{punt}. 
%rcm comment for first part of point 3
Hence, as is standard in the AdS/CFT correspondence, tuning this boundary condition for the bulk scalar field corresponds to tuning the dual coupling constant in the boundary field theory. In particular, 
we  set $\bbb=0$ since we want to study the behaviour of the critical theory (to compare to \cite{katz}).\footnote{This choice corresponds to the tuning needed to reach a QC phase transition discussed in section \ref{sec:ingredients}.  In section \ref{discuss}, we provide some preliminary remarks on tuning away from the critical point by choosing instead a nonvanishing value of $\bbb$, but leave this situation for detailed study in \cite{new}. 
%rcm more comment for point 2
Of course, setting $\bbb=0$ is also what allows us to consider irrelevant operators in the following. As is evident from eq.~\reef{boots}, the scalar would diverge near the boundary with $\bbb\ne0$ and $\Delta>3$ and hence its gravitational back-reaction would destroy the asymptotic AdS geometry.}
Further the second term in eq.~\reef{boots} corresponds to the normalizable mode, and the corresponding coefficient $\aaa$ is dual to the expectation value $\langle \mathcal O \rangle$. To fix this integration constant $\aaa$, we demand that the scalar field be regular at the horizon. As $u\to1$, the solution (\ref{solution}) has a (potential) logarithmic divergence 
%\begin{equation}
%\begin{split}
%\phi(u\to1) \rightarrow & \left(a + 4\alphasource g_\Delta(1)\right)\frac{\Gamma\left(\frac{2\Delta}{3}\right)}{\Gamma\left(\frac{\Delta}{3}\right)^2} \log(1-u^3)\\
%&-  4 \alphasource h_\Delta(1)\frac{\Gamma\left(2-\frac{2\Delta}{3}\right)}{\Gamma\left(1-\frac{\Delta}{3}\right)^2} \log(1-u^3)\,,
%\end{split}
%\end{equation}
which is eliminated by setting
%todd added factor -3/(2\Delta-3), fixed sign
\begin{equation}
\aaa = \alphasource\times \frac{12}{2\Delta -3}\left(g_\Delta(1) - \tfrac{\Gamma\left(2-\frac{2\Delta}{3}\right)\Gamma\left( \frac{\Delta}{3}\right)^2}{\Gamma\left(1-\frac{\Delta}{3}\right)^2\Gamma\left(\frac{2\Delta}{3}\right)}h_\Delta(1)\right)\,.
\labell{wake3}
\end{equation}
Note that $g_\Delta(1)$ and $h_\Delta(1)$ are both finite and can be determined by numerically evaluating the integrals in eq.~\reef{solute}. 

Figure \ref{scalar plots} shows the resulting scalar profiles for $\Delta =1.5$ and 4, in comparison to a simple power law $\phi(u) = \aaa\, u^\Delta$,  as used in \cite{katz}. The value of the coefficient $\aaa$ in the power-law profile was chosen to match that in the holographic solution so that the two profiles 
exactly agree as $u\to 0$. Then we find that for relevant operators (\ie $\Delta < 3$), the scalar profile produced by eq.~(\ref{solution}) is larger than the power-law profile in the vicinity of the horizon (\ie $u\to 1$). Further, the relative separation of the two profiles is increased as $\Delta$ is decreased (below 3). In contrast, for irrelevant operators (\ie $\Delta > 3$), the solution \reef{solution} is smaller than the power-law profile near the horizon. When the scalar operator is marginal (\ie $\Delta = 3$), in fact, the exact solution and the simple power-law are identical, so that $\phi(u)=\aaa u^3$ as shown in appendix \ref{marginal}.
\begin{figure}[!htb]
\centering 
\begin{subfigure}{0.45\textwidth}{\centering
\includegraphics[width=\textwidth]{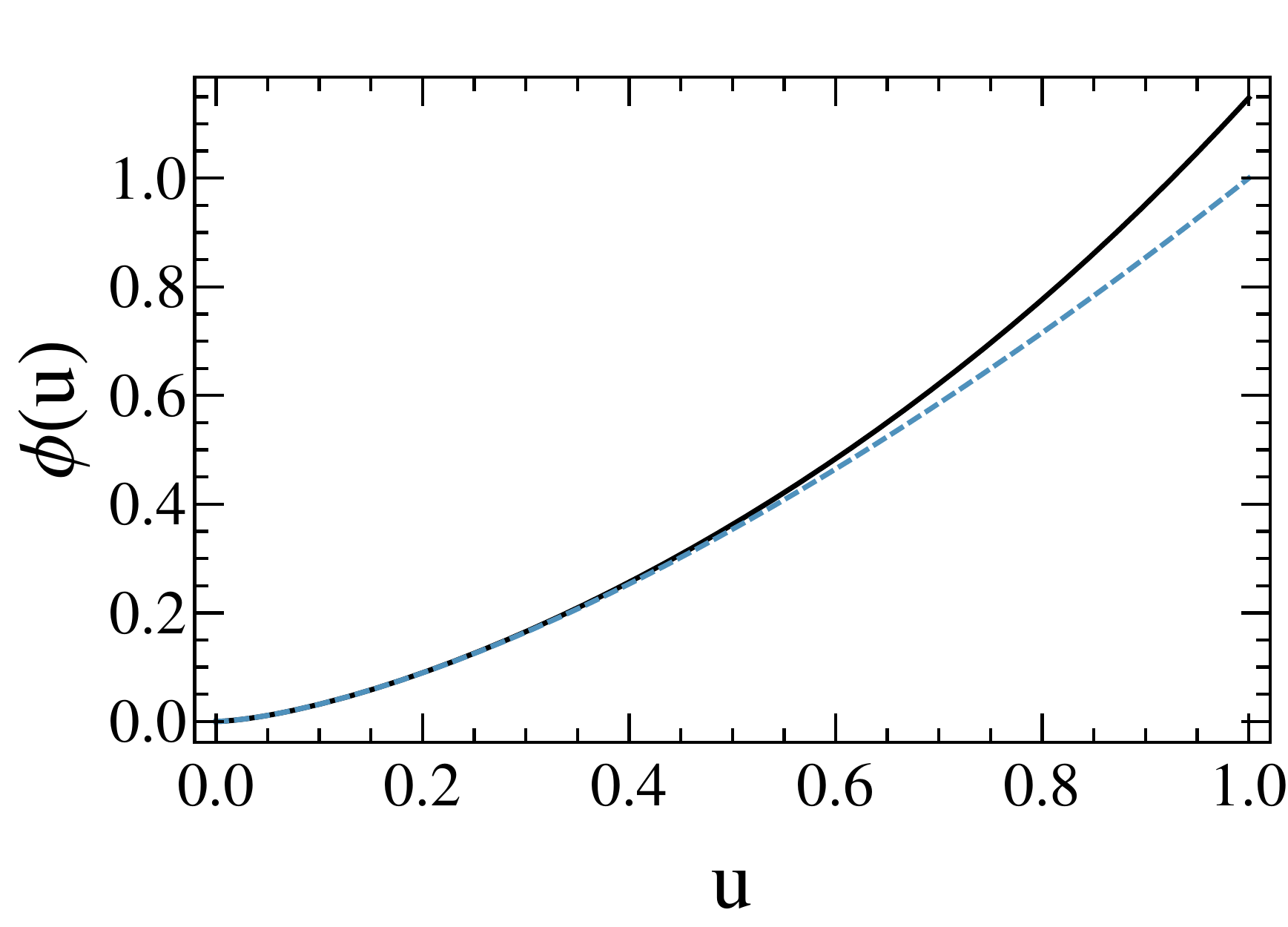}} 
\end{subfigure}
\begin{subfigure}{0.45\textwidth}{\centering
\includegraphics[width=\textwidth]{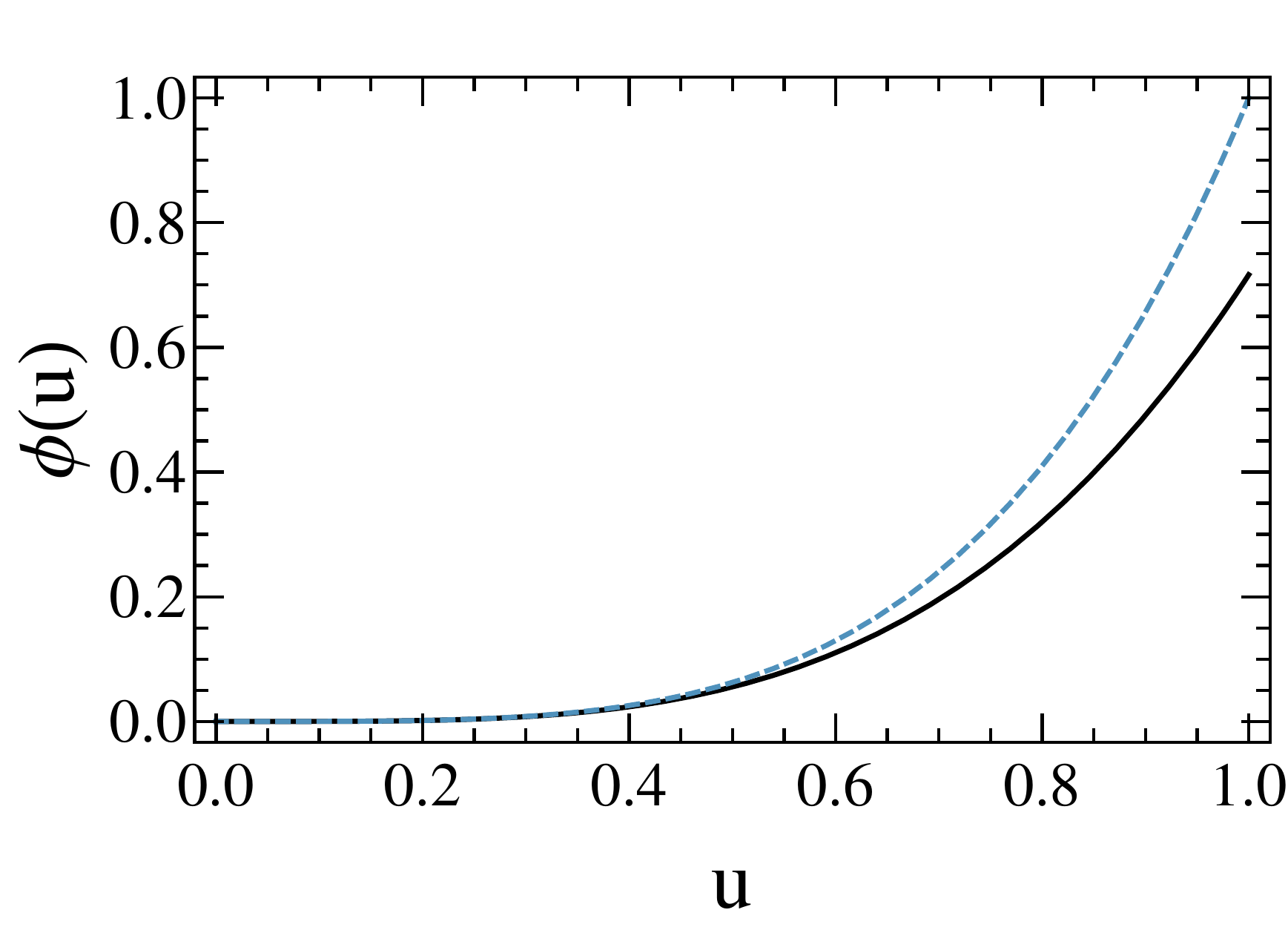}}
\end{subfigure}
\caption{The scalar profile $\phi(u)$ for $\Delta=1.5$ (left) and $\Delta = 4$ (right). The solid black line is the exact solution, while the dashed blue line is the power-law profile $\phi(u) = \aaa\, u^\Delta$, as used in \cite{katz}. To compare the two profiles, $\aaa$ is fixed to 1 so that the two profiles match to leading order as $u\to0$. }
\labell{scalar plots}
\end{figure}

Hence to leading order in our perturbative expansion, our background is the black hole metric \reef{metric} with scalar field solution \reef{solution} with $\aaa$ set as in eq.~\reef{wake3} and $\bbb=0$. As mentioned above, $\aaa$ is dual to the expectation value of the operator and using the usual holographic dictionary, we find
\beqa
\langle \mcO \rangle_T %&=& \frac{L^2}{2\pl^2}\,(2\Delta-3) \left(\frac{r_0}{L^2}\right)^\Delta  \aaa \nonumber\\
&=& \frac{\pi^2}{48}\! \left(\frac{4\pi}{3}\right)^{\!\Delta}\! (2\Delta-3)\,\aaa\ C_T\,T^\Delta\,,
\labell{expect}
\eeqa
where $\aaa\!\propto\!\alphasource$ is given in eq.~\reef{wake3}. We note that for a fixed dimension,
the expectation value above can be positive or negative depending on the sign of $\aaa\!\propto\!\alphasource$.  
Our holographic calculation recovers the expected form given in eq.~\reef{OT}. Recall our perturbative framework assumes that the 
amplitude of the bulk scalar is small ($|\langle \mcO\rangle_T|/T^\Delta \!\ll\! C_T$), which is equivalent to $|\aaa| \ll 1$ or $|\alphasource|\ll1$. 

The above expression may appear to vanish when $\Delta=d/2=3/2$, however, as noted in footnote \ref{footy7}, our scalar field solution 
eq.~\ref{solution} breaks down at this point. Hence the scalar profile and any subsequent calculations must be reconsidered for this particular value of the conformal dimension, as discussed in appendix \ref{special} --- the resulting expectation value $\langle \mcO \rangle_T$ is given in eq.~\ref{expect9}.

\subsection{Holographic conductivity} \labell{hsigma}

Next we examine the charge response, in particular the frequency-dependent conductivity, of the boundary theory in our holographic model. Note that in our perturbative approach, the scalar profile is directly proportional to the coupling $\alphasource$ and further the scalar modifies the charge response through the $\phi\,F^2$ interaction in eq.~\reef{gaugeX}, which in turn is controlled by $\alphamax$.  Therefore we will find that the charge response only depends on the product $\alpha_1\alpha_2$, not on their separate values. Thus, for example, the normalized dynamical conductivity $\sigma(\omega)/\sigma_\infty$ is only a function of two parameters, $\Delta$ and $\alpha_1\alpha_2$, as illustrated 
in figure \ref{variousX}. 
   
Given the gauge field action in eq.~\reef{gaugeX}, we can consider the stretched horizon method of \cite{Kovtun:2003wp,Brigante:2007nu}. The natural conserved current to consider charge diffusion is then
\begin{equation}
j^a = \frac{1}{4}\,  g^{a[b}n^{c]}\,\Big(1+\alphamax\phi(u)\Big)\left.F_{bc}\right|_{u=1}\,,
\end{equation}
where $n_a$ is the outward-pointing radial unit vector. The charge density then satisfies the diffusion equation \cite{Kovtun:2003wp}
\begin{equation}
\partial_t j^t = D\ \partial_i\partial_i j^t
\end{equation}
where the charge diffusion constant $D$ is given by \cite{Myers:2010pk,will-hd} 
\begin{equation}
D = \frac{3}{4 \pi T}\left(1+\alphamax\phi(1)\right)\int_0^1 \frac{du}{1+\alphamax\phi(u)}\,.\labell{diff} 
\end{equation}
The value of the scalar field at the horizon is given by
\begin{equation}
\phi(1) = \alphasource\times 8 \tfrac{\Gamma\left(2-\frac{2\Delta}{3}\right)}{\Gamma\left(1-\frac{\Delta}{3}\right)^2} h_\Delta(1)\left(\psi(\Delta/3) - \psi(1-\Delta/3) \right)\,,
\end{equation}
where $\psi(x) = \Gamma'(x)/\Gamma(x)$ is the digamma function. 

Figure \ref{diffusion} shows the diffusion constant as a function of the scaling dimension $\Delta$ of the scalar operator (while holding the combination $\alpha_1\alpha_2$ fixed). For relevant operators (\ie $\Delta <3$), the diffusion constant calculated from the exact solution is larger than for the pure power-law $\aaa u^\Delta$, while for irrelevant scalars (\ie $\Delta>3$), the ratio of the two results is reversed. As expected, the two curves cross at $\Delta=3$ where the two scalar profiles are identical. 
\begin{figure}[h]
\begin{subfigure}{0.48\textwidth}{\centering
\includegraphics[width=\textwidth]{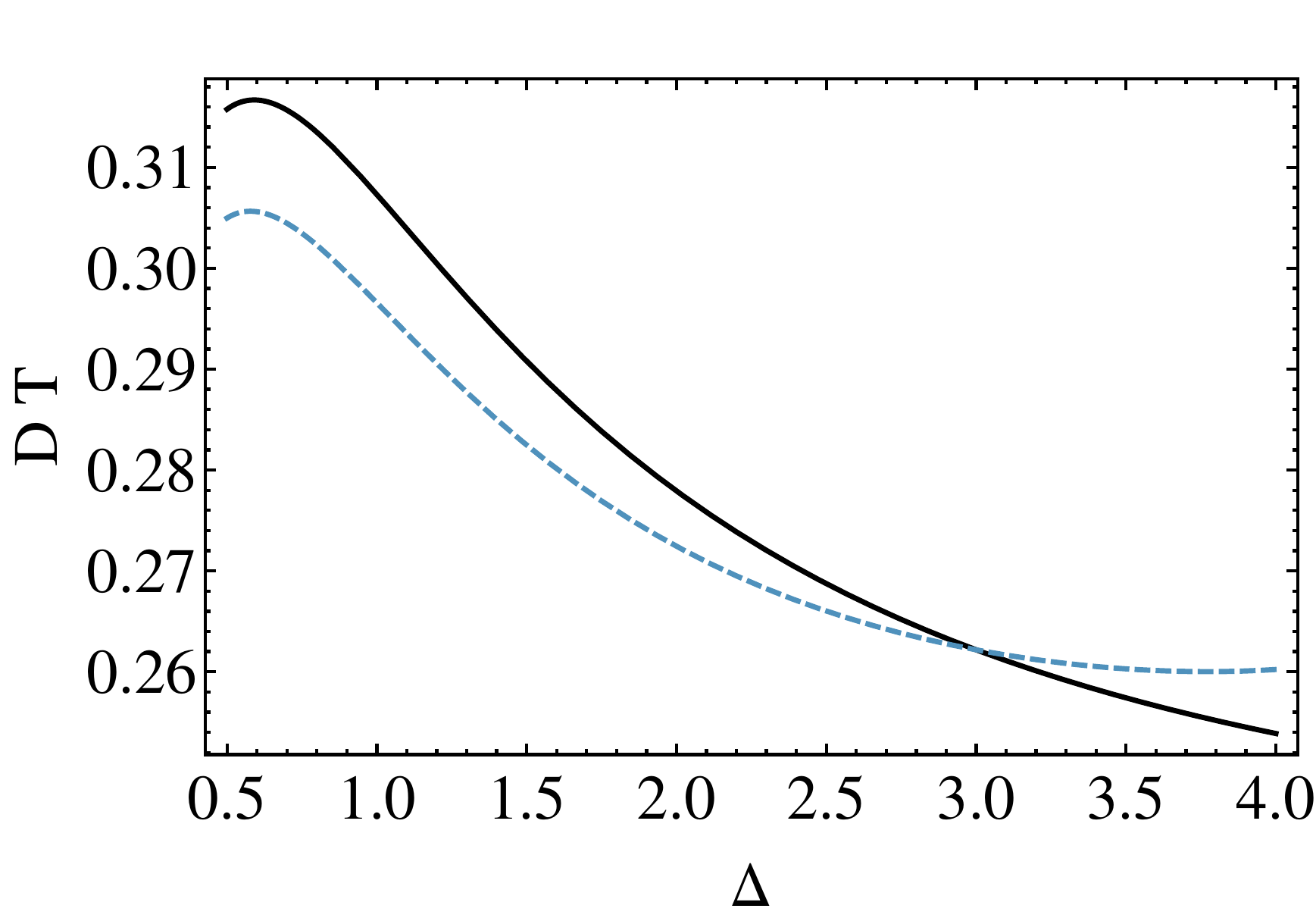}}
\end{subfigure}
\begin{subfigure}{0.48\textwidth}{\centering
\includegraphics[width=\textwidth]{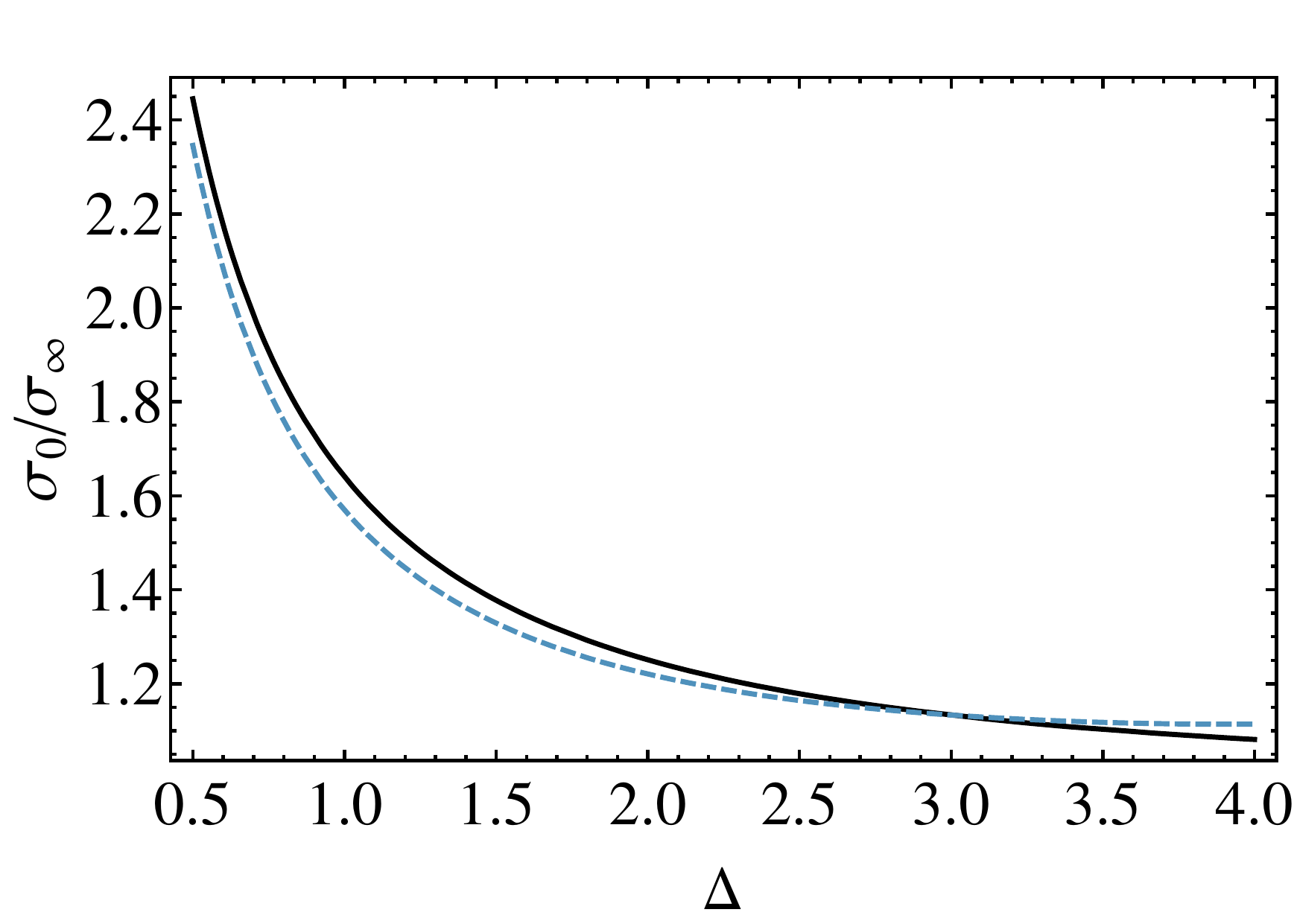}}
\end{subfigure}
\caption{On the left, we have the diffusion as a function of scaling dimension $\Delta$ with $ \alphasource\alphamax = 0.1$. The solid black line is the diffusion constant for the holographic model while for comparison, the dashed blue line is the diffusion calculated using the power-law profile $\phi(u) =\aaa \,u^\Delta$ (and with the coefficient $\aaa$ chosen to match to holographic solution for each $\Delta$). On the right we have the DC conductivity $\sigma_0 / \sigma_\infty $ as a function of the scaling dimension $\Delta$ with $\alphasource \alphamax = 0.1$. The solid black line is for the holographic model while the dashed blue line is found using $\phi(u) =\aaa\, u^\Delta$. \labell{diffusion}}
\end{figure}

The conductivity at zero frequency is given by \cite{Ritz:2008kh,Myers:2010pk}\footnote{Implicitly, we have set $(e^*)^2/\hbar=1$ here, where $e^*$ is the charge of the quantum charge carriers --- see \cite{katz}. Recall that $\sigma_\infty=\sigma(\omega/T\to\infty)=1/g_4^2$ in our holographic model.}
\begin{equation}
\sigma_0 = \frac{1+\alphamax\,\phi(1)}{ g_4^2}\,.\labell{dc}  
\end{equation}
Of course, the results shown in figure \ref{diffusion} are readily understood in terms of the behaviour of the scalar profiles illustrated in figure \ref{scalar plots}. That is, we found that the profile produced by our holographic model is smaller (larger) than the simple power-law profile near the horizon for $\Delta<3$ ($\Delta>3$).
Note that $\sigma_0$ is finite in our holographic model, even in the absence of momentum dissipation. This phenomenon is possible for systems where momentum and current are distinct, like in CFTs \cite{damle}. However,  
in general ``small-N" CFTs like the Wilson-Fisher QCPs with a finite symmetry group (\ref{WF}), it is expected that $\sigma(\omega\!\ll\! T)$ will show a weak logarithmic divergence $\log(T/\omega)$ that arises from the phenomenon of long-time tails of hydrodynamics \cite{kovtun-rev,natphys}. This is tantamount to saying that current-current correlations decay more slowly at long-times because of current conservation. It was shown \cite{simon} that these
long-time tails can be recovered in holography by including quantum corrections in the bulk, \ie they are suppressed by a factor of $1/C_T$.

The frequency-dependent conductivity is given by \cite{Myers:2010pk}
\begin{equation}
\sigma(\omega) = \frac{4 \pi T}{3i \,g_4^2\omega}\, \left.\frac{\partial_u A_y}{ A_y}\right|_{u\rightarrow0}\,,
\labell{pourZ}
\end{equation}
where  the temperature $T$ is given in eq.~(\ref{temp}) and $A_y(u,\omega)$ is the Fourier transform of (the $y$-component of) the gauge field. 
The latter profile is determined by numerically solving the gauge field equations of motion resulting from  eq.~\reef{gaugeX}, with appropriate boundary conditions at the event horizon --- see details in appendix \ref{holosig}.  

We plot the resulting $\sigma(\omega)$ as a function of real and Euclidean frequency in figures \ref{conductivity plot} and \ref{euclidean plot} for various values of the scaling dimension $\Delta$. In each case, we compare the conductivity calculated with our holographic model to that calculated with a simple power-law profile for the bulk scalar $\phi(u) = \phi_1 u^\Delta$, as in \cite{katz}. The two results are nearly in agreement. 
In particular, in figure \ref{conductivity plot}, we adjust the amplitude of the scalar profile with $\Delta = 1.5$ to fit to conductivity for Euclidean frequencies to the quantum Monte Carlo data of  \cite{katz,natphys} and we see that the two results
agree almost exactly for Euclidean frequencies $\Omega>2\pi T$ --- see further discussion in section \ref{discuss}. The largest discrepancies in all of these comparisons appear at the origin $\omega=0$, where the conductivity probes the holographic background near the event horizon. As noted above, the conductivity $\sigma_0$ in our holographic model is higher (lower) than for the power-law profile when $\Delta<3$ ($\Delta>3$).
\begin{figure}[!htb]
\begin{subfigure}{0.5\textwidth}{
\includegraphics[width = \textwidth]{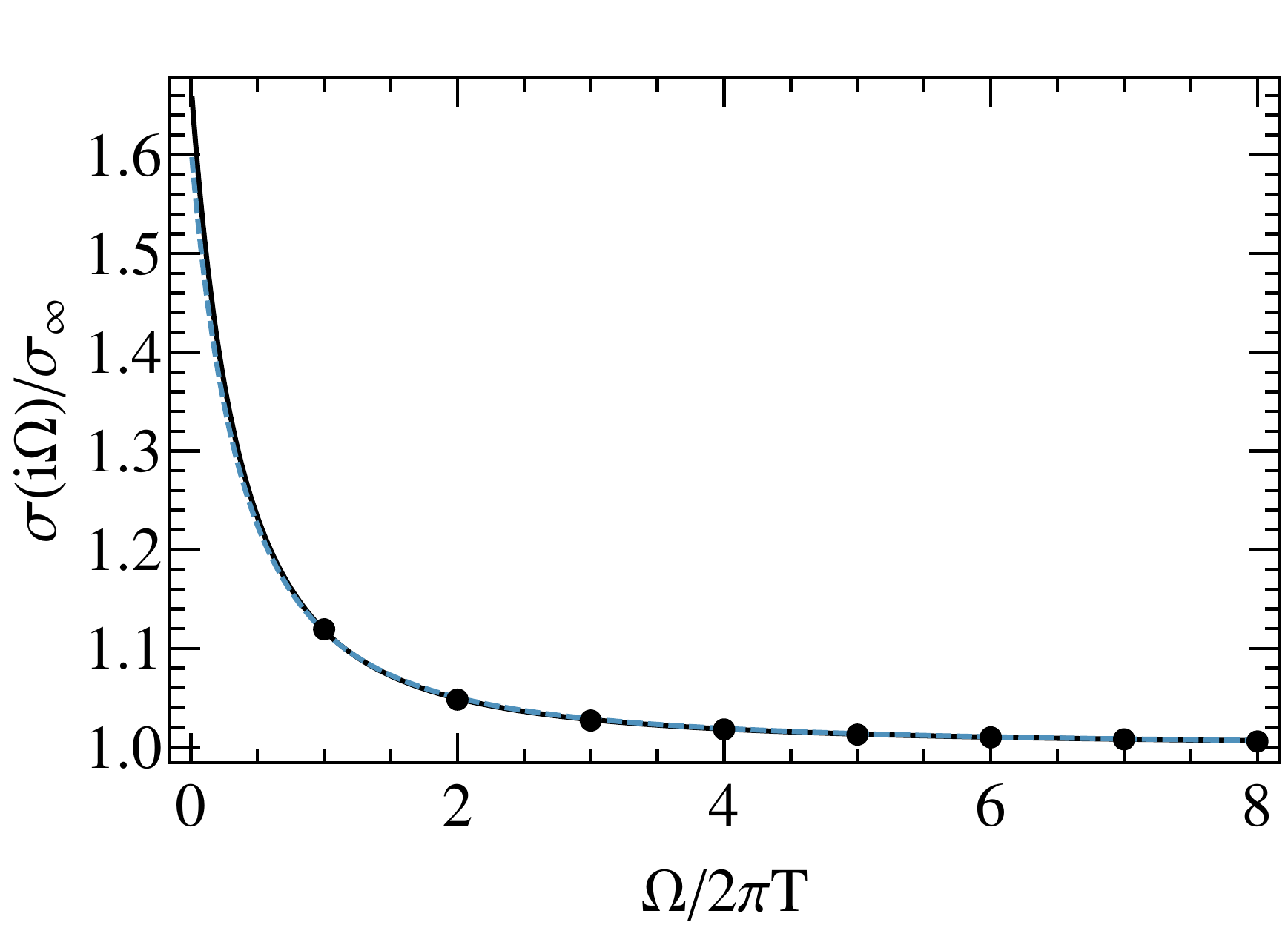}}
\end{subfigure}
\begin{subfigure}{0.5\textwidth}{
\includegraphics[width = \textwidth]{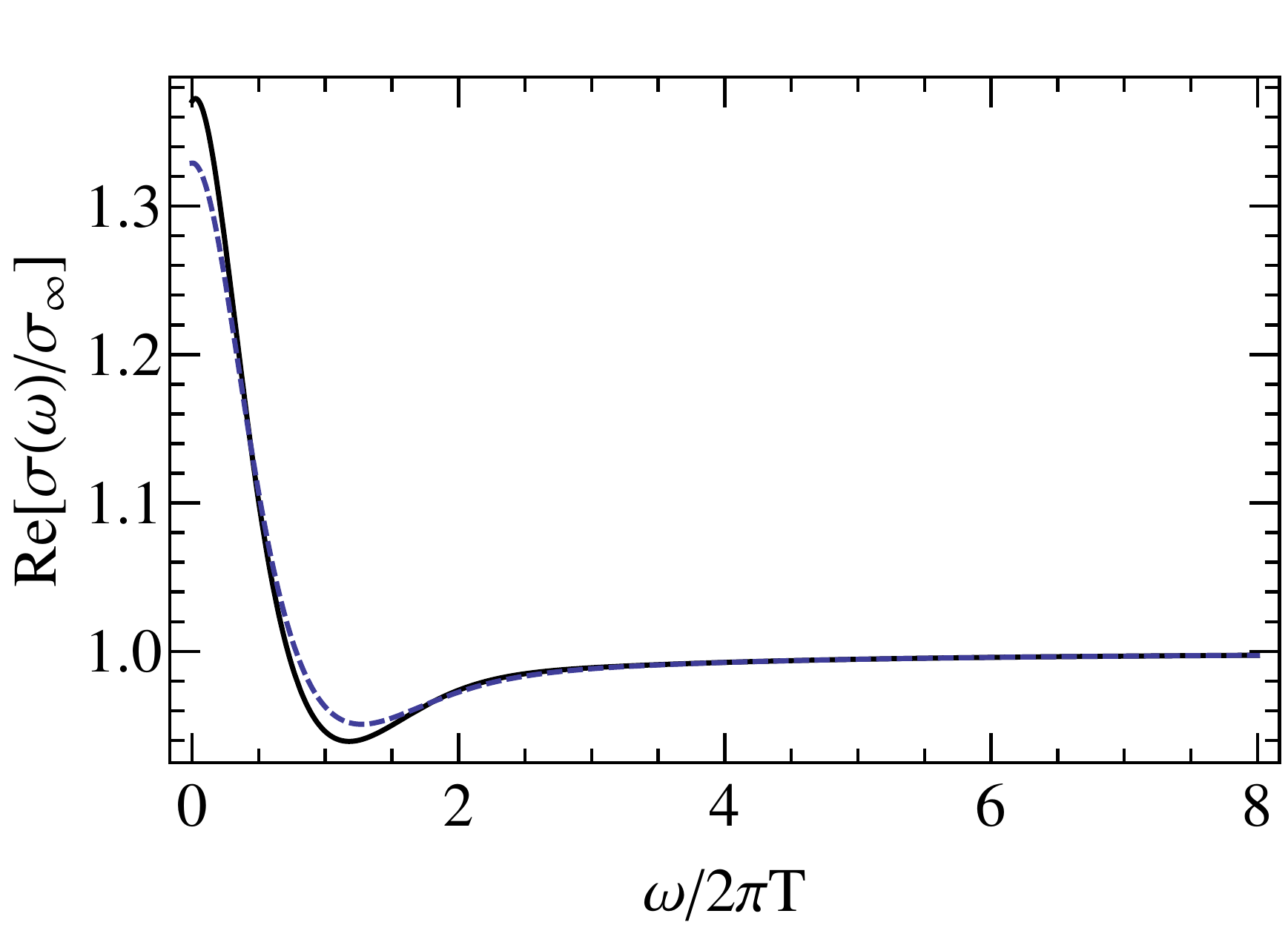}}
\end{subfigure}
\caption{Plots of the conductivity for Euclidean (left) and real (right) frequencies for $\Delta=1.5$ with $\aaa \alphamax$ fit to the quantum Monte Carlo data for the O(2) Wilson-Fisher CFT \cite{katz,natphys} (see also \cite{chen}). The solid black
line represents the conductivity using the scalar profile given in eq.~\reef{solution} with $\aaa\alphamax = 0.589$, while 
the dashed blue line represents the value for the conductivity using the simple power-law profile $\phi(u) =  \aaa\,u^\Delta$ with $\aaa\alphamax = 0.611$.  \labell{conductivity plot}} 
\end{figure}
\begin{figure}[!htb]
\begin{subfigure}{0.5\textwidth}{
\includegraphics[width=\textwidth]{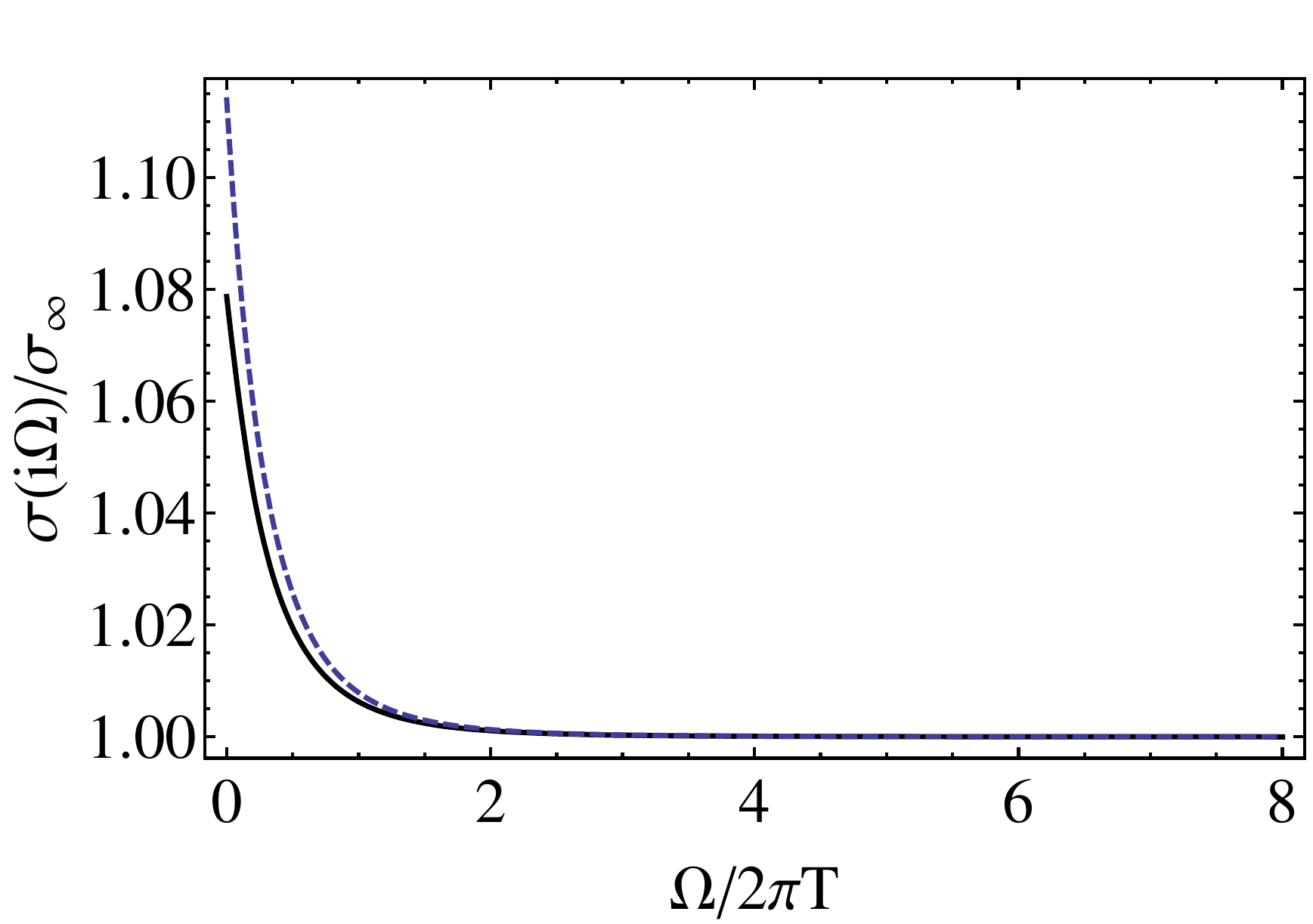}}
\end{subfigure}
\begin{subfigure}{0.5\textwidth}{
\includegraphics[width=\textwidth]{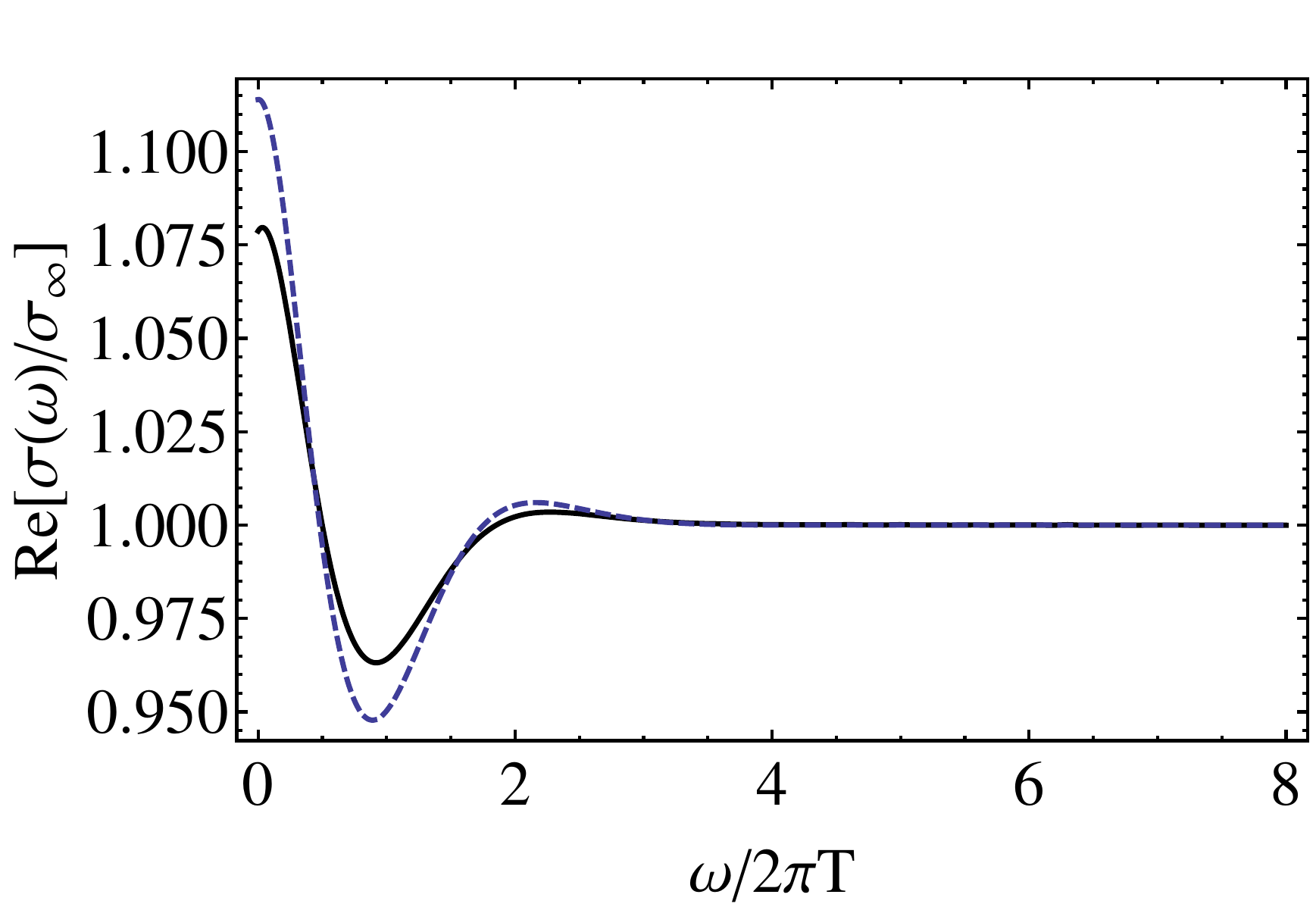}}
\end{subfigure}
\caption{Plots of the conductivity for Euclidean (left) and real (right) frequencies for $\Delta=4$. The solid black line represents the conductivity found using the scalar profile given in eq.~\reef{solution} while the dashed blue line represents the conductivity found using the simple power-law profile $\phi(u) =  \aaa\,u^\Delta$. Both plots were generated using with $\alphasource\alphamax = 0.1$.
\labell{euclidean plot} }
\end{figure}

\section{Asymptotic expansion of conductivity \& OPEs} \labell{main}  

The asymptotic expansion of the conductivity for frequencies which are large compared to the temperature is useful for many reasons.
First, it reveals important properties about the operators with low scaling dimensions. It also allows us to establish non-trivial sum 
rules, \eg \cite{Son09,sum-rules,justin2,ws,katz}. Further, it plays a role in the comparison of holographic response functions with Euclidean data for  
the conductivity, as the latter is available from Monte Carlo simulations for 
frequencies exceeding $2\pi T$, \eg \cite{natphys,katz}. 
In this section, we first obtain the expansion in our holographic model directly from the equation of motion for the gauge field dual to the current.
Then we re-derive the expansion by using the operator product expansion (OPE) of the boundary CFT. This analysis reveals fundamental properties
of our model, and the corresponding dynamical charge response. Let us also note that similar analyses of the 
modifications of the high frequency behaviour of the conductivity and viscosity 
due to scalar expectation values was made for a variety of other holographic backgrounds in \cite{justin1,justin2}.
In those studies, the scalars were chiral primaries that acquired an expectation value as a result of turning
on a chemical potential. 
%todd new refs

\subsection{High frequency expansion} \labell{freqy}

We now compute the conductivity at frequencies much greater than the temperature. Working in Euclidean frequencies,
this corresponds to evaluating  $\sigma(\omega\!=\!i\W)$ with $\W \gg T$.\footnote{Our notation $\Omega_n$ alludes to Matsubara frequencies that arise in finite temperature quantum field theory. In this case, these frequencies would be discrete multiples of $2\pi T$, however, $\Omega_n$ can be thought of as a continuous variable in the following.} 

In the following, we will calculate the high frequency asymptotics perturbatively in the dimensionless coupling $\alphamax$. Recall that this coupling controls the strength of the $\phi\,F^2$ interaction in eq.~\reef{gaugeX}, which determines how the scalar operator in the boundary modifies the conductivity. In this approach, it is convenient to first change coordinates from $u$ to $z$, where $dz/du = 1/f(u)$. The boundary, $u=0$, corresponds to $z=0$, however, the horizon $u=1$ is stretched to $z=\infty$ in these new coordinates. With this coordinate choice, the equation determining the gauge field profile --- see eq.~\reef{gauge ode} --- becomes
\beq
\big[\partial_z^2-\freq^2\big] A_y = -\frac{\alpha_2\partial_z\phi}{1+\alpha_2\phi}\,\partial_z A_y\,,
\labell{newde}
\eeq
where we have introduced the rescaled (dimensionless) Euclidean frequency
\begin{align}
 \freq = \frac{3\W}{4\pi T}\,. \labell{nook}
\end{align}
Now in our perturbative approach, we expand the gauge profile as $A_y=A^{\ssc(0)}_y+\alpha_2 A^{\ssc(1)}_y+\alpha_2^{\,2} A^{\ssc(2)}_y+\cdots$. Similarly, expanding the gauge equation \reef{newde}, the zeroth order component satisfies $\big[\partial_z^2-\freq^2\big]A^{\ssc(0)}_y=0$, and the solution (which is regular or ``in-falling'' at the horizon) is
\begin{equation}
A^{\ssc(0)}_y = e^{-\freq z}\, . \labell{leeed}
\end{equation}
Next at first order in $\alpha_2$, eq.~\reef{newde} yields
\beq
\big[\partial_z^2-\freq^2 \big]A^{\ssc(1)}_y = \freq e^{-\freq z}\,\partial_z \phi\,.\labell{guage first}
\end{equation}
This equation can be solved with the use of the following Green's function
\begin{equation}
G(z,\tilde{z}) = -\frac{1}{\freq}\Big(\sinh(\freq z)e^{-\freq \tilde z}\, \theta(\tilde z - z) + \tilde z \leftrightarrow z\Big)\,,
\labell{greeen}
\end{equation}
where $[\partial_z^2 -\freq^2] G(z,\tilde z) = \delta(z-\tilde z)$ and $G(z,\tilde z)$ vanishes at $z\to 0$ and at $z \to \infty$. The solution to eq.~(\ref{guage first}) is then given by
\begin{equation}
A^{\ssc(1)}_y = \int_0^\infty \! d\tilde z\;  G(z,\tilde z)\, \freq\, e^{-\freq \tilde z}\, \partial_{\tilde z} \phi\,  .
\end{equation}
To calculate the conductivity, we must evaluate
\begin{equation}
%langle J_x(\freq)J_x(-\freq)\rangle_T 
 \frac{\sigma(i\freq)}{\sigma_\infty} = -\frac{1}{\freq} \partial_u A_y\big|_{z=0} %(z\!=\!0) 
=  1 + \alphamax \int_0^\infty \! dz\; e^{-2 \freq z }\partial_z\phi + O(\alphamax^2)\,.  \labell{JJperb}
\end{equation}
Substituting the power series for $\phi (z) = \sum\limits_\ell  c_\ell\, z^{\alpha_\ell}$ into eq.~\reef{JJperb} yields
\begin{equation}
   \frac{\sigma(i\freq)}{\sigma_\infty} = 1 + \alphamax \sum_\ell \frac{\Gamma(\alpha_\ell+1)}{(2 \freq)^{\alpha_\ell}}\, 
c_\ell   + O(\alphamax^2) \,.\labell{general cond}
\end{equation} 
The first few terms in the near boundary expansion (\ie $u\rightarrow 0$) of the scalar field profile \reef{solution} are 
\begin{equation}
\phi(u) = \aaa u^\Delta + \frac{\aaa\Delta}{6} u^{\Delta+3} + \frac{12 \alphasource}{ (\Delta-6)(3+\Delta)} u^6 + O(u^{\Delta+6})\,.\labell{scalar series}
\end{equation}
Given this result,\footnote{As well as using $u=z\,(1-\frac14 z^3 + \frac{3}{28} z^6 + \cdots)$. \labell{footy}} we obtain 
the first few terms for the conductivity at $\freq \gg 1$:
\begin{align}
  \frac{\sigma(i \freq)}{\sigma_\infty} = 1 + \frac{\aaa \alphamax \Gamma(\Delta+1)}{(2\freq)^\Delta} - \frac{\aaa\alphamax\Delta}{12} \frac{\Gamma(\Delta+4)}{(2\freq)^{\Delta+3}} + \frac{12\alphasource \alphamax  \Gamma(7)}{(\Delta-6)(\Delta+3)(2\freq)^6} 
  + O\!\left( \frac{1}{\freq^{\Delta+6}}\right).\labell{asymp}
\end{align}
If we recall that $\aaa\propto\alpha_1$ in eq.~\reef{wake3}, we see explicitly here that in this expansion, the 
normalized conductivity is only a function of the two model parameters, $\Delta$ and $\alpha_1\alpha_2$, as well as the 
frequency $\freq=3\Omega_n/(4\pi T)$. 

One can easily extend the above analysis to second order in the coupling $\alphamax$ --- see appendix \ref{lock}. 
Here we note that at that expansion order, the leading correction to the high-frequency expansion \reef{asymp} is proportional to $(\aaa\alphamax)^2 /\freq^{2\Delta}$ and therefore the leading $1/\freq^\Delta$ term above remains unchanged. The fact that the leading term above is exact can be anticipated by the arguments in the next section which determine the coefficient of this contribution from the OPE. 
%\blue{In the appendix, it's mentioned that the calculation of $C_{JJO}$ is perturbative in $\alpha_2$, but here we say it's exact?}

\subsection{OPE analysis} \labell{anal}

To gain a deeper physical insight into the asymptotic expansion (\ref{asymp}),
we now reconstruct it using the OPE and the CFT data corresponding to our holographic model. Here, we focus on the first two terms.
The leading term $\sigma_\infty=1/g_4^2$ is simply the ground state conductivity, which obtains from the vacuum
current-current correlator (appendix \ref{ap:2pt}).  
The second term is nontrivial as it arises because the relevant operator $\mcO$, the CFT operator dual to $\phi$, appears in the $JJ$ OPE
and acquires an expectation value at $T\!>\!0$ \cite{katz}.

First, let us recall the OPE of two (conserved) currents in the CFT written in momentum space \cite{katz}
\begin{equation}
  \lim_{q\gg p} J_\mu(\bs q) J_\nu(\bs p-\bs q) =  -\sigma_{\infty} q\,I_{\mu\nu}(\bs q)\, \delta^{(3)}(\bs p) -  \frac{\cjjo \, I_{\mu\nu}(\bs q)}{q^{\Delta-1}}\,\mathcal O(\bs p) + \dotsb \labell{jj ope}
\end{equation}
where $\bs p,\bs q$ are Euclidean 3-momenta. $I_{\mu\nu}(\bs q) = \delta_{\mu\nu} - \frac{q_\mu q_\nu}{q^2}$ is the tensorial structure satisfying the conformal symmetries and the Ward identity arising from current conservation, \ie $q^\mu I_{\mu\nu}(\bs q) = 0$.  Above,
we have only included the contributions from the identity and from the scalar $\mcO$ with dimension $\Delta$. The ellipsis denotes the appearance of higher dimension operators in the OPE, \eg the stress tensor \cite{katz}.\footnote{Implicitly, we are assuming that $\mcO$ is a relevant operator with $\Delta<3$ for the stress tensor to appear as a higher dimension operator. \labell{footy2}} To obtain the asymptotic expansion of the finite temperature conductivity, we take the thermal expectation value of eq.~(\ref{jj ope}) setting $\mu=x=\nu$ and $\bs q=(\W,0,0)$ with $\W>0$:   
\begin{align}
  \langle J_x(\W)J_x(-\W)\rangle_T = -\W\left( \sigma_\infty + C_{JJO}\, \frac{\langle \mcO\rangle_T}{\W^\Delta} +\dotsb \right) \labell{hmmm}
\end{align}
where we have used $\langle \mathcal O(\bs p) \rangle_T=\delta^{(3)}(\bs p) \langle \mathcal O \rangle_T$.
Further, to connect this result to the expansion \reef{asymp}, we recall that the conductivity can be evaluated 
with the Kubo formula
\beq
\sigma(i\W)=-\frac1{\W}\,\langle J_x(\W)J_x(-\W)\rangle_T \,.
\eeq
   
Now, recall that for our holographic model, $\sigma_\infty=1/g_4^2$ and $\langle \mcO \rangle_T = B T^\Delta$, where $B$ is given in eq.~(\ref{expect}). 
We can use the results in appendix \ref{oo sec} to derive the value of the OPE coefficient $C_{JJO}$ for the boundary CFT. In particular, inserting (the $\mu=x=\nu$ component of) eq.~\reef{jj ope} in a vacuum correlator with $\mcO(-\bs p)$ yields
\begin{equation}
 \langle J_x(\bs q) J_x(\bs p - \bs q) \mcO (-\bs p)\rangle{\big|_{\text{sing.}}} = -\frac{C_{JJO}}{|\bs q|^{\Delta-1}}\, \langle \mcO(\bs p) \mcO(-\bs p)\rangle\,, 
\end{equation}
which is understood to be in the limit $|\bs q| \gg |\bs p|$. Our notation above emphasizes that this is the \emph{singular}
part, as the full three-point function also contains terms regular in $|\bs p|$ as $|\bs p|\to 0$, but these are not relevant for the OPE. 
Now comparing this expression with the holographic result in eq.~\reef{oops}, we find that the $JJ\mcO$ OPE coefficient in our model is
\begin{align}
  C_{JJO} = \frac{\alphamax}{g_4^2}\,\frac{\pl^2}{L^2}\,\frac{\Gamma(\Delta+1)}{ 2^{\Delta}(2\Delta-3) }\,, 
  \labell{polar}
\end{align}
which is proportional to $\alphamax$, as advertised previously.\footnote{Again, the pole at $\Delta=3/2$ in eq.~\reef{polar} signals that our calculations have to be reconsidered for this special value of the conformal dimension --- see appendix \ref{special}.} Substituting this expression into eq.~\reef{hmmm} then yields
\begin{equation}
  \begin{split}
    \langle J_x(\W) J_x(-\W) \rangle_T =& -\W \sigma_\infty \left(1 +\alphamax\,\frac{\pl^2}{L^2}\, \frac{\Gamma(\Delta+1)}{2^{\Delta}(2\Delta -3) } \,\frac{\langle \mcO\rangle_T}{\W^\Delta} + \cdots \right)\,.
%\\    =& -\frac{\omega}{g_4^2} \left(1 + a\alphamax\Gamma(\Delta+1) \left(\frac{2\pi T}{3 \omega}\right)^\Delta + ... \right)
  \end{split}
\end{equation}
Using the expression for $\langle \mcO\rangle_T$ in (\ref{expect}), as well as $C_T = \frac{24}{\pi^2}\,\frac{L^2}{\pl^2}$, we find 
\begin{align} \label{asym-sigma}
    \frac{\sigma(i\W)}{\sigma_\infty}  = 1 + \aaa\alphamax\Gamma(\Delta+1) \left(\frac{2\pi T}{3 \W}\right)^\Delta + \dotsb\,,
\end{align}
which matches precisely with the first two terms of eq.~(\ref{asymp}), if we recall the definition the rescaled frequency $\freq$ in eq.~\reef{nook}. We note that eq.~(\ref{asym-sigma}) can be analytically continued 
to real frequencies, $i\W \to \omega + i0^+$, so that for generic $\Delta$ both the real
and imaginary parts of $\sigma(\omega)$ will contain a $(T/\omega)^\Delta$ term at large frequencies \cite{Caron-Huot,willprl}.  

At this point, let us observe that generically we expect the stress tensor will appear in the $JJ$ OPE \reef{jj ope} and so there would be additional contributions to the asymptotic expansion \reef{general cond}, beginning at the order $1/\freq^3$. Of course, the latter would in fact be the dominant frequency-dependent contribution when $\Delta>3$. It is an `exceptional' feature of our holographic model that the vacuum correlator $\langle JJ T\rangle$ vanishes and such contributions are not present in the asymptotic expansion above. In fact, if the same holographic model was studied for $d=4$, we would find that  $\langle JJ T\rangle$ is nonvanishing and additional terms appear in the analog of eq.~\reef{jj ope}. Alternatively, the holographic model could be extended to include a new bulk interaction $C_{abcd}F^{ab}F^{cd}$, as in \cite{Myers:2010pk,natphys}.

\subsection{Fingerprints of large-$N$ factorization} \labell{finger}

We first consider the higher order terms in the high frequency expansion of the conductivity given in eq.~\reef{asymp},
which is valid to linear order in our $\alpha_2$ expansion. 
As shown in eq.~\reef{general cond}, the expansion of the scalar field controls the high frequency expansion of conductivity and the powers in the high frequency expansion matches the powers of $z$ in the expansion of $\phi(u)$. Hence, examining eq.~(\ref{scalar series}) and the translation between the $u$ and $z$ coordinates --- see footnote \ref{footy} --- we conclude that that beyond $1/\freq^\Delta$, the only powers of $1/\freq$ which will appear in the expansion of conductivity \reef{asymp} will be $\Delta +3\ell$  and $3+3\ell$ with $\ell=1,2,\cdots$.

First, let us consider the sequence of terms with $1/\freq^{\Delta_\ell}\sim(T/\W)^{\Delta_\ell}$ where $\Delta_\ell=\Delta+ 3\ell$. These contributions should arise from the thermal expectation value of a local operator with conformal dimension $\Delta_\ell$, which appears in the $JJ$ OPE in eq.~\reef{jj ope}. If $\mcO$ is a primary operator, one might naively think that these higher dimension operators are descendants of $\cO$.
For example, the operator $\partial_\mu \partial^2\mcO$ would have dimension $\Delta+3$. However, it
cannot contribute the term proportional to $(T/\W)^{\Delta+3}$ in the asymptotic expansion because its thermal expectation value vanishes by symmetry.
Indeed, $\langle \cO\rangle_T$ is space- and time-independent. The natural interpretation is that 
this asymptotic term arises from the composite operator $\join{\mcO T_{\mu\nu}}$, obtained by ``composing'' $\mcO$ and the stress tensor.
In a general CFT, such a ``composition'' (reminiscent of free theories) is not well-defined and thus one cannot interpret the 
result as a well-defined local operator. However, in the large-$N$ limit (or alternatively, the limit of large central charge $C_T$) implicit in our holographic model, such a composition is natural because of the large-$N$ factorization arising in such theories \cite{el-showk}. Similarly, one can attach a string of $\ell$ stress tensors
to $\mcO$ to obtain an operator with scaling dimension $\Delta_\ell=\Delta+ 3\ell$ for higher values of $\ell$. We note that these operators
have non-zero thermal expectation values and that in our model, their OPE coefficients with two currents are determined by $\alpha_2$.
By the same token, the same composition explains the presence of terms with powers $\Delta'_\ell=3+3\ell$, as these will correspond to
strings of $(1+\ell)$ stress tensors.  

In appendix \ref{lock}, we find that at second order in the coupling $\alpha_2$, the asymptotic expansion of the conductivity acquires a new term proportional to $(T/\Omega_n)^{2\Delta}$. Following the above discussion, it is natural to interpret this contribution as arising from the composite operator $\join{\cO^2}$. Usually these composite operators are irrelevant, however,
we observe then that when the original conformal dimension lies in the 
range $\tfrac12\le\Delta<\tfrac32$, then the conformal dimension of this new operator is $\Delta'=2\Delta<3$. That is, in this regime, our holographic model has at least 
two relevant scalar operators, and hence it describes a quantum 
\emph{multicritical} point, rather than a simple critical point. It would be interesting to further study the interplay of these two operators in the dynamics of the multicritical point using the holographic techniques established for so-called ``multi-trace'' operators, \eg \cite{multi1,multi2,multi3,newmulti}

% We note that this composition or factorization property is not general, and thus the
% asymptotic expansion of a general conformal QCP will only agree with up to $(T/\W)^\Delta$ but not for higher order terms. 

\section{Discussion} \labell{discuss}

To recap, ref.~\cite{katz} recognized the important role of the relevant operator at a quantum critical phase transition in determining 
the dynamics of the corresponding QCP. They also took some steps to investigating this question in a holographic framework. 
A shortcoming of their construction was that the dual of the relevant operator in the boundary theory was not  incorporated as a dynamical field in the bulk gravity theory. 
Of course, it is well understood that including a bulk scalar field $\phi$ with the appropriate mass, \ie $m^2 L^2 = \Delta (\Delta-3)$ will introduce a scalar operator $\cO$ with conformal dimension $\Delta$ in the boundary theory, \eg see \cite{revue}. However, for the present purposes, a weakness of holographic theories studied up to this point is that the corresponding operator will not acquire a nonvanishing expectation value at finite temperature. Hence the key innovation of our holographic model was to include a natural mechanism which ensures that $\langle \cO\rangle_T \ne 0$, as in eq.~\reef{OT}. That is, the bulk scalar is sourced to have a nontrivial profile in the dual black hole background, which then allows us to study the dynamical conductivity in a self-consistent holographic model. 
However, let us add the nontrivial observation that the conductivity obtained using our model is well-approximated
by the simple Ansatz of \cite{katz} for a wide range of parameters, as illustrated in figures \ref{conductivity plot} 
and \ref{euclidean plot}. 
We examine this point in more detail below. Further, we will also discuss below (section \ref{non critical}) how our 
holographic model provides a starting point to examine the response functions as a function 
of the relevant coupling $\lambda$ --- see eq.~\reef{punt} --- as we tune away from the QCP. 

In section \ref{sec:ingredients}, we motivated the construction of our holographic model with a discussion of QC phase transitions which involve a relevant operator, with $\Delta<3$. However, our holographic analysis easily extends to considering irrelevant boundary operators, with $\Delta>3$, as well. In the latter case, the results may be interesting to better understand the dynamical response of certain QC \emph{phases} 
(where there is no relevant scalar operator whose coupling needs to be fine-tuned). In this case,  
we could consider $\cO$ to be the leading irrelevant operator controlling RG flows down to this critical phase. The stress tensor would be the minimal dimension operator which acquires a thermal expectation value and hence one would also want to include the $C_{abcd}F^{ab}F^{cd}$ bulk interaction considered in \cite{Myers:2010pk}.
This would ensure, \eg that the stress tensor produces the leading contribution in the high-frequency expansion \reef{asymp} proportional to $1/\freq^3$,\footnote{We note that in certain CFTs, supersymmetry
will forbid this $1/\freq^3$ contribution coming from the stress tensor \cite{william9}.} whereas that coming from the irrelevant operator is higher order being proportional to $1/\freq^\Delta$. However, this contribution could still be significant when $\cO$ is nearly marginal, \ie when $\Delta$ is only slightly larger than 3. 

\subsection{Minimality and related models}

Again, the key new feature of our holographic model is that the scalar operator $\cO$ in the boundary theory acquires a nonvanishing thermal expectation value, as in eq.~\reef{OT}.  This feature was engineered by adding the new interaction in eq.~\reef{scalarX} which couples the dual scalar field $\phi$ to the Weyl curvature of the bulk geometry. This choice was motivated by the observation that the Weyl curvature vanishes in the vacuum AdS geometry but is nonvanishing in the black hole geometry \reef{metric}. Hence the resulting equation \reef{scalar equation} for the bulk scalar has no source in the AdS vacuum and the relevant solution is just $\phi=0$. However, the equation has a nonvanishing source in the black hole geometry and $\phi$ acquires a nontrivial profile in this background. As desired then,  $\langle \mathcal O \rangle_T\ne 0$ in the boundary theory. 

As noted before, previous holographic models did not reproduce this simple physical behaviour in the boundary theory. Certainly, one could imagine more complex approaches to produce the same physics and so one might think of our approach as providing the minimal holographic model with this feature.  One simple modification would be to introduce an interaction with higher powers of the Weyl curvature, however, the behaviour found in our model would not be modified in an essential way. For example, with a $\phi\, C^n$ interaction (with $n\ge2$), the leading term in the high-frequency expansion would still be proportional to $1/\freq^\Delta$ and in fact, it would still be given by exactly the same expression as in eq.~\reef{asymp} if there are no other changes to the holographic action. The effect of this new interaction would only appear at higher orders. In particular, the $1/\freq^6$ term in eq.~\reef{asymp} would be replaced by a new contribution proportional to $1/\freq^{3n}$. One defining feature of the boundary CFT which would be modified is that the three-point correlator $\langle TT\cO\rangle$ would vanish with this new bulk interaction. However, this then indicates that in general there is no direct connection between the CFT parameter controlling this three-point function and the thermal expectation value $\langle \mathcal O \rangle_T$.

\subsection{Perturbative bulk expansion}

Next we discuss the perturbative nature of our calculations, however, let us first comment on the fact that we are using a higher curvature interaction in the scalar action \reef{scalarX} to generate $\langle \cO\rangle_T \ne 0$. Similar higher curvature interactions will generically appear in string theoretic models, \eg as $\alpha^{\prime}$ corrections in the low-energy effective action \cite{gross}. However, rather than constructing explicit top-down holographic models, our approach here is to examine simple toy holographic models involving higher curvature interactions in the bulk gravity theory 
(see Refs.~\cite{Ritz:2008kh,Myers:2010pk,will-hd,Bai2013} for different such models without scalar operators). 
Our perspective is that if there are interesting universal properties which hold for all CFTs, then they should also appear in the holographic CFTs defined by these toy models as well. This approach has been successfully applied before, \eg in the discovery of the F-theorem \cite{Myers:2010xs,Myers:2010tj} and more recently, in uncovering universal behaviour in the corner entanglement entropy for $d=3$ CFTs \cite{corner1,corner2}. 

We also stress that we are only working perturbatively in the dimensionless coupling $\alphasource$ for our new interaction. 
Higher curvature actions are typically regarded as problematic because generically they lead to ``unstable'' higher derivative equations of motion. However, these issues are essentially overcome when treating the higher curvature (or more generally, higher derivative) interactions as providing ``small'' perturbative corrections to a second-order theory \cite{JZ}. Hence our perturbative approach evades this problem.

At the outset, we said that our construction of the holographic background was perturbative in the amplitude of the bulk scalar. As indicated by eq.~\reef{wake3}, this is equivalent to a perturbative expansion in terms of the dimensionless coupling $\alphasource$, which controls the strength of the $C^2$ source in the scalar wave equation \reef{scalar equation}.  
In terms of the boundary theory, we can characterize this approach as considering the regime where
the thermal expectation value of $\mathcal O$ is much smaller than the thermal energy density, \ie
 $|\langle\cO\rangle_T|/T^\Delta \! \ll\! \varepsilon/T^3$, where $\varepsilon\!=\! \langle T_{00}\rangle_T$.

In fact, we only carried out our analysis to linear order in $\alphasource$ and so the holographic background consisted of the unmodified black hole geometry along with the scalar field profile given in eqs.~\reef{metric} and \reef{solution}, respectively. The next step in extending our perturbative construction would be to include the contributions of the scalar action \reef{scalarX} in the gravitational equations of motion. The back-reaction of the scalar would then produce $O(\alphasource^2)$ perturbations in the black hole metric. Evaluating the conductivity would then extend the analysis in appendix \ref{holosig} by considering the gauge field equation of motion in this modified metric. As a result, one would then find contributions in the conductivity proportional to $\alphasource^3\alphamax$. Hence we may conclude that the full conductivity $\sigma(\omega)$ in our holographic model depends independently on the three parameters, $\Delta$, $\alphasource$ and $\alphamax$. That is, finding that the charge response in section \ref{hsigma} was a function of only $\Delta$ and the product $\alpha_1\alpha_2$ was an artifact of only carrying out our perturbative construction to first order.
Working beyond first order also suggests the possibility of obtaining bounds on the holographic 
couplings $\alphasource$ and $\alphamax$ from the boundary theory, in analogy to the bounds 
found in, \eg \cite{Myers:2010pk,Brigante:2008gz}. 
However, we leave all of these interesting research directions for future work.

\subsection{Monte Carlo data and analytic continuation} 

We are building on the holographic studies in \cite{natphys,katz} and our construction is a next step in developing holography as a useful tool in studying the real-time dynamics of QCPs. One of the successes of these previous works was using quantum Monte Carlo (QMC) to study the dynamical conductivity of the O(2) Wilson-Fisher fixed-point theory and fitting the numerical results for imaginary frequencies with a holographic model. Further the holographic results are easily analytically continued to real frequencies, which is not possible for the QMC data, which only provides $\sigma(i\W)$ for the discrete Matsubara frequencies $\W=2\pi nT$ with $n=1,2,3,\cdots$. For this fixed-point theory, the conformal dimension of the relevant operator is very close to  $\Delta=3/2$ \cite{Campostrini01,bootstrap}. Figure \ref{conductivity plot} show the results of fitting the QMC data with our holographic model with $\Delta=3/2$ and compares it to the results in \cite{katz}, which used a simple power-law profile for the bulk scalar. Both the conductivity fit for imaginary frequencies and the analytic continuation to real frequencies are almost identical for the two holographic models. Hence in this case, the two approaches do not differ in any essential way.  

\begin{figure}[!htb]
\begin{subfigure}{0.5\textwidth}{
\includegraphics[width=\textwidth]{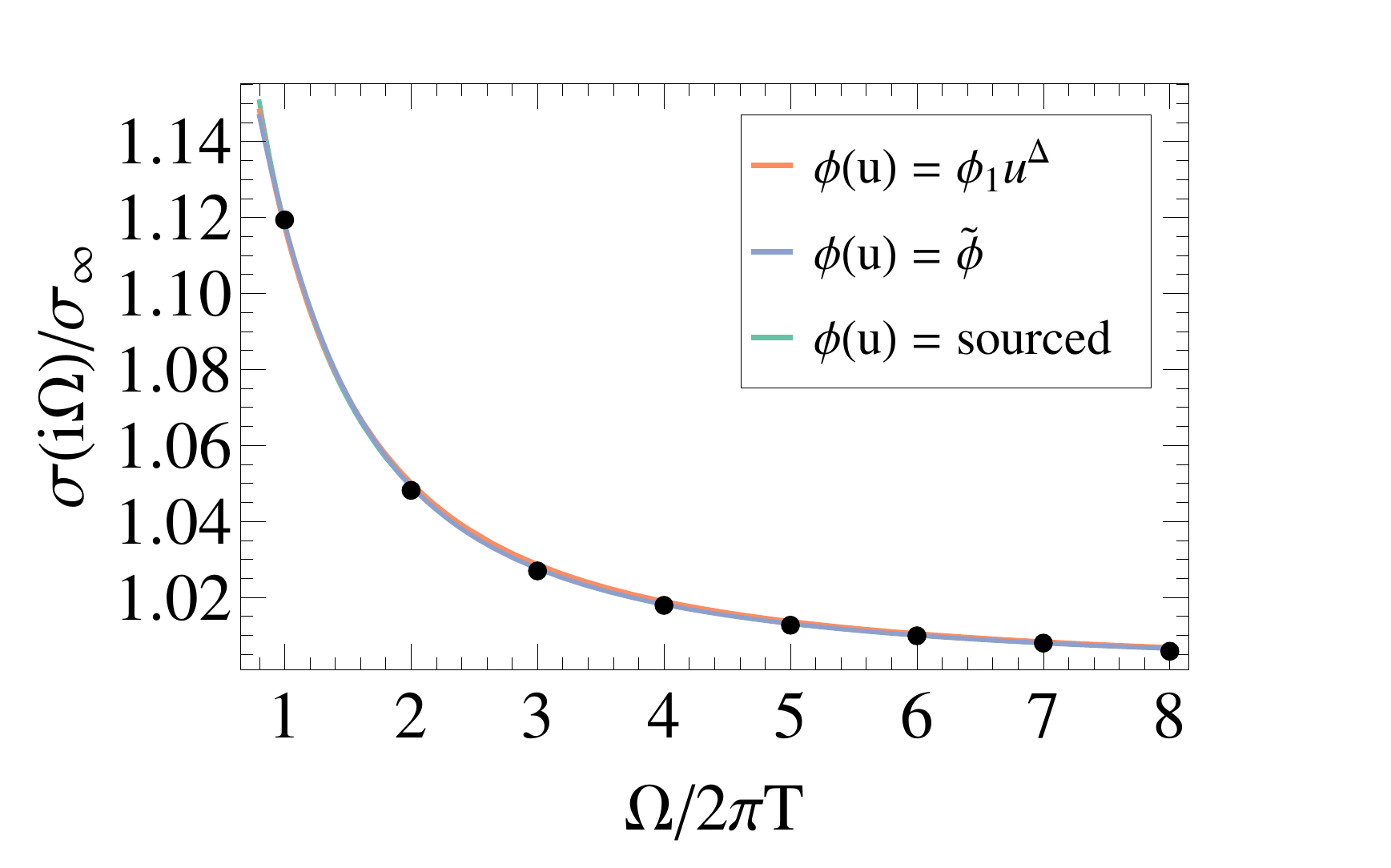}}
\end{subfigure}
\begin{subfigure}{0.5\textwidth}{
\includegraphics[width=\textwidth]{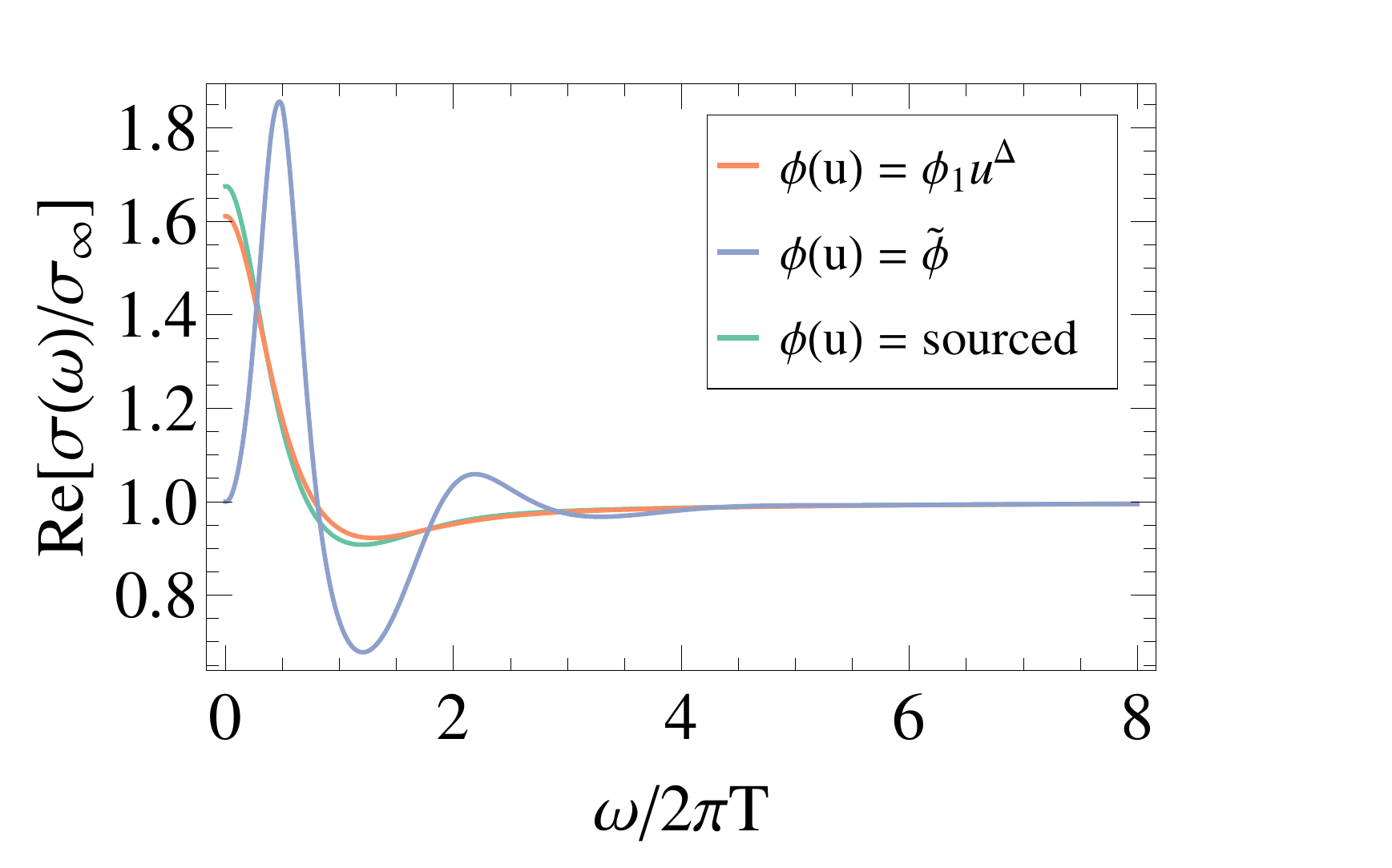}}
\end{subfigure}
\caption{Conductivity for Euclidean (left) and real (right) frequencies for $\Delta=1.5$ with $\aaa \alphamax$ fit to the quantum Monte Carlo data \cite{katz,natphys}. The fit yields
$\aaa\alphamax = 0.611,\ 0.581$ and 0.589 for the profiles proportional to $\aaa u^\Delta$, $\tilde \phi$, and that given 
by our model eq.~\reef{solution}, respectively. \labell{various}}
\end{figure} 
However, the power-law profile considered in \cite{katz} is a more or less ad hoc choice and we would like to emphasize the importance of developing a self-consistent holographic model for potential future studies. To illustrate this point, we show the
result of fitting the QMC data with holographic models constructed in the same spirit as \cite{katz} with a new simple scalar profile: 
\beq
\tilde \phi =  \aaa\,u^\Delta\left(1 + 2 u^6 - 3 u^{12}\right)\,. \labell{frump}
\eeq
As shown figure \ref{various}, the model with this new profile fits the QMC data for imaginary frequencies essentially as well as that with the $u^\Delta$ profile or our holographic model. However, as the figure also shows, evaluating the conductivity for real frequencies with the new profile yields rather different behaviour for $\omega<4\pi T$. In particular, the scalar profile in eq.~\reef{frump} was designed to yield $\sigma_0=\sigma_\infty$.

Let us consider the fit for the imaginary-frequency conductivity in more detail. As noted above, the QMC studies only yield $\sigma(i\W)$ for the discrete Matsubara frequencies $\W=2\pi nT$ with $n=1,2,3,\cdots$. In particular, the first data point appears at $\Omega_n=2\pi T$ or at $\freq=3/2$, in terms of the dimensionless frequency introduced in eq.~\reef{nook}. Now examining eq.~\reef{JJperb}, we see that the contribution of the scalar profile to $\sigma(i\freq)$ is suppressed near the horizon by the exponential factor in the integral. Roughly, we can say that $\sigma(i\freq)$ only probes to holographic background up to $z\sim \tfrac{1}{2\freq}$. Hence we might conclude that the fit to all of the QMC data points is only probing the bulk geometry up to $z\sim 1/3$ or $u\sim0.3$ in our holographic model.\footnote{Note that
$z=\frac16\,\log\!\big[\frac{1+u+u^2}{(1-u)^2}\big]+\frac{1}{\sqrt{3}}\big[\! \tan^{-1}\!\big( \frac{2u+1}{\sqrt{3}}\big)-\frac{\pi}{6}\big]$.} On the other hand, the analytic continuation of the conductivity to real frequencies clearly relies much more on the detailed structure of the holographic model, including the near horizon region. Hence it is not difficult to engineer scalar field profiles which provide a good fit to the QMC data but yield disparate (and even peculiar) results for the real-frequency conductivity. For example, beyond the example given in eq.~\reef{frump}, one can easily construct 
examples where the conductivity seems to be vortex-like rather than particle-like, in the sense discussed in \cite{Myers:2010pk}, 
\ie with $\sigma_0<\sigma_\infty$. However, this simply illustrates the hazards of applying holography in an unprincipled manner,
and we conclude that the most constrained and most reliable approach is focus on constructing self-consistent holographic models.

It might be interesting to extend this comparison to the QMC data by including the contribution of the $\langle JJ T\rangle$ coupling, \ie one would extend the gauge field action \reef{gaugeX} to include an additional interaction proportional to $C_{abcd}F^{ab}F^{cd}$, as in \cite{Myers:2010pk}. As noted above, this new coupling would modify the high-frequency expansion \reef{asymp} of the conductivity by introducing a new contribution proportional to $1/\freq^3$. Including these contributions may improve the fit to the QMC data. However, a priori, it is not clear if extending the calculations to higher orders in the $\alphasource$ expansion will produce equally important modifications of the conductivity. 
Of course, our model can be easily adapted with other conformal QCPs, such as the Ising CFT in $d=2+1$.  
%where the conformal dimension of the relevant operator takes any allowed value, \ie $1/2\le\Delta\le3$.  
It is likely that the stress tensor contributions will become more important as the conformal dimension of $\cO$ moves closer to 3.

\subsection{Tuning away from criticality}\label{non critical} 

Throughout the main text, we were considering a critical boundary theory which required setting the coefficient of the non-normalizable mode in eq.~\reef{boots} to zero. 
As was commented above, this coefficient $\bbb$ is dual to the coupling to the scalar operator $\cO$ in the boundary theory, as in eq.~\reef{punt}. More precisely, we have 
\beq
\lambda =\left(\frac{r_0}{L^2}\right)^{3-\Delta} \bbb = \left(\frac{4\pi T}{3}\right)^{\!3-\Delta} \bbb\,.
\labell{newdef}
\eeq
Hence setting $\bbb=0$ corresponds to the tuning needed to reach a QC phase transition as discussed in section \ref{sec:ingredients}. However, our holographic model then also provides a starting point to examine the response functions as a function of the relevant coupling $\lambda$ as we tune away from the QCP. 
To study the off-critical behaviour of the boundary theory, we simply need to extend our analysis to scalar profiles \reef{solution} 
having nonvanishing $\bbb$. 

As in the main text, we would still calculate perturbatively in the amplitude of the scalar field and so our analysis would be limited to the regime where $|\bbb| \sim |\lambda| / T^{3-\Delta}\ll 1$. 
We must also assume that $\cO$ is a relevant operator, \ie $\Delta<3$. For $\Delta>3$, the non-normalizable mode of the bulk scalar diverges asymptotically, \eg see eq.~\reef{boots}, and as a result, the back-reaction of the scalar field cannot be controlled for $\bbb\ne0$. In order for $\phi$ to be regular at the black hole horizon, the coefficient $\aaa$ must be chosen as
\begin{equation}
\aaa = \aaa\big|_{crit}  - \bbb \ \times\ \frac{\Gamma\left(2-\frac{2\Delta}{3}\right)\Gamma\left(\frac{\Delta}{3}\right)^2}{\Gamma\left(1-\frac{\Delta}{3}\right)^2\Gamma\left(\frac{2\Delta}{3}\right)}\,,
\end{equation}
where $\aaa|_{crit}$ is the value given in eq.~\reef{wake3}. Hence as might be expected, the boundary theory responds linearly to the introduction of a small coupling $\lambda$. For example, the shift in the expectation value of the scalar operator becomes\footnote{We also expect that $\langle \mcO \rangle_{T=0}\ne0$ away from the QCP, however, our perturbative analysis does not capture this contribution which would be nonanalytic in the coupling $\lambda$.}
\beq
\langle \mcO \rangle_T 
- \langle \mcO \rangle_T\big|_{crit}=- {c_\Delta} \ C_T\,T^{2\Delta-3}\,\lambda \,,
\labell{expectnew}
\eeq
where $\langle \mcO \rangle_T\big|_{crit}$ is given by eq.~\reef{expect} and $c_\Delta$ is a numerical coefficient depending only on the conformal dimension. 
%is given by\footnote{This coefficient vanishes for $\Delta=3/2$ reminding us that this is a special case which requires special attention --- see appendix \ref{special1}.}
%\beq
%\gamma(\Delta)=\frac{\pi^2}{144}\! \left(\frac{4\pi }{3}\right)^{\!2\Delta-3} \frac{\Gamma\left(2-\frac{2\Delta}{3}\right)\Gamma\left(\frac{\Delta}{3}\right)^2}{\Gamma\left(1-\frac{\Delta}{3}\right)^2\Gamma\left(\frac{2\Delta}{3}-1\right)}\,.
%\eeq

\begin{figure}[!htb] \centering
  \begin{subfigure}{0.48\textwidth}{\centering
      \includegraphics[width=\textwidth]{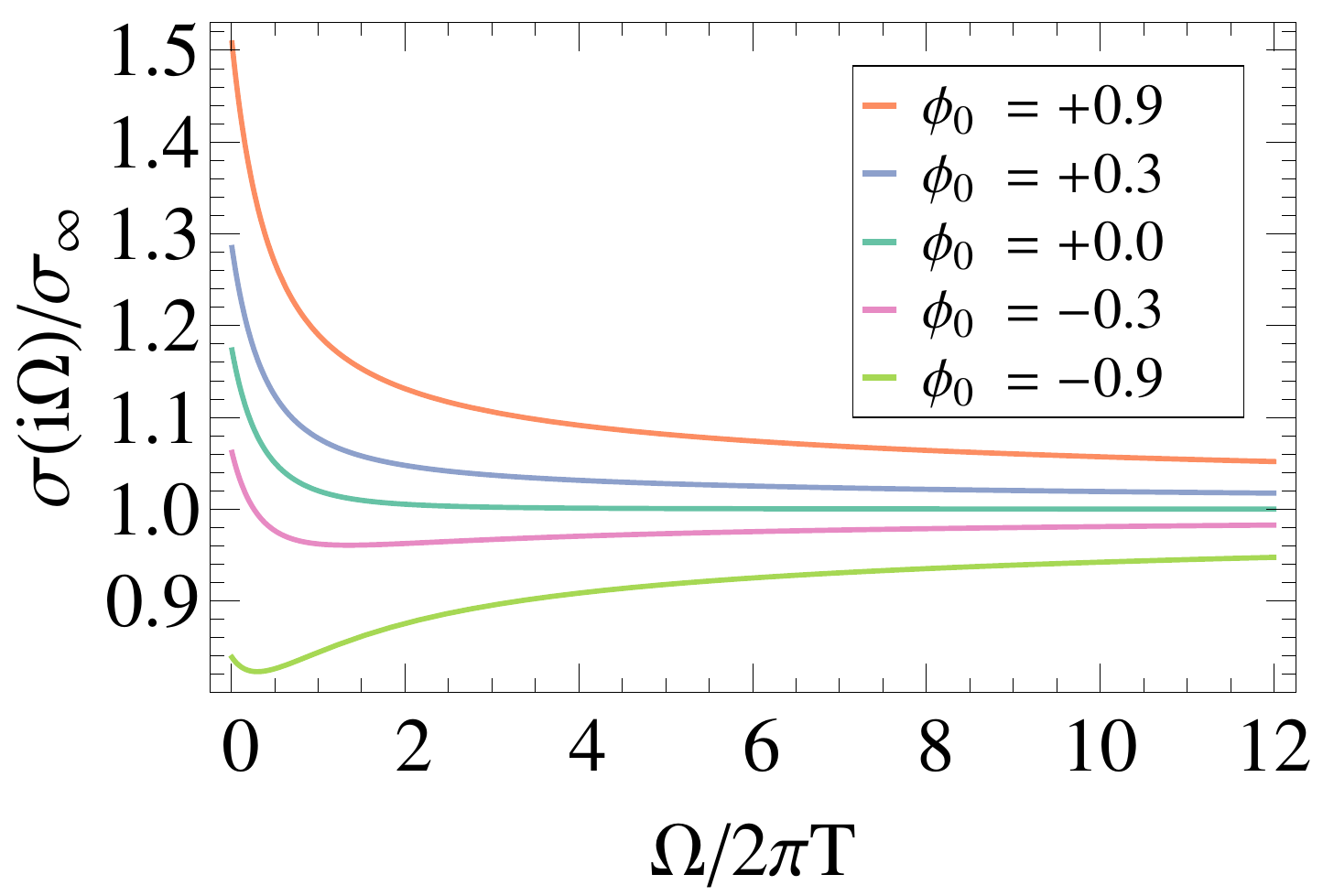}}
  \end{subfigure} \hspace{0.2cm}   
  \begin{subfigure}{0.48\textwidth}{\centering
      \includegraphics[width=\textwidth]{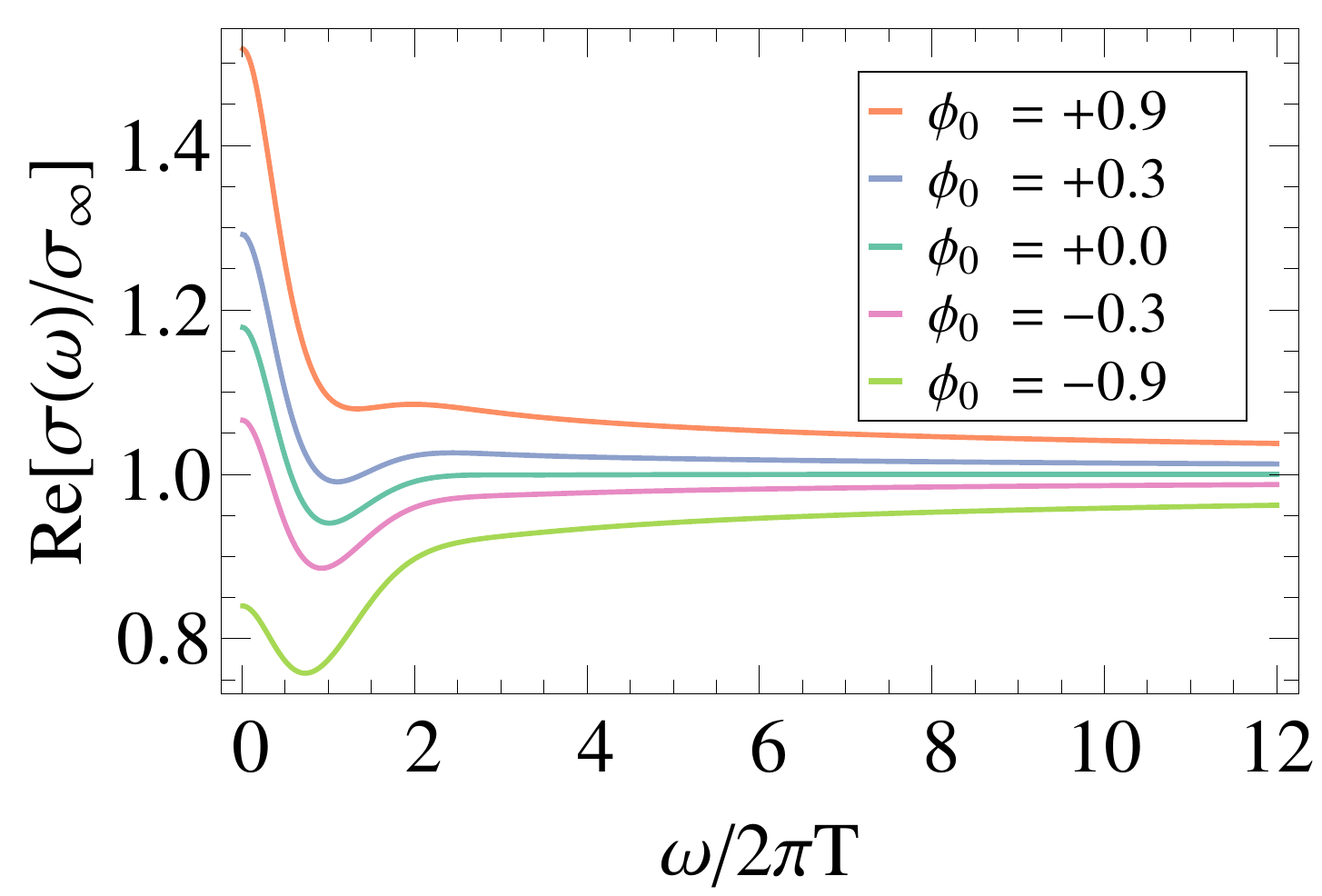}}
  \end{subfigure}
  \caption{{\bf Detuning from the QCP} -- Conductivity at Euclidean (left) and real (right) frequencies at various detuning strengths
$\bbb\propto\lambda$. We fixed $\Delta=2.5$ and $\alphasource\alphamax =0.1$.\labell{critical cond}}
\end{figure}
Given the new scalar profile, it is straightforward to again evaluate the dynamical conductivity, as described in appendix \ref{holosig}.
Figure \ref{critical cond} shows the response of the conductivity to variations of $\bbb$. One might note the similarity of the plot for imaginary frequencies to the QMC results, shown in figure 6(a) of \cite{natphys}
and also in \cite{gazit14}. The extension of the analysis of the high-frequency expansion given in section \ref{freqy} is also straightforward.    
In particular, turning on both coefficients in the near-boundary expansion \reef{boots} of the bulk scalar, the leading terms in 
the asymptotic expansion of the conductivity take the form
\begin{align}
  \frac{\sigma(i\W)}{\sigma_\infty}=1 + \alphamax\, b(\Delta)\frac{\lambda\ \ }{\W^{3-\Delta}} + \alphamax\,\frac{a(\Delta) \alphasource T^\Delta+\tilde{a}(\Delta) \lambda T^{2\Delta-3}}{\W^\Delta} +  \cdots\,.
\end{align} 
With $\lambda=0$, the second term proportional to $\alphamax$ is precisely the $1/\freq^\Delta$ term in eq.~\reef{asymp}. Hence we see that tuning away from criticality introduces a small shift in the $1/\W^\Delta$ contribution but it also generates a new term proportional to $1/\W^{3-\Delta}$ which is completely independent of the temperature. 
%rcm more comment for point 3
Let us emphasize that the above off-critical behaviour applies for $|\lambda| / T^{3-\Delta}\ll 1$. In terms of the phase diagram illustrated in figure \ref{fig:1}, we are studying the theory deep in the ``fan'' where the physics is still dominated by the QCP. We plan to investigate the off-critical response 
further in \cite{new}, with the goal of shedding light on the response functions in the entire phase
diagram near a quantum critical point.

\section*{Acknowledgments} 
We would like to thank A. Buchel, S.~Hartnoll, C.~Herzog, E.~Katz, A.~L. Fitzpatrick, A. Lucas, P. McFadden, and S. Sachdev 
for useful discussions. Research at Perimeter Institute is supported by the Government of Canada through the Department of Innovation, Science and Economic Development and by the Province of Ontario through the Ministry of Research \& Innovation. WWK was supported in part by a postdoctoral fellowship from NSERC. TS is supported in part by the Ontario Graduate Scholarship. RCM and TS are also supported in part by an NSERC Discovery grant. RCM is also supported by research funding from the Canadian Institute for Advanced Research and from the Simons Foundation through the ``It from Qubit" Collaboration. 

\appendix

\section{Vacuum correlation functions}\labell{oo sec}     
In this appendix, we provide some of the details of calculating various vacuum correlators in our holographic model, which are used in section \ref{main}. In order to calculate correlation functions, we will be 
working with Euclidean time, \ie the time coordinate for Euclidean spacetime is given by
the Wick rotation $t_E=-it$.

\subsection{Two-point functions} \labell{ap:2pt}

To evaluate the two-point correlation functions, we begin with the `free part' of the Euclidean bulk action 
\begin{equation}
S = \int d^{4} x \sqrt {g} \left[ \frac{1}{2 \pl^2} \left(-R - \frac{6}{L^2}+(\nabla_a\phi)^2 +m^2 \phi^2\right) + \frac{1}{4  g_{4}^2} F_{ab}F^{ab} \right]\,,
\labell{free}
\end{equation}
\ie the Euclidean version of eqs.~\reef{gamma0}, \reef{scalarX} and \reef{gaugeX} with $\alpha_1=0=\alpha_2$. We will be working with Poincar\'e coordinates in the AdS vacuum
\begin{equation}
ds^2 = g_{ab}\, dx^a dx^b = \frac{L^2}{z^2} \left( dz^2 + \delta_{\mu \nu}\,dx^\mu dx^\nu \right)\,, \labell{vack}
\end{equation}
where $\delta_{\mu\nu}$ is the three-dimensional Euclidean metric on $\mathbb{R}^3$. As in the main text, we will use Latin indices ($a,b$) to refer to bulk directions and Greek indices ($\mu,\nu$) to refer to boundary directions. Also, points in the AdS bulk will have no special emphasis $x$ but points on the boundary will denoted in bold $\bs x$. Of course, the
asymptotic boundary is reached with $z\to 0$. %The coordinates $\bs x $ span $\mathbb R^3$, while $z>0$ and $z\to0$ is the AdS boundary.
As usual \cite{revue}, the two-point functions, $\langle \mcO (\bs p)\, \mcO(-\bs p)\rangle $ and $\langle J_\mu(\bs p) J_\nu(-\bs p)\rangle$, will be calculated from the boundary term arising in evaluating the free on-shell action. 

Using the scalar equation of motion
$\left(\nabla^2 - m^2\right)\phi =0$ and Stokes' theorem, the scalar terms in the action \reef{free} reduce to the boundary term 
\begin{equation} 
S_\text{scalar} = - \frac{1}{2\pl^2}\int d^3 x \sqrt{g}g^{zz}\phi \,\partial_z\phi\, \Big|_{z=\epsilon} \labell{skales}
\end{equation}
where $z=\epsilon$ is a UV regulator surface. In order to evaluate this on-shell action \reef{skales}, we write the bulk solutions as
\beq
\phi(z,\bvect x)=\int \frac{d^3k}{(2\pi)^3}\, e^{i \bvect k \cdot  \bvect x}\,K_\Delta(z ,\bs k)\,\phi_0(\bs k) \labell{bb1}
\eeq
where $\phi_0(\bs k)$ of the Fourier transform of the boundary profile of the scalar field and $\bvect k \cdot \bvect x \equiv \delta_{\mu\nu} k^\mu x^\nu=  \Omega\, t_E + k^x x + k^y y$. This expression also uses the bulk-boundary propagator:
\begin{equation}
K_\Delta(z ,\bs k) =\epsilon^{3-\Delta} \frac{z^{3/2}K_{\Delta - 3/2}(|\bs k| z)}{\epsilon^{3/2}K_{\Delta - 3/2}(|\bs k| \epsilon)} \label{gg1}
\end{equation}
where we have introduced a UV cutoff $\epsilon$, and where $K_{\Delta - 3/2}(|\bs k| z)$ is modified Bessel function of the second kind. The expression for the action \reef{skales} then becomes
\begin{equation}
S_\text{scalar} = -\left.\frac{L^2}{2\pl^2}\int\! \frac{d^3 k}{(2\pi)^3}\, \frac{\left(\frac{z}{\epsilon}\right)^{3/2}K_{\Delta - 3/2}(|\bs k| z)\,\partial_z\left[\left(\frac{z}{\epsilon}\right)^{3/2} K_{\Delta - 3/2}(|\bs k| z) \right]}{\epsilon^{-2(3-\Delta)}z^2\ K_{\Delta - 3/2}(|\bs k| \epsilon)^2} \right|_{z\rightarrow\epsilon}\phi_0(\bs k) \phi_0(- \bs k).
\end{equation}
The expansion of this expression is divergent as $\epsilon \rightarrow 0$, but all of the divergent terms are analytic in $k$ and can be removed by adding local counterterms \cite{revue,Chowdhury:2012km}. After evaluating the remaining expression and using
\begin{equation}
\langle \mcO(\bs k) \mcO(-\bs k) \rangle %= \frac{\delta}{\delta \phi_0(\bs k)}\frac{\delta}{\delta\phi_0(-\bs k)} S_\text{scalar}
= - \frac{\delta^2\,S_\text{scalar}}{\delta \phi_0(\bs k)\,\delta\phi_0(-\bs k)} \,,
\end{equation}
we find
\begin{equation}
\begin{split}
\langle \mcO(\bs k) \mcO(-\bs k) \rangle =& \epsilon^{3-2\Delta}\,\frac{L^{2}}{\pl^2} \left( 3 - \Delta - \frac{|\bs k| \epsilon\, K_{\Delta-5/2}(|\bs k|\epsilon)}{K_{\Delta - 3/2}(|\bs k|\epsilon)}\right)\\
\simeq &\cdots + (2\Delta -3)\,\frac{L^{2}}{\pl^2}\,\frac{ \Gamma(3/2-\Delta)}{\Gamma(\Delta - 3/2)}\ \left(\frac{|\bs k|}{2}\right)^{2\Delta -3} \labell{book7}
\end{split}
\end{equation}
where the ellipsis represents the (power law) divergent terms which are removed by local counterterms. Note that in all of the momentum space correlation functions that we write, there is an implicit $(2 \pi)^3 \delta^{(3)}(\sum_n{\bs k}^{(n)})$ factor from conservation of momentum. Instead of writing this factor repeatedly, we write the correlation functions to explicitly have momentum conservation and we drop the $\delta$-function term. As already commented in the main text and is clear from the above expression, $\Delta\!=\! d/2\!=\! 3/2$ and indeed any half-integer value of $\Delta$, are special cases \cite{mcfadden}. 

Similarly, we can evaluate the free gauge action (\ref{free}) on-shell using the equation of motion $\nabla_a\left( F^{ab}\right) = 0$ and Stokes' theorem to find
\begin{equation}
S_\text{gauge} = - \frac{1}{2g_4^2} \int d^3 x \sqrt{g} g^{zz}g^{\mu\nu} A_\mu \, \partial_z A_\nu\, \Big|_{z=\epsilon}\,.
\end{equation}
Implicitly, we have chosen the standard gauge where $A_u=0$ and $\nabla^\mu A_\mu=0$. Then we write the bulk gauge field as
\beq
A_\mu(z,\bs x) = \int \frac{d^3k}{(2\pi)^3}\, e^{i \bvect k \cdot  \bvect x}\,G_{\mu \rho}(z,\bs k)A^\rho_0(\bs k)\,, \labell{bb2}
\eeq
where the bulk-boundary gauge propagators are given by
\beq
G_{\mu \nu}(z,\bs k) =  e^{-|\bs k|z}\ I_{\mu \nu}(\bs k)\,, \qquad {\rm with}\quad I_{\mu\nu} = \delta_{\mu\nu} - 
\frac{\bs k_\mu \bs k_\nu}{|\bs k|^2}\,. \labell{gg2}
\eeq
Now using
\begin{equation}
\langle J_\mu(\bs k) J_\nu(-\bs k) \rangle = - \frac{\delta^2\, S_\text{gauge}}{\delta A_0^\mu\ \delta A_0^\nu} \ ,
\end{equation}
we find
\begin{align}
\langle J_\mu(\bs k) J_\nu(-\bs k) \rangle = \frac{1}{g_4^2}\left.\int\! \frac{d^3 k}{(2\pi)^3}\, %\sqrt{-g}g^{zz}g^{\rho\sigma} 
\delta^{\rho\sigma} G_{\rho\mu}(z, \bs k)\, \partial_z G_{\sigma\nu}(z,\bs k)\right|_{z\rightarrow 0} = - \frac{|\bs k|}{g_4^2}\tij(\bs k)\,.
\end{align}
The above equation shows that $\sigma_\infty=1/g_4^2$.

\subsection{Calculation of $\langle J J \mathcal O \rangle$}\labell{jjo sec}
We are interested in calculating  $\langle J J \mathcal O \rangle$ for our holographic model where we have 
added the interaction 
\begin{equation}
S_\text{int} =  \frac{\alphamax}{4g_4^2}\int d^{4} x \sqrt{g}\, \phi\, F_{ab}F^{ab}\,.\labell{jjo int}
\end{equation}
Of course, this (vacuum) correlator vanishes for the boundary theory dual the free bulk action \reef{free}. However, with the above interaction, $\langle J J \mathcal O \rangle$ is given by a single process --- see figure \ref{witt1}.\footnote{Let us emphasize that we are only considering the classical theory in the bulk, \ie the ``planar'' limit of the boundary theory. In principle, quantum processes in the bulk would modify this result but these corrections should be suppressed in the regime where we are studying the theory.}
\begin{figure}[h]
\centering
\def\svgwidth{175pt}
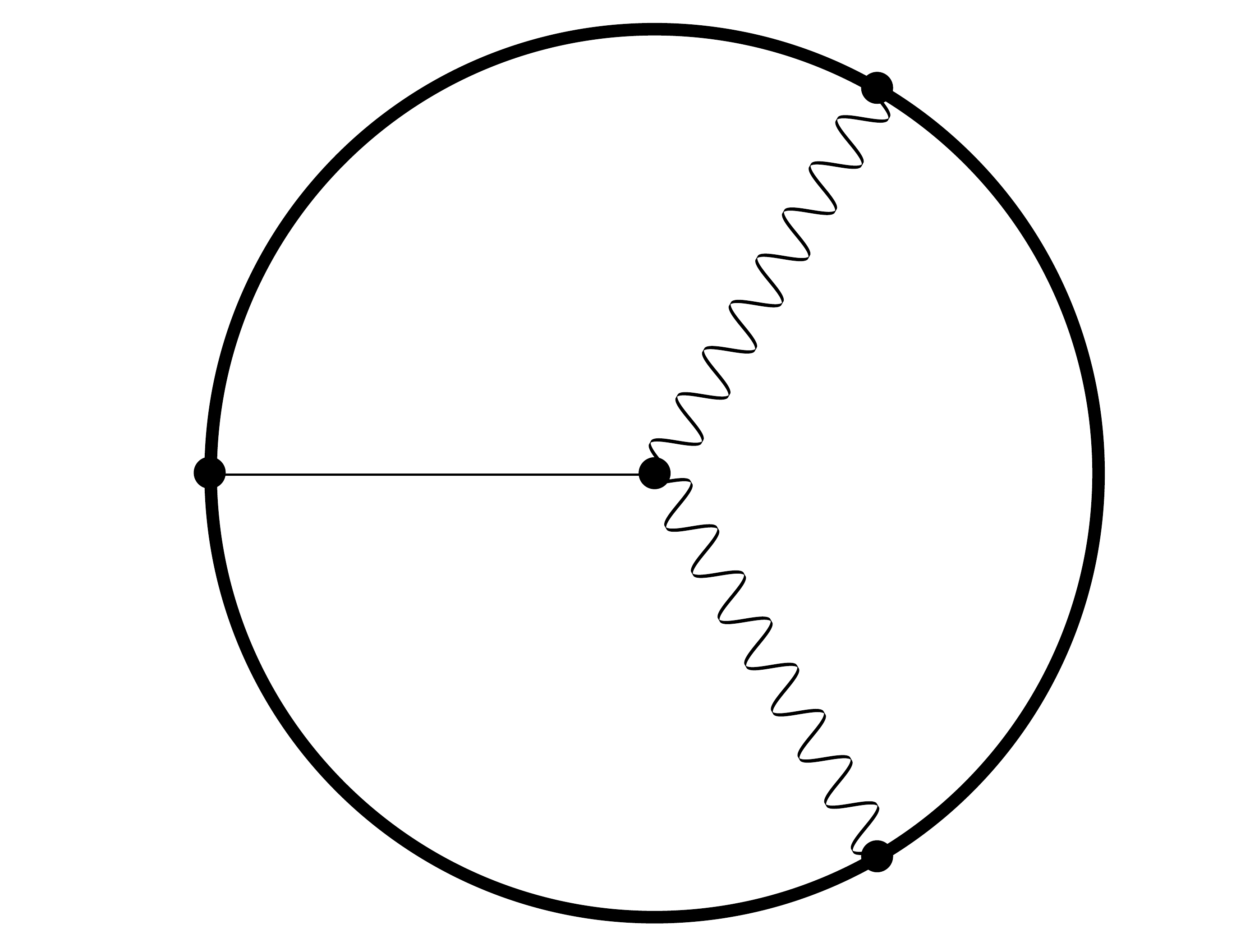
\caption{Witten diagram illustrating the tree-level process contributing to $\langle JJ\mathcal O\rangle$. \labell{witt1}}
\end{figure}
To calculate the three-point function at this tree-level order, we only need  evaluate eq.~(\ref{jjo int}) on-shell. Hence we substitute eqs.~\reef{bb1} and \reef{bb2} to find
\begin{equation}
\begin{split}
S_\text{int} = \frac{\alphamax}{2g_4^2}\int d^{4} x \sqrt{g} \int &\frac{d^3 k\, d^3 p\, d^3 q}{(2\pi)^9} e^{i \bs x\cdot(\bs p+\bs k+\bs q)} K_\Delta(z,\bs q) \phi_0(\bs q)\\
\times\ \ & g^{ab}g^{\mu\nu}\partial_a G_{\mu \rho}(z,\bs k) \partial_{[b} G_{\nu]\sigma}(z,\bs p) A^\rho_0(\bs p)A^\sigma_0(\bs k),
\end{split}
\end{equation}
and then the correlation function is given by
\begin{equation}
\langle J_\mu(\bs p_1) J_\nu(\bs p_2) \mcO(\bs p_3)\rangle  = - \frac{\delta}{\delta \phi_0(\bs p_3)} 
\frac{\delta}{\delta  A_0^\mu(\bs p_1)} \frac{\delta}{\delta A_0^\nu(\bs p_2)} S_\text{int}\,. \labell{jjo def}
\end{equation}
With the propagators $K_\Delta$ and $G_{\mu\nu}$ given in eqs.~\reef{gg1} and \reef{gg2}, 
a straightforward calculation then yields 
\begin{multline}
\langle J_\mu(\bs p_1) J_\nu(\bs p_2) \mcO (\bs p_3)\rangle = - \frac{\alphamax}{g_4^2} 
{\int_0^\infty}\!  dz\, \frac{|\bs p_3|^{(\Delta -3/2)} z^{3/2} K_{\Delta - 3/2}(|\bs p_3| z)e^{-(|\bs p_1|+|\bs p_2|)z}}{\Gamma(\Delta - 3/2)\,2^{\Delta - 1/2}} %(2\pi)^3 \delta^3(\bs p_1+\bs p_2+\bs p_3)
\\\!\! \times \left[|\bs p_1| |\bs p_2| \left(\delta_{\mu\nu} - \frac{p_{1\mu} p_{1\nu}}{|\bs p_1|^2} 
- \frac{p_{2\mu} p_{2\nu}}{|\bs p_2|^2} + \frac{(\bs p_1\cdot \bs p_2) p_{1\mu} p_{2\nu}}{|\bs p_1|^2 |\bs p_2|^2}\right) 
- (\bs p_1 \cdot \bs p_2) \delta_{\mu \nu} + p_{2\mu} p_{1\nu} \right]\! ,\label{general jjo}
\end{multline}
where again we have an implicit $\delta$-function on the right-hand side imposing $\sum_{a=1}^3 \bs p_a=0$.
To apply this result in section \ref{main}, we choose $\bs p_1 = \bs q$, $\bs p_2 = \bs p - \bs q$ and 
$\bs p_3 = - \bs p$ where $\bs q =(\Omega,0,0)$ and $\bs p=(\tilde\Omega,0,0)$; we also assume $\Omega,\tilde\Omega>0$. Now in the limit that $|\bs q| \gg | \bs p|$, the $(x,x)$-component of eq.~(\ref{general jjo}) reduces to
\begin{equation}
\langle J_x(\bs q) J_x(\bs p - \bs q) \mcO(-\bs p)\rangle = 
- \frac{\alphamax}{g_4^2}\frac{\Omega^2\,|\bs p|^{\Delta - 3/2}}{\Gamma(\Delta - 3/2)2^{\Delta - 3/2}} \int  dz  z^{3/2} K_{\Delta - 3/2}(|\bs p| z)\,e^{-2\,\Omega\, z}\,. \label{integral}
\end{equation}
After performing the remaining $z$-integral \cite{Grads:2007}, we find
\begin{equation}
\begin{split}
\langle J_x(\bs q) J_x(\bs p - \bs q) \mcO (-\bs p)\rangle\sing =&- \frac{\alphamax}{g_4^2} \frac{\Gamma(3/2-\Delta)}{\Gamma(\Delta - 3/2)} \left(\frac{|\bs p|}{2}\right)^{2\Delta -3} \frac{\Gamma(\Delta+1)}{2^\Delta \Omega ^{\Delta-1}}\\
=& -\frac{\alphamax}{g_4^2}\,\frac{\pl^2}{L^2}\,\frac{\Gamma(\Delta+1)}{ 2^\Delta(2\Delta - 3) \Omega ^{\Delta-1}}\, \langle \mcO(\bs p) \mcO(-\bs p)\rangle\,,
\end{split}\labell{oops}
\end{equation}
where we have used eq.~\reef{book7} to relate the final result to the two-point function of the scalar operator. Our notation here indicates that we are calculating the singular or nonanalytic part of the three-point function. In particular, it also contains contributions
which are analytic in $|\bs p|$  
as $|\bs p|\to0$ \cite{mcfadden}, but these will not contribute to the OPE in section \ref{anal}. 
We note that our result \ref{oops} appears to be problematic for half-integer conformal dimensions, \ie $\Delta=\frac{1}{2}+n$ with
$n$ is a non-negative integer. Extra care is required in these special cases \cite{mcfadden}. We refrain from describing the necessary calculations here, however,
we refer the interested reader to appendix \ref{special1} for further discussion on $\Delta=3/2$. 

\subsection{Calculation of $\langle TT\mathcal O\rangle$}

Again, the $\langle TT\mathcal O\rangle$ correlator vanishes for the boundary theory dual the free bulk action \reef{free}. However, we included the interaction 
\begin{equation}
S_\text{source} = - \alphasource \frac{L^2}{\pl^2} \int d^4 x\sqrt{g}\, \phi \,C_{abcd}C^{abcd}\label{source}
\end{equation}
in our holographic action, which has the effect of generating a nonvanishing three-point function.  Again,
there is a single (classical) process contributing to $\langle TT\mathcal O \rangle$, shown in figure \ref{witt2}.
\begin{figure}[ht]
\centering
\def\svgwidth{200pt}
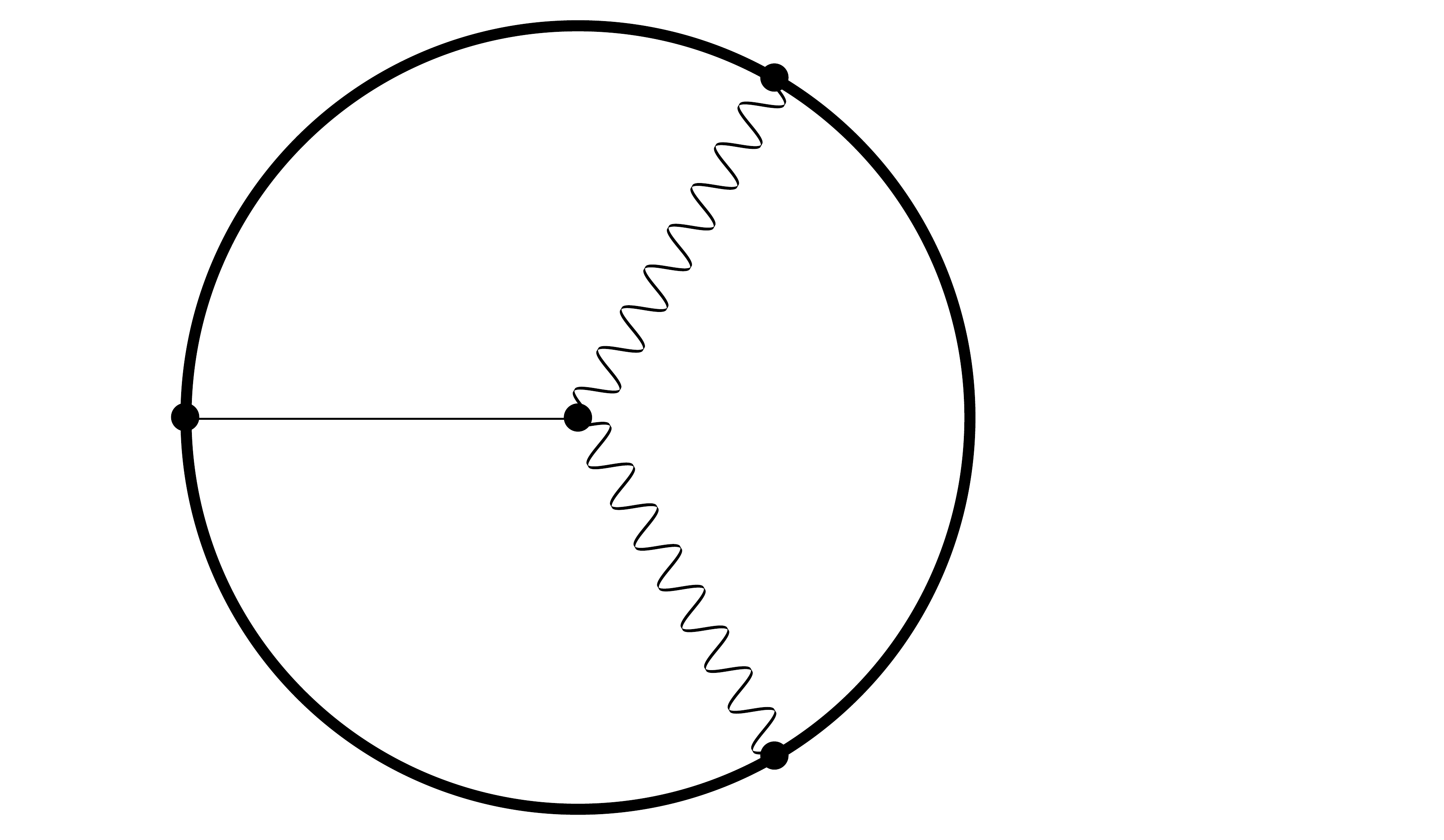
\caption{Witten diagram  illustrating the tree-level process contributing to  $\langle TT\mathcal O\rangle$.\labell{witt2}}
\end{figure}
Of course, the boundary stress tensor is dual to metric perturbations around the AdS vacuum \reef{vack}. Following \cite{Chowdhury:2012km}, we normalize the perturbations with
\begin{equation}
g_{ab}(z,\bs x) = \frac{L^2}{z^2}\,\delta_{ab} + h_{ab}(z,\bs x)\,,
\end{equation}
where we recall that we are working in Euclidean time.
Choosing the standard gauge where $h_{\mu z}=0$ and $\delta^{\sigma\mu}\partial_\sigma h_{\mu\nu} = 0$, we may write the on-shell metric perturbations as
\begin{equation}
h_{\mu\nu}(z,\bs x) = \frac{L^2}{z^2}\int\frac{d^3k}{(2\pi)^3}\,e^{i \bs k \cdot \bs x}\, e^{-|\bs k|z}(1+|\bs k|z)\,h^0_{\mu\nu}(\bs k)
%\qquad\epsilon_\mu(\bs k) \epsilon_\nu(\bs k) .
\end{equation}
where the polarization tensor is transverse and traceless, \ie $k^\mu h^0_{\mu\nu}(\bs k)=0=
\delta^{\mu\nu}h^0_{\mu\nu}(\bs k)$. 
%[$\epsilon(\bs k)\cdot \bs k=0=\epsilon(\bs k)\cdot\epsilon(\bs k)$]
In order to evaluate the on-shell action, it is convenient to use the expansion of the Weyl tensor in terms of the metric perturbations given in \cite{Chowdhury:2012km}. 
Then the source action (\ref{source}) evaluates to 
\begin{equation}
\begin{split}
S_\text{source} =& -\alphasource \frac{L^2}{2\pl^2} \int d^4 x\sqrt{g} \int\frac{d^3k\,d^3p\,d^3q}{(2\pi)^9} e^{i \bs x \cdot (\bs p+\bs k+\bs q)} K_{\Delta - 3/2}(|\bs q| z) \,\phi_0(\bs q)\\
&\times\ \ \bigg[ \frac{z^4}{L^4}(|\bs k|^2 R_k + \ddot R_k ) ( |\bs p|^2 R_p + \ddot R_p)\,{\rm tr}\big(h^0(\bs k)\cdot h^0(\bs p)\big)
\\
%\left(\epsilon_k\cdot \epsilon_p\right)^2 \right.\\
&\qquad\ - \frac{4 z^4}{L^4}\dot R_k \dot R_p\Big(\bs k \cdot \bs p\,{\rm tr}\big(h^0(\bs k)\cdot h^0(\bs p)\big) - \bs p\cdot h^0(\bs k)\cdot
 h^0(\bs p)\cdot \bs k\Big)\bigg]
%(\bs p \cdot \epsilon_k)(\bs k \cdot \epsilon _p))
\end{split}
\end{equation}
where $R_k(z) =  e^{-|\bs k|z}(1+|\bs k|z)$, $\dot R_k(z)=\partial_z R_k(z)$ and the polarization tensors are contracted with the flat boundary metric, \ie ${\rm tr}\big(h^0(\bs k)\cdot h^0(\bs p)\big)=\delta^{\mu\nu}\delta^{\rho\sigma}h^0_{\mu\rho}(\bs k)h^0_{\nu\sigma}(\bs p)$.
The desired boundary correlator would then be given by the variation
\begin{equation}
\langle T_{\mu\nu}(\bs p_1) T_{\rho\sigma}(\bs p_2) \mcO(\bs p_3)\rangle = - \frac{\delta}{\delta \phi_0(\bs p_3)} \frac{\delta}{\delta  h_0^{\mu\nu}(\bs p_1)} \frac{\delta}{\delta h_0^{\rho\sigma}(\bs p_2)} S_\text{source}\,. \labell{tto def}
\end{equation}
Analogously to our calculation of the $\langle J_\mu\, J_\nu \,\mcO \rangle$ correlation function, we let $\bs p_1 = \bs q$, $\bs p_2 = \bs p - \bs q$ and $\bs p_3 = - \bs p$ where $\bs q=(\Omega,0,0)$ and $\bs p=(\widetilde\Omega,0,0)$. Further, we consider the limit $|\bs q| \gg |\bs p|$, which yields
\begin{equation}
\langle T_{xy}(\bs q) T_{xy}(\bs p - \bs q) \mcO(-\bs p)\rangle = - \frac{\alphasource L^2\Omega^6 |\bs p|^{\Delta - 3/2}}{\pl^2\,\Gamma(\Delta - 3/2) 2^{\Delta -9/2}} \int dz z^{7/2}\,K_{\Delta - 3/2}(|\bs p| z)\,e^{-2\Omega z}\,.
\end{equation}
This equation is of the same form as eq.~(\ref{integral}) and so we perform the final $z$ integral in the same way. The final three-point function can be expressed in terms of $\langle \mathcal O \mathcal O \rangle$ using eq.~\reef{book7} to yield
\begin{equation}
\langle T_{xy}(\bs q) T_{xy}(\bs p - \bs q) \mcO(-\bs p)\rangle \sing = \alphasource \, \frac{\Gamma(\Delta +3)}{2^{\Delta-1} (2\Delta - 3) \Omega^{\Delta-3}}
\,\langle \mathcal O(\bs p) \mathcal O(-\bs p)\rangle\,,
\end{equation}
where as before we focus on the singular (non-analytic) part in $|\bs p|$. As for the $\langle JJ\cO\rangle$ correlator, we note that this result is valid for
scaling dimensions different from $\Delta=\frac{1}{2}+n$, where
$n$ is a non-negative integer.

\section{Holographic dynamical conductivity} \labell{holosig}

Here, we describe some of the details for the calculation of the dynamical conductivity $\sigma(\omega)$ in section \ref{hsigma}. In particular, we must solve the
the equations of motion for the gauge field resulting from eq.~\reef{gaugeX}, 
\begin{equation}
\nabla_a\left[(1+\alphamax\phi)F^{ab} \right]=0.\labell{maxwell}\,.
\end{equation}
Following \cite{Myers:2010pk}, we choose the standard gauge where $A_u=0$ and $\nabla^\mu A_\mu=0$  and expand $A_\mu$ in momentum space
\begin{equation}
A_\mu(u,t,x,y) = \int \frac{d^3q}{(2\pi)^3}\, e^{i \bvect q \cdot  \bvect x}\, {A_\mu}(u,\bvect q)\,,
 \end{equation}
where in real time, as usual, $\bvect q \cdot \bvect x \equiv - \omega t + q^x x + q^y y$. We can calculate the transverse component $A_y$ (setting $q^y=0$) in order to find the conductivity, which is then given by
\begin{equation}
\sigma(\omega) = \frac{4 \pi T}{3i \,g_4^2\omega}\ \left.\frac{\partial_u A_y}{ A_y}\right|_{u\rightarrow0}\,.
\labell{pour}
\end{equation}
Recall that the temperature $T$ is given in eq.~(\ref{temp}). 

Now we wish to solve for $A_y(u,\bvect q)$, the radial profile  of the gauge field, in the background given by the black hole metric \reef{metric} and the scalar profile \reef{solution}, with $\aaa$ set as in eq.~\reef{wake3} and $\bbb=0$. Since we are only interested in the frequency dependence, we consider the Fourier transform of eq.~(\ref{maxwell}) inserting the above momentum and then take the limit $q^x \to 0$,  
% \begin{equation}
% \partial_u\Big((1+\alphamax\phi(u))\,(1-u^3)\,\partial_u {A_y}\Big) + \left(\frac{3 \omega}{4 \pi T}\right)^2 \frac{1+ \alphamax\phi(u)}{1-u^3}\, {A_y} = 0\,,
% \labell{gauge ode}
% \end{equation}
\begin{equation}
\partial_u\Big(X_1(u)\,f(u)\,\partial_u {A_y}\Big) + \left(\frac{3 \omega}{4 \pi T}\right)^2 \frac{X_1(u)}{f(u)}\, {A_y} = 0\,,
\labell{gauge ode}
\end{equation}
where $X_1(u)=1+\alpha_2 \phi(u)$ using the notation of \cite{Myers:2010pk}, and $f(u)=1-u^3$. 
We solve this equation numerically with $\phi(u)$ given in eq.~(\ref{solution}). 
However, we must first determine the boundary conditions at the horizon (\ie $u=1$): We take the ansatz
\begin{equation}
A_y(u,\bs q) = (1-u)^b \ F(u),\labell{scalar ansatz}
\end{equation}
where $F(u)$ is assumed to be regular at $u=1$. To alleviate the notation,
we leave the $\bs q$ dependence of $F$ implicit.
 Without any loss of generality, we set $F(1) = 1$. 
Substituting into eq.~(\ref{gauge ode}) and taking the limit $u\to 1$, we find
\begin{equation}
(1-u)^{(b-1)}\ \frac{3 X_1(1)}{16 \pi^2T^2} \ (16 b^2\pi^2T^2+\omega^2)= 0\,.
\end{equation}
In order for $F(u)$ to be regular we require that 
\begin{align}
  b = -i\frac{\omega}{4 \pi T}\,. 
\end{align}
Now looking at the next-to-leading order 
term in eq.~(\ref{gauge ode}), we find
\begin{equation}
\frac{3(i \omega - 2 \pi T)X_1(1)(4 \pi T F'(1) - i \omega) + 6 i \pi T \omega X_1'(1)}{8 \pi^2T^2}=0\,.
\end{equation}
Following \cite{will-hd}, we may write the desired boundary condition for $F'(1)$ as
\begin{equation}
F'(1) = -b \left[ 1+ \frac{1}{1+2b } \frac{X_1'(1)}{X_1(1)} \right]\,.
\end{equation}
With this condition and the choice of $b$ fixed above, we now have the two boundary conditions needed to solve eq.~\reef{gauge ode} for the profile $A_y(u,\bs q)$ and then evaluate the corresponding conductivity \reef{pour}. We note that these calculations 
can also be carried out for imaginary frequencies by setting $\omega\to i\Omega_n$.

\section{Scalar profile for special $\Delta$ }\labell{special}

In this appendix, we consider the scalar profile and conductivity for some special values of 
$\Delta$, the conformal dimension of the operator $\mcO$ dual to $\phi$. In particular, we show that the profile takes 
a simple power-law form when $\Delta=3$, the marginal case. We also consider the solution for $\Delta = 6$, which sits on boundary of the values where eq.~(\ref{solution}) is no longer valid, \ie $\Delta \ge 6$.
We also comment on $\Delta=3n$ for integer $n>2$.  
Finally, we examine the case $\Delta=3/2$ where eq.~\reef{solution} also fails because the two independent solutions given there  are actually identical.

\subsection{$\Delta=3$}\labell{marginal}

When the scaling dimension of the scalar operator is $\Delta=3$, the bulk scalar field is massless. In this case, the scalar wave equation (\ref{scalar ode}) reduces to
\begin{equation}
 u^4\,\partial_u\!\left(\frac{(1-u^3)\,\partial_u\phi(u)}{u^2}\right) + 12\alphasource u^6=0\,.
\end{equation}
The solution for latter has a simple closed form:
\begin{equation}
\phi(u) = -\frac{4 \alphasource  }{3}\,(1-u^3) + c_1  + \frac{4 \alphasource - c_2}{3}\,\log(1-u^3)\,,
\end{equation}
where $c_1$ and $c_2$ are integration constants. For $\phi(u)$ to have the desired boundary conditions, 
\ie $\phi \sim u ^3$ near the asymptotic boundary $u\to 0$ and regularity at the horizon, we must choose $c_1 = 4 \alphasource / 3$ and $c_2 = 4 \alphasource$. With this choice, the solution reduces to
\begin{equation}
\phi(u) = \frac{4 \alphasource }{3}\, u^3.
\end{equation}
Hence, we see that the scalar field has a simple power law profile for the case $\Delta=3$, which is precisely the scalar field profile used in \cite{katz}. 

Substituting $\Delta=3$ into the high frequency expansion of the conductivity (\ref{general cond}), we find
\begin{equation}
\frac{\sigma(i\freq)}{\sigma_\infty} = 1 + \frac{8 \alphasource \alphamax}{(2 \freq)^3} - \frac{720 \alphasource \alphamax}{(2\freq)^6} + O\left(\frac{1}{\freq^9}\right)\,.
\end{equation}
The first two terms of the series match the asymptotic expansion obtained using a WKB analysis in \cite{will-hd}.
Here the two series of higher order terms discussed in section \ref{finger} have collapsed to a single series because the conformal weight of the scalar operator $\cO$ matches that of the stress tensor. However, we should recall that we expect in a typical three-dimensional CFT the stress tensor will appear in the $JJ$ OPE \reef{jj ope} and so there would be additional contributions to the asymptotic expansion \reef{general cond}, beginning at the order $1/\freq^3$  --- see sections \ref{anal} and \ref{finger}. 

\subsection{$\Delta=6$}\labell{special2}
The point where the scalar operator has scaling dimension $\Delta = 6$ is a special case because it sits on the border line of where 
the solution given in eq.~(\ref{solution}) is no longer valid, \ie $\Delta \ge 6$. This situation is also distinguished by the fact that the source term in the bulk scalar equation \reef{scalar equation} and the normalizable mode \reef{boots} have precisely the same asymptotic decay, \ie $u^\Delta=u^6$ --- see comments below. We will see in the following that this leads to additional logarithmic factors appearing in the radial profile of the scalar.\footnote{We explicitly verified that if the power of the source term in eq.~\reef{scalar equation} is replaced by $u^3$, analogous logarithmic factors appear for $\Delta=3$. Further, with the $u^3$ source term, the particular solution diverges for $\Delta>3$ in analogy to the divergences discussed below for $\Delta>6$.} Substituting $\Delta=6$ into eq.~\reef{scalar ode} yields
\begin{equation}
u^4\, \partial_u\! \left(\frac{(1-u^3)\,\partial_u\phi(u)}{u^2}\right) - 18\, \phi(u) + 12 \alphasource u^6 = 0\,. \labell{scalar ode6}
\end{equation}
and we find the general solution to be
\begin{equation}
\begin{split}
\phi(u) =&  \frac{2-u^3}{u^3}\,c_1 + \frac{4 + (2-u^3) \log{(1-u^3)}}{3u^3}\,c_2\\
&\quad-\frac{4\alphasource}3\,\bigg[\frac{(2-u^3) (\log(1-u^3) - 2 {\rm Li}_2(u^3))}{u^3}+6-u^3  \\
&\qquad\qquad\qquad-  \frac{6 \big(2u^3 + (2-u^3)\log(1-u^3)\big)}{u^3}\,\log u\bigg]\,,
\end{split}
\end{equation}
where ${\rm Li}_2(z)$ is the dilogarithm. In the near-boundary limit $u\to 0$, the non-normalizable mode dominates with $\phi(u)\to \frac{2}{u^3}\left( c_1+ \frac{2}3\,c_2\right)+\cdots$. For this model to accurately represent a QCP, this term must vanish. Thus we set $c_1 =  -\frac{2}{3} c_2 $ to eliminate this boundary divergence. Further, there is a potential logarithmic divergence as we approach the black hole horizon, \ie $u\to 1$. In order to remove this singularity at the horizon, we must set $c_2 = 4\alphasource$. With these choices, the solution reduces to
\begin{equation}
\phi(u) =-\frac{4\alphasource}3\,\bigg[4-u^3 -\frac{2(2-u^3) }{u^3}\, {\rm Li}_2(u^3)  
-  \frac{6 \big(2u^3 + (2-u^3)\log(1-u^3)\big)}{u^3}\,\log u\bigg]\,. \labell{ramble1}
\end{equation}
The leading two terms in the near-boundary expansion for $\phi(u)$ are given by
\begin{equation}
\phi(u\to0)= -\frac{2\alphasource}{27}\,u^6\, (18\, \log u +1 ) + O(u^9\log u)\,. \labell{pro}
\end{equation} 
Surprisingly, we see that the leading asymptotic behaviour has a puzzling logarithmic enhancement with $u^6\, \log u$. 
%At this point, one might ask what replaces the expression for the expectation value in eq.~\reef{expect}. However, properly evaluating $\langle\cO\rangle_T$ would require reconsidering the  holographic renormalization of this model. This interesting question is beyond the scope of the present paper --- however, see comments below. Further, 
However, given this scalar profile \reef{pro}, it is straightforward to determine the high frequency expansion of the conductivity following the analysis in section \ref{main}.
To leading order, we find 
\begin{equation}
\frac{\sigma(i \freq)}{\sigma_\infty} = 1 + \frac{16\alphasource\alphamax}{3}\,\frac{180\gamma_E-451 + \log(2\freq) }{(2\freq)^6} + \cdots \labell{wallop}
\end{equation}
where $\gamma_E$ is Euler's constant. Hence there is a logarithmic enhancement in the expected $1/\freq^6$ contribution. We leave the interesting question of connecting this result to the OPE analysis in section \ref{anal} for future study. However, we note again that for typical three-dimensional CFTs, this contribution would still be dominated by a $1/\freq^3$ term coming from the appearance of the stress tensor in the $JJ$ OPE;
our model does not contain such a contribution.  

%\subsection{$\Delta=3n$} \labell{specialN}

The two previous cases $\Delta=3,6$ are part of a more general trend valid for $\Delta = d\, n= 3n$, with 
integer $n>0$. With such a choice of scalar dimension, the equation of motion for the scalar admits a ``simple'' solution.
For instance, for $\Delta=9$ we find using the methods described above
\begin{multline}
  \phi(u)= \frac{4 \alpha_1}{3 u^6} \Big[u^3 \left(u^6-27 u^3+36+54 \left(u^3-2\right) \log u\right) \\
-6 \left(u^6-6 u^3+6\right) \left(\text{Li}_2\!\left(u^3\right)+3 \log u\, \log \left(1-u^3\right)\right)\Big]\,,
\labell{ramble2}
\end{multline}
where $\text{Li}_2(z)$ is the dilogarithm, which also appeared for the $\Delta=6$ case.
The small-$u$ expansion reads:
\begin{align}
  \phi(u)= \frac{\alpha _1}{3}\, u^6 + \frac{\alpha _1 }{225} \,u^9\, (180 \log u+43) + O(u^{12}\log u)  \,.
\end{align}
This will lead to the first subleading term in the asymptotic conductivity to go as $(T/\Omega_n)^6$,
irrespective of the fact that the scalar has $\Delta=9$. This contribution comes from the particular solution 
of eq.~\reef{scalar ode} rather than the homogeneous solution, \ie it is driven by the source term in the scalar field equation. Analogous behaviour will also hold for $\Delta=12,15,\cdots$.
 
We now discuss the general scalar profile eq.~\reef{solution} for general irrelevant scalar operators of large $\Delta$.
When the scalar field solution was introduced, we noted that if the boundary operator became too irrelevant, \ie for $\Delta\ge 6$, the profile given in eq.~\reef{solution} was no longer valid.  In particular, the function $g_\Delta (u)$ in eq.~\reef{solute} diverges for these values of the conformal dimension. To better understand the physical significance of this divergence,  we can introduce a UV cut-off surface at $u=\varepsilon\ll1$. With this cut-off, $g_\Delta (u)$ becomes
%todd fixed
\begin{equation}
g_\Delta(u) =  \int_{\varepsilon}^{u} dy\,y^{5-\Delta}\, \hyperf{1-\frac\Delta 3}{1-\frac\Delta 3}{2-\frac{2\Delta}{3}}{y^3}\,.
\labell{eppy}
\end{equation}
Now applying the usual boundary conditions, we would set $\bbb=0$ and $\aaa$ would be fixed as in eq.~\reef{wake3}. However, for that latter quantity, one finds
%todd added 12 / (2\Delta -3) UV limit is very apparent from integral now, fixed sign
\begin{equation}
\aaa\simeq -\frac{12\alphasource}{2\Delta-3}\,g_\Delta(1) \simeq - \frac{12\alphasource}{(\Delta-6)(2\Delta-3)}\,\bigg(\frac{3}{4\pi T\delta}\bigg)^{\Delta-6} \,,\labell{hat}
\end{equation}
where we have written the dominant contribution in terms of $\delta$, the physical short-distance cut-off in the boundary theory, using $\varepsilon=\frac{4\pi}3T\delta$.  For example then, the expectation value $\langle \cO\rangle_T$ in 
eq.~\reef{expect} diverges in the limit that the cut-off is removed, \ie $\delta\to0$. Therefore the holographic solution 
only really makes sense with a finite UV cut-off in this regime. 
% As noted above, for the special
% dimensions $\Delta=3n$, with integer $n>2$, well-behaved solutions to the equation of motion
% for $\phi(u)$ can be found. 

One might contrast the above treatment with the fact that eqs.~\reef{ramble1} and \reef{ramble2} provide perfectly finite solutions for $\Delta=6$ and 9, respectively.  In fact, finite solutions can be generated for general $\Delta\ge6$ by simply shifting the lower endpoint of the integral defining $g_\Delta(u)$ in eq.~\reef{solute}. Here, we would hold the endpoint fixed at some finite value of $y$, rather than tying it to the UV cut-off surface as in eq.~\reef{eppy}, which amounts to shifting $\aaa$ by a (divergent) constant. While this procedure yields a finite solution, it obscures the physical interpretation the holographic model by concealing the divergence in the expectation value  $\langle \cO\rangle_T$.
We leave this point for ulterior study.
    
%The fact that $\aaa$ diverges in the $\delta\to0$ limit is also a signal that our calculations are breaking down in this regime. Recall that our construction of the background was perturbative in the amplitude of the bulk scalar field. Here, however, the amplitude becomes arbitrarily large and so we can no longer ignore the back-reaction of the scalar on the background spacetime geometry.
 
\subsection{$\Delta=3/2$}\labell{special1}
As noted previously, the scalar field solution \reef{solution} breaks down at $\Delta=d/2=3/2$ because the two independent solutions appearing there reduce to the same function. With $\Delta=3/2$, the scalar wave equation (\ref{scalar ode}) becomes
\begin{equation}
u^4 \partial_u \left(\frac{(1-u^3)\,\partial_u\phi(u)}{u^2}\right) + \frac{9}{4} \phi(u) + 12 \alphasource u^6 = 0\,. \labell{scalar ode15}
\end{equation}
The general solution of this equation can be written as
%todd 4's -> 8's
\begin{equation}
\begin{split}
\phi(u) = &\tfrac{2}{\pi}\, u^{3/2}\, K(u^3)\, (\aaa - 8 \alphasource\,\tilde g_{3/2}(u)) 
\\ & + \ \tfrac{2}{3}\, u ^{3/2}\, K ( 1-u^3)\, (\bbb + 8 \alphasource \,\tilde h_{3/2}(u))
\end{split}
\end{equation}
where $K(k^2)=F(\varphi=\frac\pi2,k)$  is the complete elliptical integral of the first kind.
As our notation above suggests, there is a simple relationship between $K(u^3)$ and the hypergeometric function appearing in eq.~\reef{solution} for $\Delta=3/2$: $\tfrac2 \pi K(u^3) =  \hyperf{\frac12}{\frac12}{1}{u^3}$. Further we can write $\tfrac{2}{3}K(1-u^3)\simeq\log u\ \hyperf{\frac12}{\frac12}{1}{u^3}+\cdots$, were the ellipsis denotes terms polynomial in $u^3$. The functions $\tilde g_{3/2}(u)$ and $\tilde h_{3/2}(u)$ provide the particular solution of eq.~(\ref{scalar ode15}) with
%todd fixed and removed denominator
\begin{equation}
\tilde g_{3/2}(u) = \int^{u}_0dy\,y^{7/2}\, K(1-y^3)\quad{\rm and}\quad
\tilde h_{3/2}(u) = \tfrac 3 \pi \int^{u}_0dy\,y^{7/2}\, K(y^3)\,.
\end{equation}
In order for the dual boundary theory to be conformal, we set $\bbb=0$ which removes the logarithmic divergence as $u\to0$ arising from $K(1-u^3)$. Approaching the black hole horizon with $u\to 1$, the functions $\tilde g_{3/2}(u)$, $\tilde h_{3/2}(u)$ and $K(1-u^3)$ are all finite, but $K(u^3)$ is logarithmically divergent. Therefore regularity at the horizon requires $\aaa = 8 \alphasource\, \tilde g_{3/2}(1).$\footnote{Numerically, we find $\tilde g_{3/2}(1)= 0.4112$.}
%todd - the last sentence 4 --> 8 + footnote (value got halved)

The result for $\langle \mcO \rangle_T$ given in eq.~\reef{expect} is no longer valid in this special case, \eg substituting $\Delta=3/2$ there yields a vanishing expectation value. Rather in this special case, one has to revisit the holographic renormalization procedure to evaluate the scalar expectation value --- see, \eg \cite{tedX}:
\beq
\langle \mcO \rangle_T = -\frac{L^2}{2\pl^2}\,\left(\frac{r_0}{L^2}\right)^{3/2}  \aaa 
= -\frac{\pi^{7/2}}{18\sqrt{3}}\,\aaa\ C_T\,T^{3/2}\,.
\labell{expect9}
\eeq
Given the profile of the bulk scalar, the conductivity is calculated as described in section \ref{hsigma} and appendix \ref{holosig} and the results differ little from those for nearby values of $\Delta$. 
Hence, \eg $\sigma(\omega)$ remains a smooth function of the conformal dimension in the vicinity of $\Delta=3/2$. We can also use the above profile to evaluate the high-frequency expansion of the conductivity as in section \ref{main}. Here we need the Taylor expansion of $\phi(u)$ near the asymptotic boundary:
\begin{equation}
\phi(u) =  \aaa \, u^{3/2} +  \frac{1}{4}\,\aaa \,u^{9/2} - \frac{16}{27}\,\alphasource\, u^6 + O(u^{15/2})\,. \labell{poll}
\end{equation}
We note that this expansion precisely matches that given in eq.~\reef{scalar series} upon substituting $\Delta=3/2$.
Then from eq.~(\ref{JJperb}), the first few terms in the expansion of the conductivity for $\freq\gg1$ are
\begin{equation}
\begin{split}
\frac{\sigma(i\freq)}{\sigma_\infty} = & 1 + \frac{3\sqrt\pi  }{4}\,  \frac{ \aaa\alphamax }{(2\freq)^{3/2}}
 + \frac{ 945\sqrt\pi   }{256}\,\frac{ \aaa\alphamax }{(2\freq)^{9/2}}  - \frac{1280}{3 } \frac{ \alphasource\alphamax }{(2\freq)^{6}}+\cdots\,.
\end{split}
\end{equation}
Again, the results here precisely matches the expansion in eq.~\reef{asymp} upon substituting $\Delta=3/2$. Let us add that an interesting feature that appears at $\Delta=3/2$ is that when we move away from the critical point in the boundary theory by turning on $\bbb$, the leading term in this expansion is enhanced by a logarithmic factor similar to that in eq.~\reef{wallop}. This extra logarithmic factor arises because with nonvanishing $\bbb$, the boundary expansion \reef{poll} of the bulk scalar contains a new term proportional to $\bbb\,u^{3/2}\log u$. As commented before, special care is required in evaluating the two- and three-point functions when $\Delta=3/2$ \cite{mcfadden}.

\section{$O(\alphamax^{\,2})$ corrections to conductivity} \labell{lock}

We saw in section \ref{freqy} that to first order in $\alphamax$, the leading correction in the high frequency expansion \reef{asymp} of the conductivity appeared at order $1/\freq^\Delta$. Here we extend the perturbative analysis presented in that section to order $\alphamax^{\,2}$ to see how the expansion will be modified at this order. At this order, eq.~(\ref{newde}) yields
\begin{equation}
\big[\partial_z  - \freq^2 \big]A\up2_y = \partial_z A\up0_y\,\phi\, \partial_z \phi  - \partial_z A\up1_y \,\partial_z\phi\,. \labell{newde2}
\end{equation}
Using the Green's function in eq.~\reef{greeen}, we then find 
\begin{equation}
A\up2_y = \int_0^\infty d\tilde z \,G(z,\tilde z)\,\left(
\partial_{\tilde z} A\up0_y\,\phi\, \partial_{\tilde z} \phi  - \partial_{\tilde z} A\up1_y \,\partial_{\tilde z}\phi
\right)
\end{equation}
and taking the limit where we approach the asymptotic boundary, \ie $z\to0$, the derivative of this expression yields
\begin{equation}
\left.\partial_z A\up2_y \right|_{z=0} = \int_0^\infty d\tilde z\, e^{-\freq \tilde z}\left(\freq  e^{-\freq \tilde z}\,\phi\, \partial_{\tilde z} \phi + \partial_{\tilde z}A\up1_y\, \partial_{\tilde z}\phi \right)\,, 
\end{equation}
where we have used the leading order solution \reef{leeed} above.

We only wish to identify the leading correction that this makes to the high frequency expansion \reef{asymp}. For simplicity, we will substitute the profile: $\phi \sim \aaa\, z^\Delta$. Note that we are using $z^\Delta$ rather than $u^\Delta$ here, but these two profiles only differ at order $z^3$ --- see footnote \ref{footy}. We note \emph{en passant} that
our analysis thus applies to the simple ansatz of \cite{katz}. With this scalar profile, we find
\begin{equation}
\begin{split}
\left.\partial_z A\up1_y\right|_{z=0} = &-\frac{\aaa \freq\, \Gamma(\Delta+1)}{(2\freq)^\Delta}\\
\left.\partial_z A\up2_y\right|_{z=0} = &\frac{\aaa^2\,\freq  \left(\Delta\,\Gamma(2\Delta) - 2^\Delta\, \Gamma(\Delta + 1)^2\right)}{(2\freq)^{2\Delta}} 
\end{split}
\end{equation}
Eq.~\reef{JJperb} then yields
\begin{equation}
\frac{\sigma(i \freq)}{\sigma_\infty} = 1 + \frac{\aaa\alphamax\,\Gamma(\Delta+1)}{(2\freq)^\Delta} -\frac{(\aaa\alphamax)^2\left(\Delta\,\Gamma(2\Delta)-2^\Delta\,\Gamma(\Delta+1)^2\right)}{(2\freq)^{2\Delta}} + O(\alphamax^{\,3}) \,,
\end{equation} 
as the first three contributions in the $\alphamax$-expansion. Of course, the first two terms precisely match those found in eq.~\reef{asymp}. We might note that the new $O(\alphamax^{\,2})$ correction implies that the existence of a new primary operator with conformal dimension $2\Delta$. By the reasoning considered in section \ref{finger}, $\join{\mcO^2}$ is the obvious candidate. As a further comment, we observe that the coefficient of the $1/\freq^{2\Delta}$ term vanishes for $\Delta \simeq 2.58$. It would be interesting to better understand the physical significance of this vanishing.

\bibliography{references}

\end{document}

%% file: Witten_phiJJ.pdf_tex
%% Creator: Inkscape 0.91_64bit, www.inkscape.org
%% Accompanies image file 'Witten_phiJJ.pdf' (pdf, eps, ps)
\begingroup%
  \makeatletter%
  \providecommand\color[2][]{%
    \errmessage{(Inkscape) Color is used for the text in Inkscape, but the package 'color.sty' is not loaded}%
    \renewcommand\color[2][]{}%
  }%
  \providecommand\transparent[1]{%
    \errmessage{(Inkscape) Transparency is used (non-zero) for the text in Inkscape, but the package 'transparent.sty' is not loaded}%
    \renewcommand\transparent[1]{}%
  }%
  \providecommand\rotatebox[2]{#2}%
  \ifx\svgwidth\undefined%
    \setlength{\unitlength}{804.08552246bp}%
    \ifx\svgscale\undefined%
      \relax%
    \else%
      \setlength{\unitlength}{\unitlength * \real{\svgscale}}%
    \fi%
  \else%
    \setlength{\unitlength}{\svgwidth}%
  \fi%
  \global\let\svgwidth\undefined%
  \global\let\svgscale\undefined%
  \makeatother%
  \begin{picture}(1,0.7557885)%
    \put(0,0){\includegraphics[width=\unitlength,page=1]{Witten_phiJJ.pdf}}%
    \put(-0.00330344,0.36598662){\color[rgb]{0,0,0}\makebox(0,0)[lb]{\smash{$\mathcal{O}_\Delta$}}}%
    \put(0.7204705,0.72555229){\color[rgb]{0,0,0}\makebox(0,0)[lb]{\smash{$J_\mu$}}}%
    \put(0.7204705,0.00938566){\color[rgb]{0,0,0}\makebox(0,0)[lb]{\smash{$J_\nu$}}}%
    \put(0.64087698,0.24816623){\color[rgb]{0,0,0}\makebox(0,0)[lb]{\smash{$G_{\nu \sigma}$}}}%
    \put(0.64087698,0.4867717){\color[rgb]{0,0,0}\makebox(0,0)[lb]{\smash{$G_{\mu \rho}$}}}%
    \put(0.30959823,0.40605273){\color[rgb]{0,0,0}\makebox(0,0)[lb]{\smash{$K_\Delta$}}}%
  \end{picture}%
\endgroup%

%% file: Witten_phiCC.pdf_tex
%% Creator: Inkscape 0.91_64bit, www.inkscape.org
%% Accompanies image file 'Witten_phiCC.pdf' (pdf, eps, ps)
\begingroup%
  \makeatletter%
  \providecommand\color[2][]{%
    \errmessage{(Inkscape) Color is used for the text in Inkscape, but the package 'color.sty' is not loaded}%
    \renewcommand\color[2][]{}%
  }%
  \providecommand\transparent[1]{%
    \errmessage{(Inkscape) Transparency is used (non-zero) for the text in Inkscape, but the package 'transparent.sty' is not loaded}%
    \renewcommand\transparent[1]{}%
  }%
  \providecommand\rotatebox[2]{#2}%
  \ifx\svgwidth\undefined%
    \setlength{\unitlength}{1044.37302246bp}%
    \ifx\svgscale\undefined%
      \relax%
    \else%
      \setlength{\unitlength}{\unitlength * \real{\svgscale}}%
    \fi%
  \else%
    \setlength{\unitlength}{\svgwidth}%
  \fi%
  \global\let\svgwidth\undefined%
  \global\let\svgscale\undefined%
  \makeatother%
  \begin{picture}(1,0.58189802)%
    \put(0,0){\includegraphics[width=\unitlength,page=1]{Witten_phiCC.pdf}}%
    \put(-0.00254339,0.28178107){\color[rgb]{0,0,0}\makebox(0,0)[lb]{\smash{$\mathcal{O}_\Delta$}}}%
    \put(0.55470592,0.5586185){\color[rgb]{0,0,0}\makebox(0,0)[lb]{\smash{$T_{\rho\sigma}$}}}%
    \put(0.55470592,0.00722623){\color[rgb]{0,0,0}\makebox(0,0)[lb]{\smash{$T_{\mu\nu}$}}}%
    \put(0.48729705,0.19106858){\color[rgb]{0,0,0}\makebox(0,0)[lb]{\smash{$G_{\mu\nu, \alpha\beta}$}}}%
    \put(0.48729705,0.37477613){\color[rgb]{0,0,0}\makebox(0,0)[lb]{\smash{$G_{\rho\sigma, \gamma\delta}$}}}%
    \put(0.23836642,0.31262884){\color[rgb]{0,0,0}\makebox(0,0)[lb]{\smash{$K_\Delta$}}}%
  \end{picture}%
\endgroup%

%% file: paper.bbl
\providecommand{\href}[2]{#2}\begingroup\raggedright\begin{thebibliography}{10}

\bibitem{katz}
E.~Katz, S.~Sachdev, E.~S. Sorensen, and W.~Witczak-Krempa, {\it {Conformal
  field theories at nonzero temperature: Operator product expansions, Monte
  Carlo, and holography}},  {\em Phys. Rev.} {\bf B90} (2014), no.~24 245109,
  [\href{http://arxiv.org/abs/1409.3841}{{\tt arXiv:1409.3841}}].

\bibitem{book}
S.~Sachdev, {\em Quantum Phase Transitions}.
\newblock Cambridge University Press, England, 2~ed., 2011.

\bibitem{fisher1990}
M.~P.~A. Fisher, G.~Grinstein, and S.~M. Girvin, {\it Presence of quantum
  diffusion in two dimensions: {Universal} resistance at the
  superconductor-insulator transition},  {\em Phys. Rev. Lett.} {\bf 64} (Jan.,
  1990) 587.

\bibitem{damle}
K.~{Damle} and S.~{Sachdev}, {\it {Nonzero-temperature transport near quantum
  critical points}},  {\em Phys. Rev. B} {\bf 56} (Oct., 1997) 8714--8733,
  [\href{http://arxiv.org/abs/cond-mat/9}{{\tt cond-mat/9}}].

\bibitem{Maldacena}
J.~M. Maldacena, {\it {The Large N limit of superconformal field theories and
  supergravity}},  {\em Adv.Theor.Math.Phys.} {\bf 2} (1998) 231--252,
  [\href{http://arxiv.org/abs/hep-th/9711200}{{\tt hep-th/9711200}}].

\bibitem{selfdual}
C.~P. Herzog, P.~Kovtun, S.~Sachdev, and D.~T. Son, {\it {Quantum critical
  transport, duality, and M-theory}},  {\em Phys.Rev.} {\bf D75} (2007) 085020,
  [\href{http://arxiv.org/abs/hep-th/0701036}{{\tt hep-th/0701036}}].

\bibitem{Myers:2010pk}
R.~C. Myers, S.~Sachdev, and A.~Singh, {\it {Holographic Quantum Critical
  Transport without Self-Duality}},  {\em Phys. Rev.} {\bf D83} (2011) 066017,
  [\href{http://arxiv.org/abs/1010.0443}{{\tt arXiv:1010.0443}}].

\bibitem{sum-rules}
D.~R. Gulotta, C.~P. Herzog, and M.~Kaminski, {\it {Sum Rules from an Extra
  Dimension}},  {\em JHEP} {\bf 1101} (2011) 148,
  [\href{http://arxiv.org/abs/1010.4806}{{\tt arXiv:1010.4806}}].

\bibitem{ws}
W.~Witczak-Krempa and S.~Sachdev, {\it {The quasi-normal modes of quantum
  criticality}},  {\em Phys.Rev.} {\bf B86} (2012) 235115,
  [\href{http://arxiv.org/abs/1210.4166}{{\tt arXiv:1210.4166}}].

\bibitem{ws2}
W.~{Witczak-Krempa} and S.~{Sachdev}, {\it {Dispersing quasinormal modes in
  (2+1)-dimensional conformal field theories}},  {\em PRB} {\bf 87} (Apr.,
  2013) 155149, [\href{http://arxiv.org/abs/1302.0847}{{\tt arXiv:1302.0847}}].

\bibitem{natphys}
W.~Witczak-Krempa, E.~Sorensen, and S.~Sachdev, {\it {The dynamics of quantum
  criticality via Quantum Monte Carlo and holography}},  {\em Nature Phys.}
  {\bf 10} (2014) 361, [\href{http://arxiv.org/abs/1309.2941}{{\tt
  arXiv:1309.2941}}].

\bibitem{chen}
K.~{Chen}, L.~{Liu}, Y.~{Deng}, L.~{Pollet}, and N.~{Prokof'ev}, {\it
  {Universal Conductivity in a Two-Dimensional Superfluid-to-Insulator Quantum
  Critical System}},  {\em Physical Review Letters} {\bf 112} (Jan., 2014)
  030402, [\href{http://arxiv.org/abs/1309.5635}{{\tt arXiv:1309.5635}}].

\bibitem{will-hd}
W.~{Witczak-Krempa}, {\it {Quantum critical charge response from higher
  derivatives in holography}},  {\em Phys. Rev. B} {\bf 89} (Apr., 2014)
  161114, [\href{http://arxiv.org/abs/1312.3334}{{\tt arXiv:1312.3334}}].

\bibitem{justin2}
J.~R. David and S.~Thakur, {\it {Sum rules and three point functions}},  {\em
  JHEP} {\bf 11} (2012) 038, [\href{http://arxiv.org/abs/1207.3912}{{\tt
  arXiv:1207.3912}}].

\bibitem{willprl}
W.~Witczak-Krempa, {\it {Constraining quantum critical dynamics: 2+1D Ising
  model and beyond}},  {\em Phys. Rev. Lett.} {\bf 114} (2015) 177201,
  [\href{http://arxiv.org/abs/1501.03495}{{\tt arXiv:1501.03495}}].

\bibitem{airport}
A.~Buchel, J.~Escobedo, R.~C. Myers, M.~F. Paulos, A.~Sinha, and M.~Smolkin,
  {\it {Holographic GB gravity in arbitrary dimensions}},  {\em JHEP} {\bf 03}
  (2010) 111, [\href{http://arxiv.org/abs/0911.4257}{{\tt arXiv:0911.4257}}].

\bibitem{Torii:2001pg}
T.~Torii, K.~Maeda, and M.~Narita, {\it {Scalar hair on the black hole in
  asymptotically anti-de Sitter space-time}},  {\em Phys. Rev.} {\bf D64}
  (2001) 044007.

\bibitem{Winstanley:2002jt}
E.~Winstanley, {\it {On the existence of conformally coupled scalar field hair
  for black holes in (anti-)de Sitter space}},  {\em Found. Phys.} {\bf 33}
  (2003) 111--143, [\href{http://arxiv.org/abs/gr-qc/0205092}{{\tt
  gr-qc/0205092}}].

\bibitem{Buchel:2007vy}
A.~Buchel, S.~Deakin, P.~Kerner, and J.~T. Liu, {\it {Thermodynamics of the
  N=2* strongly coupled plasma}},  {\em Nucl. Phys.} {\bf B784} (2007) 72--102,
  [\href{http://arxiv.org/abs/hep-th/0701142}{{\tt hep-th/0701142}}].

\bibitem{Buchel:2013lla}
A.~Buchel, L.~Lehner, R.~C. Myers, and A.~van Niekerk, {\it {Quantum quenches
  of holographic plasmas}},  {\em JHEP} {\bf 05} (2013) 067,
  [\href{http://arxiv.org/abs/1302.2924}{{\tt arXiv:1302.2924}}].

\bibitem{new}
A.~Lucas, R.~C. Myers, T.~Sierens, and W.~Witczak-Krempa, {\it {Modelling
  quantum critical responses using holography: General dimensions and
  observables. In preparation}}, .

\bibitem{Kovtun:2003wp}
P.~Kovtun, D.~T. Son, and A.~O. Starinets, {\it {Holography and hydrodynamics:
  Diffusion on stretched horizons}},  {\em JHEP} {\bf 10} (2003) 064,
  [\href{http://arxiv.org/abs/hep-th/0309213}{{\tt hep-th/0309213}}].

\bibitem{Brigante:2007nu}
M.~Brigante, H.~Liu, R.~C. Myers, S.~Shenker, and S.~Yaida, {\it {Viscosity
  Bound Violation in Higher Derivative Gravity}},  {\em Phys. Rev.} {\bf D77}
  (2008) 126006, [\href{http://arxiv.org/abs/0712.0805}{{\tt
  arXiv:0712.0805}}].

\bibitem{Ritz:2008kh}
A.~Ritz and J.~Ward, {\it {Weyl corrections to holographic conductivity}},
  {\em Phys. Rev.} {\bf D79} (2009) 066003,
  [\href{http://arxiv.org/abs/0811.4195}{{\tt arXiv:0811.4195}}].

\bibitem{kovtun-rev}
P.~{Kovtun}, {\it {Lectures on hydrodynamic fluctuations in relativistic
  theories}},  {\em Journal of Physics A Mathematical General} {\bf 45} (Nov.,
  2012) 3001, [\href{http://arxiv.org/abs/1205.5040}{{\tt arXiv:1205.5040}}].

\bibitem{simon}
S.~Caron-Huot and O.~Saremi, {\it {Hydrodynamic Long-Time tails From Anti de
  Sitter Space}},  {\em JHEP} {\bf 11} (2010) 013,
  [\href{http://arxiv.org/abs/0909.4525}{{\tt arXiv:0909.4525}}].

\bibitem{Son09}
P.~{Romatschke} and D.~T. {Son}, {\it {Spectral sum rules for the quark-gluon
  plasma}},  {\em Phys. Rev. D} {\bf 80} (Sept., 2009) 065021,
  [\href{http://arxiv.org/abs/0903.3946}{{\tt arXiv:0903.3946}}].

\bibitem{justin1}
J.~R. David, S.~Jain, and S.~Thakur, {\it {Shear sum rules at finite chemical
  potential}},  {\em JHEP} {\bf 03} (2012) 074,
  [\href{http://arxiv.org/abs/1109.4072}{{\tt arXiv:1109.4072}}].

\bibitem{Caron-Huot}
S.~{Caron-Huot}, {\it {Asymptotics of thermal spectral functions}},  {\em Phys.
  Rev. D} {\bf 79} (June, 2009) 125009,
  [\href{http://arxiv.org/abs/0903.3958}{{\tt arXiv:0903.3958}}].

\bibitem{el-showk}
S.~{El-Showk} and K.~{Papadodimas}, {\it {Emergent spacetime and holographic
  CFTs}},  {\em Journal of High Energy Physics} {\bf 10} (Oct., 2012) 106,
  [\href{http://arxiv.org/abs/1101.4163}{{\tt arXiv:1101.4163}}].

\bibitem{multi1}
O.~Aharony, M.~Berkooz, and E.~Silverstein, {\it {Multiple trace operators and
  nonlocal string theories}},  {\em JHEP} {\bf 08} (2001) 006,
  [\href{http://arxiv.org/abs/hep-th/0105309}{{\tt hep-th/0105309}}].

\bibitem{multi2}
E.~Witten, {\it {Multitrace operators, boundary conditions, and AdS / CFT
  correspondence}},  \href{http://arxiv.org/abs/hep-th/0112258}{{\tt
  hep-th/0112258}}.

\bibitem{multi3}
M.~Berkooz, A.~Sever, and A.~Shomer, {\it {'Double trace' deformations,
  boundary conditions and space-time singularities}},  {\em JHEP} {\bf 05}
  (2002) 034, [\href{http://arxiv.org/abs/hep-th/0112264}{{\tt
  hep-th/0112264}}].

\bibitem{newmulti}
O.~Aharony, G.~Gur-Ari, and N.~Klinghoffer, {\it {The Holographic Dictionary
  for Beta Functions of Multi-trace Coupling Constants}},  {\em JHEP} {\bf 05}
  (2015) 031, [\href{http://arxiv.org/abs/1501.06664}{{\tt arXiv:1501.06664}}].

\bibitem{revue}
O.~Aharony, S.~S. Gubser, J.~M. Maldacena, H.~Ooguri, and Y.~Oz, {\it {Large N
  field theories, string theory and gravity}},  {\em Phys. Rept.} {\bf 323}
  (2000) 183--386, [\href{http://arxiv.org/abs/hep-th/9905111}{{\tt
  hep-th/9905111}}].

\bibitem{william9}
W.~Witczak-Krempa and J.~Maciejko, {\it {Exact dynamical responses of
  interacting quantum critical points with emergent supersymmetry}},
  \href{http://arxiv.org/abs/1510.06397}{{\tt arXiv:1510.06397}}.

\bibitem{gross}
D.~J. Gross and J.~H. Sloan, {\it {The Quartic Effective Action for the
  Heterotic String}},  {\em Nucl. Phys.} {\bf B291} (1987) 41--89.

\bibitem{Bai2013}
S.~{Bai} and D.-W. {Pang}, {\it {Holographic charge transport in 2+1 dimensions
  at finite N}},  {\em International Journal of Modern Physics A} {\bf 29}
  (Apr., 2014) 1450061, [\href{http://arxiv.org/abs/1312.3351}{{\tt
  arXiv:1312.3351}}].

\bibitem{Myers:2010xs}
R.~C. Myers and A.~Sinha, {\it {Seeing a c-theorem with holography}},  {\em
  Phys. Rev.} {\bf D82} (2010) 046006,
  [\href{http://arxiv.org/abs/1006.1263}{{\tt arXiv:1006.1263}}].

\bibitem{Myers:2010tj}
R.~C. Myers and A.~Sinha, {\it {Holographic c-theorems in arbitrary
  dimensions}},  {\em JHEP} {\bf 01} (2011) 125,
  [\href{http://arxiv.org/abs/1011.5819}{{\tt arXiv:1011.5819}}].

\bibitem{corner1}
P.~Bueno, R.~C. Myers, and W.~Witczak-Krempa, {\it {Universality of corner
  entanglement in conformal field theories}},  {\em Phys. Rev. Lett.} {\bf 115}
  (2015) 021602, [\href{http://arxiv.org/abs/1505.04804}{{\tt
  arXiv:1505.04804}}].

\bibitem{corner2}
P.~Bueno and R.~C. Myers, {\it {Corner contributions to holographic
  entanglement entropy}},  {\em JHEP} {\bf 08} (2015) 068,
  [\href{http://arxiv.org/abs/1505.07842}{{\tt arXiv:1505.07842}}].

\bibitem{JZ}
J.~Z. Simon, {\it {Higher Derivative Lagrangians, Nonlocality, Problems and
  Solutions}},  {\em Phys. Rev.} {\bf D41} (1990) 3720.

\bibitem{Brigante:2008gz}
M.~Brigante, H.~Liu, R.~C. Myers, S.~Shenker, and S.~Yaida, {\it {The Viscosity
  Bound and Causality Violation}},  {\em Phys. Rev. Lett.} {\bf 100} (2008)
  191601, [\href{http://arxiv.org/abs/0802.3318}{{\tt arXiv:0802.3318}}].

\bibitem{Campostrini01}
M.~Campostrini, M.~Hasenbusch, A.~Pelissetto, P.~Rossi, and E.~Vicari, {\it
  Critical behavior of the three-dimensional $\mathrm{XY}$ universality class},
   {\em Phys. Rev. B} {\bf 63} (May, 2001) 214503.

\bibitem{bootstrap}
F.~{Kos}, D.~{Poland}, and D.~{Simmons-Duffin}, {\it {Bootstrapping the O( N )
  vector models}},  {\em Journal of High Energy Physics} {\bf 6} (June, 2014)
  91, [\href{http://arxiv.org/abs/1307.6856}{{\tt arXiv:1307.6856}}].

\bibitem{gazit14}
S.~{Gazit}, D.~{Podolsky}, and A.~{Auerbach}, {\it {Critical Capacitance and
  Charge-Vortex Duality Near the Superfluid-to-Insulator Transition}},  {\em
  Physical Review Letters} {\bf 113} (Dec., 2014) 240601,
  [\href{http://arxiv.org/abs/1407.1055}{{\tt arXiv:1407.1055}}].

\bibitem{Chowdhury:2012km}
D.~Chowdhury, S.~Raju, S.~Sachdev, A.~Singh, and P.~Strack, {\it {Multipoint
  correlators of conformal field theories: implications for quantum critical
  transport}},  {\em Phys. Rev.} {\bf B87} (2013), no.~8 085138,
  [\href{http://arxiv.org/abs/1210.5247}{{\tt arXiv:1210.5247}}].

\bibitem{mcfadden}
A.~{Bzowski}, P.~{McFadden}, and K.~{Skenderis}, {\it {Implications of
  conformal invariance in momentum space}},  {\em Journal of High Energy
  Physics} {\bf 3} (Mar., 2014) 111,
  [\href{http://arxiv.org/abs/1304.7760}{{\tt arXiv:1304.7760}}].

\bibitem{Grads:2007}
I.~S. Gradshteyn and I.~M. Ryzhik, {\em Table of Integrals, Series, and
  Products}.
\newblock Elsevier Inc., Burlington, MA, USA, 2007.

\bibitem{tedX}
H.~Casini, D.~A. Galante, and R.~C. Myers, {\it {Comments on Jacobson's
  "Entanglement equilibrium and the Einstein equation"}},
  \href{http://arxiv.org/abs/1601.00528}{{\tt arXiv:1601.00528}}.

\end{thebibliography}\endgroup
